\documentclass[12pt]{article}
\usepackage[dvips]{graphicx}
\usepackage{amsmath,amssymb}

\begin{document}
\title{ Long-range  correlations of neutrinos in  hadron reactions   and 
neutrino diffraction II:~ neutrino } 
\author{ K. Ishikawa and  Y. Tobita}

\maketitle
\begin{picture}(0,0)(0,0)
 \put(320,230){\makebox{EPHOU-12-008}}
\end{picture}
\begin{center}
  Department of Physics, Faculty of Science, \\
Hokkaido University, Sapporo 060-0810, Japan
\end{center}
\begin{abstract}
In this II, a probability to detect  the neutrino produced in a   
high-energy pion decay   is shown to receive the large finite-size
 correction. 
The neutrino interacts extremely weakly with matters  and 
 is  described with a many-body wave function together with the pion and
 charged lepton. This wave function slowly approaches to an asymptotic form,
which is probed by the neutrino.     The whole process is described by 
an S-matrix of a finite-time interval, which couples with states of
  non-conserving kinetic energy, and the final 
states of  a broad spectrum specific to  a relativistic invariant 
system contribute to  the positive semi-definite correction similar to 
 diffraction of waves through a hole. 
This diffraction component for the neutrino becomes long range and
 stable  under changes of the pion's energy. 
  Moreover, 
it has a universal   form that depends on the absolute neutrino mass.
Thus   a new  method   of measuring the absolute neutrino mass is suggested.
\end{abstract}

\newpage

\section{Introduction}

  The method of computing the finite-size correction developed in I 
\cite{Ishikawa-Tobita I} is
applied to the probability to detect  neutrinos in pion decay. 
   Since a neutrino, charged lepton, and pion are described by a 
many-body wave function that follows  Schr\"{o}dinger  equation,  
the kinetic energy  at a finite time is not a constant and takes a wide 
range of values, as was shown in I.
Consequently  these  waves include broad spectrum and show a diffraction 
phenomenon that is non-uniform in space-time.
Since the speed of light is an accumulation
point of the relativistic waves, a two-point correlation function has 
  the light-cone singularity and the probability to 
detect the neutrino is subject to a 
 finite-size correction. The neutrinos  are 
very light and propagate with almost the light speed, hence 
the correction becomes unusual and  depends, in fact, on  
absolute neutrino masses which are not found from  oscillation
experiments. 
Thus the  correction becomes extremely large in its magnitude and size
and  has universal properties for the neutrino.

Tritium beta decays \cite{Tritium}~have
been used for determining the 
absolute value but an existing upper bound for an effective electron 
neutrino mass-squared  is of the order  of $  2 ~[\text{eV}^2/c^4]$ and the mass 
is $ 0.3 - 1.3 ~ [\text{eV}/c^2]$ from cosmological
observations~\cite{WMAP-neutrino}. 
In these  neutrino experiments, higher precision and more statistics
have been achieved and will be improved more.

Weak decay processes have been studied using an S-matrix of plane waves 
with asymptotic 
boundary conditions, 
where 
particles in the initial and final states are regarded as free waves
without correlations \cite{Sakai-1949,Jack,Ruderman,LSZ,Low}. 
A decay rate, average life time, and 
various  distributions of charged leptons  have been computed, and 
perfect agreements with experiments have been obtained
\cite{Anderson}. These facts proved that the standard theory of electro-weak
interaction is correct.

Neutrinos   have almost the light speed and are   detected at a distant 
position, hence they are similar to an  
electrostatic potential of a moving body, which has a finite-size
correction in a form of a retarded potential. 
Due to its extremely small mass, the probability to detect the neutrino 
may  depend on the distance  between a source and  observation positions. 
To study position-dependent probabilities, the
standard S-matrix of plane waves that satisfies a boundary condition at $t=\pm\infty$, which gives  the values at the asymptotic
regions,   is
useless.    An S-matrix that satisfies boundary conditions  at a 
finite-time interval T, $S[\text T]$, and has a position dependence
is appropriate. Because boundary
conditions for $S[\text T]$ are different from those of $S[\infty]$, it
reveals different properties form those of  $S[\infty]$.
We compute   the 
finite-size corrections
of transition probabilities with $S[\text T]$ expressed by wave packets.
The wave packets vanish at a position ${\vec x}$ if the
 distance between ${\vec x}$ and the center position ${\vec X}$, 
$|{\vec x}-{\vec X}| \rightarrow \infty$, and satisfy the boundary 
conditions of the experiments \cite{LSZ,Low}. Hence they are appropriate to study 
the finite-size effect and  are used here \cite{Ishikawa-Shimomura,Ishikawa-Tobita-ptp,Ishikawa-Tobita}.

 A pion is produced in a proton reaction first and decays next. The whole process is
expressed in Fig.\,1 of I. 
The first reaction  caused by strong interactions was studied in I  
and the second reaction  caused by weak interactions is studied in II. 
From  I, a particle in the final state retains the wave nature  in the
 region of $r \leq l_0$, where $r$ is a distance between the initial
 sates and final state and  a coherence length  $l_0$ in a high-momentum
 region, $|\vec{p}| \gg m$,  is given by
\begin{eqnarray}
l_0=\left({2|\vec{p}\,| \hbar c \over m^2}\right),
\end{eqnarray}
where $m$ is the particle's mass and becomes for a
neutrino of mass $1$ [\text{eV}/$c^2$] and energy $1$ [\text{GeV}]
\begin{eqnarray}
l_0^{neutrino}={2 \hbar c \over 1^2}\times 10^{18}[\text{GeV}^{-1}]=  10^{2}-10^{3}
 [\text{m}],\label{neutrino-coherence }
\end{eqnarray} 
Thus $l_0$ of neutrino are macroscopic lengths.
Hence a measurement  of neutrino  process at a near-detector 
region may  depend on a  distance  
$|{\vec X}-{\vec X}^{(i)}|$, where ${\vec X}$ and ${\vec X}^{(i)}$ are
the positions of the nucleus in the detector and target.

Detecting  neutrinos at ${\vec X}$ of a distance 
 ${|{\vec X}-{\vec X}^{(i)}| \gg l_0}$ is studied in a usual manner. 
Since the neutrino in this
region is expressed by a plane wave that behaves like a free
particle, the neutrino's number and flux behave like those of classical 
particles and the production and detection of the neutrino are treated
separately. The neutrino flux is determined  by its distribution
functions of the  number and velocity that are determined by the decay
process   and is  used for calculation of the  scattering processes.

 Measurements  in
${|{\vec X}-{\vec X}^{(i)}| \leq l_0}$, on the other hand, are not treated by 
particle dynamics  but by wave dynamics.  The entire processes should
be taken into account and be studied either by  
time-dependent Schr\"{o}dinger  equations or by position-dependent amplitudes 
defined with wave packets. The wave packets are necessary to measure 
the quantities in this region. Since the particles
are identified  based on  signals generated  by their 
interactions with  a nucleus, atom, or larger system of
matter, this  unit of detector has a finite size  and is expressed by 
wave packets.     
Using
wave packet representations, the amplitude of detecting  neutrinos 
in a   pion decay  is
computed{\samepage\footnote{The general arguments about
the wave packet scattering are given in
\cite{Goldberger,newton,taylor,Sasakawa}. In these works, large wave
packets were  considered and small wave packets and finite-size
corrections are studied here.}}. 
The probability to detect the neutrino depends
on the distance $|{\vec X}-{\vec X}^{(i)}|$,  hence  a 
naive neutrino flux is not defined uniquely. The uniform 
flux defined by the classical particles is necessarily constant and is 
not used here. It might be reasonable to define the flux here based on the
number of events that the neutrino gives rise to.  The probabilities
can be computed relatively easily with a
set of simple  
wave packets such as Gaussian wave packets 
and are equivalent in more general wave packets. 
Although they    depend upon not only  the distance but also the 
wave packet size, they    have  universal properties. So it would make
sense to define  a  neutrino flux from the probability computed with
the Gaussian wave packet. Thus we define the neutrino flux 
in this region from the probability to detect the neutrino. 
 The idea is similar 
to define the electric field from a force that a point particle receives. 
The finite-size   effects are actually large in 
the processes of detecting neutrinos at a macroscopic distance.
Thus 
the corrections are important when the theoretical values are compared with the
experimental values.

\begin{figure}[t]
 \includegraphics[scale=0.3]{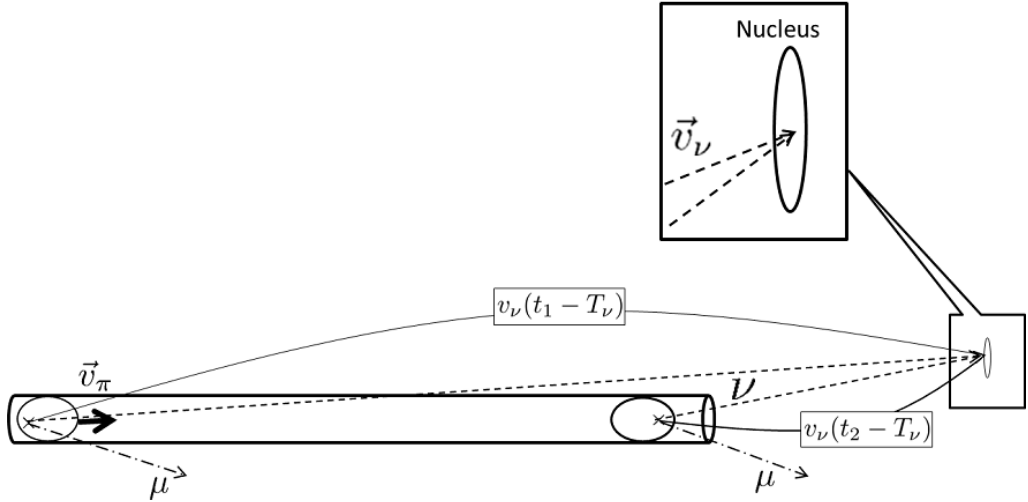}
\caption{The geometry of the pion decay region and the neutrino
 detector. The neutrino is produced in the pion decay and is detected.
Since the decay position is not fixed to one value but is arbitrary, the
 amplitude to detect the neutrino is the overlap between the superposed 
initial wave
 and a final state expressed by the wave packet of the small size.
 The neutrino observed by the detector at $T_{\nu}$  shows an
 interference pattern.  }
\label{fig:geo}
\end{figure}

It is instructive to see how a single neutrino interference is observed. A neutrino 
produced in a pion decay propagates a finite distance before it is
detected.  
A position where a neutrino is produced  varies,  so a neutrino wave 
at the detector is a  superposition of those waves
that are produced at different space-time positions. If these space-time
positions are inside of  one pion, as in Fig.\,$\ref{fig:geo}$, the
waves keep  their coherence and  reveal  interference patterns. A
  condition for the interference phenomenon to occur, for  one
  dimensional motion of  
 the pion that keeps a     coherence within  the region
 $\sqrt{\sigma_{\pi}}$ 
and  a 
 velocity   $\vec{v}_{\pi}$ is obtained in the following manner. Let a
 neutrino be produced either at  
time $t_1$ or $t_2$  from the  pion prepared  at $\text{T}_{\pi}$ and travel
for some period and 
be finally  detected  at $\text{T}_{\nu}$,   
then the waves  overlap if 
\begin{eqnarray}
|(c(\text{T}_{\nu}-t_1)+v_{\pi}(t_1-\text{T}_{\pi}))-(c(\text{T}_{\nu}-t_2)+v_{\pi}(t_2-\text{T}_{\pi}))|
 \leq \sqrt{ \sigma_{\pi}},
\label{coherence-condition}
\end{eqnarray}
is met, where the speed of light
 is used for the speed of neutrino  $v_{\nu}=c$.
So  Eq.\,$(\ref{coherence-condition})$ is one of the necessary conditions for the 
neutrino interference to occur in the one-dimensional space. 
For a plane wave of pion $\sigma_{\pi}=\infty$ and the above condition
is satisfied. For a high-energy pion of a finite $\sigma_{\pi}$, its
 speed is close to the speed of light  and the 
left hand side of Eq.\,$(\ref{coherence-condition})$ becomes 
$c(m_{\pi}^2/2E_{\pi}^2)(t_1-t_2)$. Hence this
condition Eq.\,$(\ref{coherence-condition})$ is written in the form, 
 $c(t_1-t_2) \leq 
\sqrt{\sigma_{\pi}}(2E_{\pi}^2/m_{\pi}^2)$.
 When this length $c(t_1-t_2)$ is a macroscopic size, the
interference  phenomenon  occurs at a macroscopic length. 
We  estimated the 
lengths of these particles 
in Appendix  of I and confirmed  that this condition in three-dimensional space 
is fulfilled  in a macroscopic
distance.  

An amplitude to detect a neutrino 
is  an  integral of a  product of wave functions of a  pion,
muon, and neutrino and  a velocity of   a space-time position of
transition of  the pion to  the muon  has a component of  the light
velocity and the neutrino has almost the light velocity. Hence they
overlap in wide space-time area  and  the amplitude and probability 
to detect the neutrino get a contribution  of 
  a large interference effect.  
A neutrino flux becomes distinct from the naive value obtained 
from the standard method of using an incoherent probability.

This paper is organized in the following manner. 
In section 2,   we
study  an amplitude of detecting a neutrino  in a  pion decay process 
and  compute a position-dependent probability 
in section 3.
For a rigorous calculation of 
the position-dependent probability, a correlation  function is
introduced.  Using an expression with the correlation function and its
singular structure at a light-cone region,  the finite-size correction
is computed  in section 4. Implication to neutrino experiments, features
of the finite-size corrections, and 
summary and prospects 
are given in section 5, 6, and 7. 


\section{Position-dependent amplitude of neutrino }
  Position-dependent probabilities of detecting a neutrino and a charged
  lepton  
in a pion decay process 
are computed  with a  Schr\"{o}dinger  equation at a finite time $t$ and an
  S-matrix of a finite-time interval T. They have various properties
  different from those at the infinite time. Especially the final states in
  ultraviolet energy regions couple and give a finite   and  universal
contribution. 

A neutrino or a charged lepton  is detected 
with a detector that is 
located in a macroscopic distance from the position of the initial
pion. Since their properties are determined from their 
probabilities in experiments,  the whole
process is studied. The process  is described with a many-body wave 
function of the pion and decay products. The neutrino and charged lepton
in the final state  is highly correlated
with  the pion in the initial state and have a kinetic energy 
different from the initial value at a finite $t$ due to a finite
interaction energy. Hence the many-body
state retains wave natures and  a finite-size  
correction to the probability to detect a neutrino 
is expected. The
scattering matrix of the finite-time interval  $S[{\text T}]$ is
applied and  a finite-size correction is computed    using the 
wave packet representation. We find   that the  position-dependent 
probability  is computed in a unique manner and reveals  unusual
finite-size
 correction which is understood
as neutrino diffraction.    
\subsection{Leptonic decay of the pion }
A leptonic decay of a pion is described with  the weak Hamiltonian  
\begin{eqnarray}
& &H_{1}=g \int d{\vec x}\, {\partial_{\mu}
 }\varphi(x)J_{V-A}^{\mu}(x)=-igm_{\mu}\int d{\vec x} \, \varphi(x)J_5(x),
\label{weak-hamiltonian}\nonumber\\
& &J_{V-A}^{\mu}(x)=\bar
 \mu(x)\gamma^{\mu}(1-\gamma_5)\nu(x),J_5(x)=\bar
 \mu(x)(1-\gamma_5)\nu(x),
\end{eqnarray}
where ${\varphi(x)}$, $\mu(x)$, and $\nu(x)$ are the pion field, muon
field, and neutrino field \cite{J-J}. In the above equations, the
interaction Hamiltonian was  expressed in a form of a  derivative of 
the pion field.   
$g$ is the coupling
strength, $J_{V-A}^{\mu}(x)$, and $J_5(x)$ are a
leptonic charged $V-A$ current, and a leptonic  pseudoscalar. The
coupling strength is expressed with Fermi coupling $G_F$ and a pion decay 
constant $f_\pi$,
\begin{eqnarray} 
g={G_F \over \sqrt 2}f_{\pi}.
\end{eqnarray}
An expression of the weak Hamiltonian in term of   $(V-A)\times(V-A)$ charged current interaction gives
the equivalent results to a muon  mode. In section 6, an electron mode
will be studied.  

A pion decay process is described by   a    Schr\"{o}dinger equation
\begin{eqnarray}
i\hbar {\partial \over \partial t}|\Psi(t)\rangle= H
 |\Psi(t)\rangle,\ H=H_0+H_1,
\end{eqnarray}   
 where $H_0$ stands for the free Hamiltonian. The solution is  
\begin{eqnarray}
 |\Psi(t) \rangle = e^{-i\frac{H}{\hbar}t }|\Psi(0) \rangle,
\end{eqnarray}
and satisfies 
\begin{eqnarray}
& & i\hbar {\partial \over \partial t }\langle \Psi(t)|H|\Psi(t) \rangle =
  \langle \Psi(t)|[H,H]|\Psi(t) \rangle=0, \\
& &i\hbar {\partial \over \partial t }\langle \Psi(t)|H_0|\Psi(t) \rangle
 = \langle \Psi(t)|[H_0,H]|\Psi(t) \rangle \neq 0.
\end{eqnarray}
The total energy is conserved but the kinetic energy is not 
conserved except the case 
\begin{eqnarray}
H_0  |\Psi(t) \rangle =E |\Psi(t) \rangle.
\end{eqnarray}

A wave function of   
the initial condition 
\begin{align}
 |\Psi(t) \rangle|_{t=\text{T}_{\pi}}=|\text {pion} (t)\rangle|_{t=\text{T}_{\pi}}=e^{-i\frac{E_{\pi}}
{\hbar} \text{T}_{\pi}}|{\vec p}_{\pi},\text{T}_{\pi} \rangle,
\end{align}
evolves with 
\begin{eqnarray}
 |\Psi(t) \rangle = e^{-i\frac{H}{\hbar}t}|\Psi(0) \rangle,
\end{eqnarray}
and  is studied hereafter. A solution in the first order in $H_1$ is 
a superposition of a pion state
 and a muon and neutrino state  
\begin{eqnarray}
& &|\Psi(t) \rangle= |\text {pion} (t)\rangle+|\text {muon,~neutrino}(t)
 \rangle\label{state-vector},
\end{eqnarray}
 and each component is expressed as 
\begin{align}
&|\text{pion} (t) \rangle = a(t)  |\text {pion},{\vec p}_{\pi} (t)\rangle, \\
&|\text {muon,~neutrino}(t) \rangle =\int_{\text{T}_{\pi}}^{t} {d t' \over i\hbar}H_1(t') |\text{pion}
 (t')\rangle, \label{lowest-order mu-nu state}
\end{align}
where $a(t)=1+O(g^2)$.   
The wave function Eq.\,$(\ref{lowest-order mu-nu state} )$ is a superposition of states and is written as 
\begin{align}
&|\text {muon,~neutrino}(t) \rangle= g  e^{-i\frac{E_{\pi}}{\hbar}t}\int
  d{\vec p}_{\mu} d{\vec p}_{\nu}\sqrt{m_{\mu} m_{\nu}
 \over E_{\mu}({\vec p}_{\mu})E_{\nu}({\vec p}_{\nu})} \nonumber\\
&\times \frac{e^{-i{\omega t}/ \hbar}-1}{\omega}\delta^{(3)}({\vec p}_{\pi}-{\vec
 p}_{\mu}-{\vec p}_{\nu})\times ({p_{\pi}})_{\mu}\bar \mu({\vec
 p}_{\mu})\gamma^{\mu}(1-\gamma_5)\nu({\vec p}_{\nu})|{\vec
 p}_{\mu},{\vec p}_{\nu} \rangle\label{wave-function-t},
\end{align}
where $\omega =E_{\mu}+E_{\nu}-E_{\pi}$ and   $|{\vec
 p}_{\mu},{\vec p}_{\nu} \rangle $ is a two-particle state composed of 
the muon
 and neutrino of momenta $ {\vec
 p}_{\mu}$ and  $\ {\vec p}_{\nu} $.

At the  infinite t, the oscillation function is approximated with
\begin{eqnarray}
\frac{e^{-i{\omega t}/\hbar}-1}{\omega}=-2\pi i\delta(\omega),
\label{energy-conservation}
\end{eqnarray}
and the two particle state has the energy $E_{\pi}$
\begin{align}
&|\text {muon, neutrino} (t)\rangle= -i g  e^{-i\frac{E_{\pi}}{\hbar}t}\int
  d{\vec p}_{\mu} d{\vec p}_{\nu} \sqrt{m_{\mu} m_{\nu} \over E_{\mu}({\vec p}_{\mu})E_{\nu}({\vec p}_{\nu})}\\
& \times(2\pi)\delta^{(4)}(p_{\pi}-p_{\mu} -p_{\nu}) 
\times {(p_{\pi})}_{\mu}\bar \mu({\vec
 p}_{\mu})\gamma^{\mu}(1-\gamma_5)\nu({\vec p}_{\nu})|{\vec
 p}_{\mu},{\vec p}_{\nu} \rangle.  \nonumber
\end{align} 
The norm of this state  is proportional to  $\text{T}$ and is given by, 
\begin{align}
&\langle   \text {muon,~neutrino} (\text{T})|\text
 {muon,~neutrino}(\text{T}) \rangle =\Gamma \text{T},
\end{align}
where $\Gamma$ is the average decay rate \cite{Dirac,Schiff-golden}.
A neutrino and muon exist with  a same probability that is computed  from 
the norm of wave function. 
We should bear in mind
that this is the probability that the decay products exist and a
connection  with an observation is not clear. 
Since an observation of a neutrino is made with a detector that is 
located at a different position and a distance between
 positions of the pion and  detector  is
usually large, the probability to detect the neutrino  
at the detector is defined according a geometry of the experiments with
the wave function at a finite t, which is affected by a finite time and 
retarded effect.

At a  finite t, the oscillation function is broad in $\omega$
 and Eq.\,$(\ref{energy-conservation})$ does not hold
\begin{eqnarray}
{e^{-i{\omega t}/\hbar}-1 \over \omega} \neq -2\pi i\delta(\omega),
\end{eqnarray}
and the two particle state Eq.\,$(\ref{wave-function-t})$ has the 
continuous kinetic energy $E_{\mu}+E_{\nu}$.
 The wave  of  the entire process in this region is described by 
\begin{eqnarray}
|\text { pion}(t) \rangle + |\text {muon, neutrino} (t)\rangle.
\end{eqnarray}
Because the two particle state composed of the muon and neutrino at 
a finite $t$ has continuous kinetic energy, this state is  different from two
free  particles, which  has a constant kinetic energy, and is similar to 
the waves around a potential of finite energy. 
 The wave retains  a wave
nature that varies with a time and the 
probability to
detect a particle in the final state should reflect this wave nature.   
The S-matrix  of observing the particle at a finite time $t$ 
 satisfies the boundary condition at a finite-time
interval T, hence this is different from the standard S-matrix
$S[\infty]$.  We write this as $S[\text T]$. $S[\infty]$ commutes with
$H_0$ but $S[\text T]$ satisfies,  from the proof given in I,
\begin{eqnarray}
[S[\text T],H_0] =i\left({\partial \over
 \partial \text{T}}\Omega_{-}(\text{T})\right)^{\dagger}  \Omega_{+}(\text{T})-i\Omega_{-}^{\dagger}(\text{T}){\partial \over \partial\text{T}
 }\Omega_{+}(\text{T}).
\label{commutation-relation-S(T)  }
\end{eqnarray}
Hence the states of non-conserving kinetic energy couple.   
The neutrino is measured from its collisions with nucleus in a detector,
hence  the wave function of this final state in a scattering process
 is a   wave packet of the
nuclear size.  Using the wave packets, we express  $S[\text T]$ 
following  a reduction formula of
\cite{LSZ}. This  scattering amplitude  depends  on the
position as well as the momentum.
 The physics of neutrinos
in this  area  is studied with the complete set of 
wave packets \cite{Ishikawa-Shimomura}.  
  Since all the informations of the
 wave function $|\text {muon,~neutrino}(t)
\rangle$ are included in  matrix elements, as far as a
complete set of $|\mu ,\nu \rangle $ is used,  these matrix
elements have a complete information.  Wave packets have been applied to
study neutrino flavor  oscillations before
\cite{Kayser,Giunti,Nussinov,Kiers,Stodolsky,Lipkin,Akhmedov,Asahara},
whereas  $S[\text T]$ and the probabilities that depend on T are studied
in the present paper. The finite-size correction  appears in  systems of
one flavor  and  many flavors.

Using  a  wave function of an initial pion located 
at a position ${\vec
X}_\pi$, a neutrino at a  position ${\vec
X}_{\nu}$ and a  muon,  we express the amplitude to detect a neutrino at
a finite distance.  
Since a muon  is not observed and all the states are summed over,
that is expressed with a  plane
wave. This transition amplitude is  
\begin{eqnarray}
T=\int d^4x \, \langle {\mu},{\nu}   |H_{1}(x)| \pi \rangle,
\end{eqnarray}
where the pion and neutrino states are described in terms of    wave packets 
of central values of momenta and  coordinates and the widths
in the form   
\begin{eqnarray}
|\pi \rangle=   | {\vec p}_{\pi},{\vec X}_{\pi},\text{T}_{\pi}  \rangle,\ 
|\mu ,\nu \rangle=   |\mu,{\vec p}_{\mu};\nu,{\vec p}_{\nu},{\vec X}_{\nu},\text{T}_{\nu}          \rangle.
\end{eqnarray}
The pion is also expressed in term of  a wave packet here for a
completeness,
and the   particle states are defined with  the matrix elements, 
\begin{align}
&\langle 0|\varphi(x)|{\vec p}_{\pi},{\vec X}_{\pi},\text{T}_{\pi}  \rangle
= N_{\pi}\rho_{\pi}\int d{\vec k_\pi} \, e^{-{\sigma_{\pi} \over 2}({\vec
 k}_\pi-{\vec p}_{\pi})^2 -i\left(E({\vec
 k}_\pi)(t-\text{T}_{\pi}) - {\vec
 k}_\pi\cdot({\vec x}-{\vec X}_{\pi})\right)}
\nonumber \\
&\approx 
N_{\pi}\rho_{\pi}\left({2\pi \over \sigma_{\pi}}\right)^{\frac{3}{2}}e^{-{1 \over 2 \sigma_{\pi} }\left({\vec x}-{\vec X_{\pi}}-{\vec v}_{\pi}(t-\text{T}_{\pi})\right)^2-i\left(E({\vec p}_{\pi})(t-\text{T}_{\pi}) - {\vec
 p}_{\pi}\cdot({\vec x}-{\vec X}_{\pi})\right)} ,
\label{pion-wf}\\
&\langle \mu,{\vec p}_{\mu};\nu,{\vec p}_{\nu},{\vec X}_{\nu},\text{T}_{\nu}|\bar
 \mu(x) (1 - \gamma_5) \nu(x) |0 \rangle \nonumber\\
&= \frac{N_{\nu}}{(2\pi)^{{3}}}\int d{\vec k}_{\nu}e^{-{\sigma_{\nu} \over 2}({\vec k}_{\nu}-{\vec p}_{\nu})^2}\left({m_{\mu}
 \over E({\vec p}_{\mu})}\right)^{\frac{1}{2}}\left({m_{\nu} \over E({\vec
 k}_{\nu})}\right)^{\frac{1}{2}}  \bar u({\vec p}_{\mu}) (1 -\gamma_5) \nu
 ({\vec k}_{\nu})\nonumber\\
 &\times e^{i\left(E({\vec p}_{\mu})t-{\vec
 p}_{\mu}\cdot{\vec x}\right)+ i\left(E({\vec k}_{\nu})(t-\text{T}_{\nu})-{\vec
 k}_{\nu}\cdot({\vec x}-{\vec X}_{\nu})\right)},\label{mu-nu-wf}
\end{align}
where
\begin{align}
& N_{\pi} = \left(\frac{\sigma_{\pi}}{\pi}\right)^{\frac{3}{4}},~N_{\nu}
 = \left(\frac{\sigma_{\nu}}{\pi}\right)^{\frac{3}{4}},\rho_{\pi}=
\left(\frac{1}{2E_{\pi}(2\pi)^3}\right)^{\frac{1}{2}}.
\end{align}

In the above equation,  the pion's life time is ignored.
The sizes, $\sigma_{\pi}$ and
$\sigma_{\nu}$,  in Eqs.\,$(\ref{pion-wf})$ and 
  $(\ref{mu-nu-wf})$   are sizes of the pion wave packet and  of the 
neutrino wave packet. Minimum wave packets are used in majorities  of the
present  paper but non-minimum wave packets are studied and it is shown
that  main  results are the same.\footnote{For  non-minimal wave packets which  have larger uncertainties, 
Hermite  polynomials of ${\vec k}_{\nu}-{\vec p}_{\nu}$ are multiplied to
the right-hand side of
Eq.\,$(\ref{mu-nu-wf})$ . A completeness of the wave packet states is
also satisfied for the
non-minimum case \cite{Ishikawa-Shimomura} and a total probability and  
a probability at a finite distance and time are  the same as far as
the wave packet is almost symmetric. This condition is guaranteed in
the high energy neutrino which this paper studies, but may not be so 
in the low energy neutrino.  We will 
confirm in the text and 
appendix that the universal long-range correlation of the 
present work is independent of  the wave 
packet shape as far as the wave packet is invariant under the time
inversions. Low energy neutrinos such as solar or reactor neutrinos will
be presented in the next paper.} From the result of Appendix of I,
the coherence length of the pion produced in proton nucleon collision is
short hence the pion is in the asymptotic
region in a wide area. The pion  behaves like a free particle there and can
be  treated   with  a plane wave or a wave packet.  A mean free 
path of the  pion was estimated in the Appendix of I and is
used as a size of wave packet of the initial pion in a decay process in
II.

From   Appendix of I, a size of pion wave packet is of the 
order of $0.5-1.0$\,[m] and  a momentum spreading is small. So ${\vec
k}_{\pi}$ is integrated easily, and is 
replaced with  its  central value ${\vec p}_\pi$ and the final 
expression  of Eq.\,$(\ref{pion-wf})$ is obtained. 
For  neutrinos, a size of wave packet is the nucleus. Hence to study
neutrino interferences, we use the  nuclear size for $\sigma_{\nu}$. 

The
amplitude $T$ for  one pion to decay into a neutrino and a muon  is
written in the form
\begin{eqnarray}
T =  \int d t d{\vec x}\, T(t,{\vec x}),\ T(t,{\vec x})=\int  d{\vec
 k}_{\nu}T(t,{\vec x},k_{\nu}),
\label{amplitude}
\end{eqnarray}
where
\begin{align}
 T(t,{\vec x},k_{\nu}) &=ig m_{\mu} N'' 
e^{-{1 \over 2 \sigma_{\pi} }\left({\vec x}-{\vec X_{\pi}}-{\vec v}_{\pi}(t-\text{T}_{\pi})\right)^2
-i\left(E({\vec p}_{\pi})(t-\text{T}_{\pi}) - {\vec
 p}_{\pi}\cdot({\vec x}-{\vec X}_{\pi})\right) + {i\left(E(\vec{p}_{\mu})t-{\vec p}_{\mu}\cdot{\vec
		x}\right)}}  \nonumber \\
 &\times \bar{u}({\vec p}_{\mu}) (1 - \gamma_5)
 \nu({\vec k}_{\nu})e^{i\left(E({\vec k}_{\nu})(t-\text{T}_{\nu})-{\vec
			 k}_{\nu}\cdot({\vec x}-{\vec X}_{\nu})\right)
 -\frac{\sigma_{\nu}}{2}({\vec k}_{\nu}-{\vec p}_{\nu})^2}, \nonumber \\
 N''&=N_{\pi}N_{\nu}\left({2\pi \over \sigma_{\pi}}\right)^{\frac{3}{2}}N_0,~N_0= \left(\frac{m_{\mu}m_{\nu}}{
 E_{\mu}E_{\nu}}\right)^{\frac{1}{2}}\times\frac{\rho_{\pi}}{(2\pi)^3}.
\end{align}
 $T(t,{\vec x},k_{\nu})$ depends upon the 
coordinates $(t,{\vec x})$ explicitly 
and is  not invariant  under the translation.  
      
The amplitude Eq.\,($\ref{amplitude}$) has the term from the region of
${\vec k}_{\nu} \approx {\vec p}_{\nu}$,
which  is computed with the S-matrix of plane waves, and another term from 
the region of
${\vec k}_{\nu} \neq {\vec p}_{\nu}$, in which  the energy 
is not conserved. The latter   term  is not computed with the 
S-matrix of plane waves
and is the finite-size correction  
that vanishes at $ \text{T}
 \rightarrow \infty$. We compute both with $S[\text{T}]$ in the following.

\subsection{Wave of observed neutrino : small angular velocity of a
  center motion }
Integrating  the neutrino momentum in Eq.\,$(\ref{amplitude})$ 
with the  Gaussian integral, we   find a  coordinate representation of
the neutrino wave function. 

For a not so large $|t-\text{T}_{\nu}|$ region, ${\vec k}_{\nu}$ is integrated 
around the central momentum ${\vec p}_{\nu}$, and  
the integrand $T(t,{\vec x})$ becomes,
 \begin{eqnarray}
T(t,{\vec x})&=&igm_{\mu}\tilde N   e^{-{1 \over 2 \sigma_{\pi} }\left({\vec x}-{\vec X_{\pi}}-{\vec v}_{\pi}(t-\text{T}_{\pi})\right)^2
-iE({\vec p}_{\pi})(t-\text{T}_{\pi}) +i {\vec
 p}_{\pi}\cdot({\vec x}-{\vec X}_{\pi})+iE({\vec p}_{\mu})t-i{\vec
 p}_{\mu}\cdot{\vec x}}\nonumber\label{neutrino-position-amplitude}
\\
& &\times\bar u({\vec p}_{\mu}) (1 - \gamma_5) \nu ({\vec p}_{\nu})
e^{i\phi(x)}  e^{-{1 \over 2\sigma_{\nu} }({\vec x}-{\vec X}_{\nu} -{\vec
v}_{\nu}(t-\text{T}_{\nu}))^2},
\end{eqnarray}
where $\tilde N$ is a  normalization factor, ${\vec v}_{\nu}$ is a
neutrino velocity, and $\phi$ is a  phase of neutrino
wave function. They are given in the form  
\begin{eqnarray}
& &\tilde N=N_{\pi} N_{\nu}\left(\frac{2\pi}{\sigma_{\pi}}\right)^{\frac{3}{2}}\left(\frac{2\pi}{\sigma_{\nu}}\right)^{\frac{3}{2}}N_0,\\
& &\phi(x)=E({\vec
 p}_{\nu})(t-\text{T}_{\nu})-{\vec  p}_{\nu}\!\cdot\!({\vec x}-{\vec X}_{\nu}).
\label{neutrino-phase} 
\end{eqnarray}

The neutrino wave function evolves with time in a specific manner. At
$t=\text{T}_{\nu}$, the wave is given in the form  
\begin{eqnarray}
\psi_{\nu}(\text{T}_{\nu},{\vec x})=e^{i\phi(x)-{1 \over 2\sigma_{\nu} }({\vec x}-{\vec X}_{\nu})^2},
\end{eqnarray}
which is localized around the position ${\vec X}_{\nu}$ and has the
phase $\phi=-{\vec p}_{\nu}\cdot({\vec x}-{\vec X}_{\nu})$. At a time $t <\text{T}_{\nu}$, the wave 
function becomes  
\begin{eqnarray}
\psi_{\nu}(t,{\vec x})=e^{i\phi(x)-{1 \over 2\sigma_{\nu} }({\vec
 x}-{\vec X}_{\nu}-{\vec v}_{\nu}(t-\text{T}_{\nu}))^2},
\end{eqnarray}
which is localized around the position
\begin{eqnarray}
{\vec x}_G={\vec X}_{\nu}+{\vec v}_{\nu}(t-\text{T}_{\nu}),
\label{center-coordinate}
\end{eqnarray} 
and has the phase $\phi(x)$.  
Now this  phase  $\phi(x)$ is written at a  position ${\vec r}={\vec
x}-{\vec x}_G$ in the form, 
\begin{eqnarray}
\phi(x)=\bar \phi_G+\phi({\vec r}),
\end{eqnarray}
where 
\begin{eqnarray}
\bar \phi_G
\label{phase}
&=&E({\vec p}_{\nu})(t-\text{T}_{\nu})-{\vec p}_{\nu}\cdot{\vec
 v}_{\nu}(t-\text{T}_{\nu})\\
&=&{E^{2}_{\nu}(\vec p_{\nu})-{\vec p}_{\nu}^{~2} \over E_{\nu}({\vec
 p}_{\nu})}(t-\text{T}_{\nu})= {m_{\nu}^2 \over E_{\nu}(\vec{p}_{\nu})}
 (t-\text{T}_{\nu}),\nonumber \\
\phi({\vec r})&=&-{\vec p}\cdot{\vec r}.
\end{eqnarray}
A  phase at the center,  $\bar \phi_G$, has a typical form of the relativistic
particle and is   proportional to the  mass squared and
inversely proportional to
the  energy. Since the position is moving with the time, the phase from
both components are cancelled     and becomes extremely small.

When the position is moving with the light velocity in the parallel
direction of the momentum ${\vec p}_{\nu}$ instead of Eq.\,($\ref{center-coordinate}$) 
\begin{eqnarray}
{\vec x}={\vec X}_{\nu}+{\vec c}(t-\text{T}_{\nu}),\ |{\vec c}\,|=1,
\end{eqnarray}
the phase  is given by
\begin{eqnarray}
\bar \phi_c(t - \text{T}_\nu)=E({\vec p}_{\nu})(t-\text{T}_{\nu})-{\vec p}_{\nu}\!\cdot{\vec
 c}\,(t-\text{T}_{\nu})
=\frac{m_{\nu}^2}{2 E_{\nu}(\vec{p}_{\nu})} (t-\text{T}_{\nu}),\label{light-phase}
\end{eqnarray}
and becomes a half of $\bar{\phi_G}$. 

If  the coordinate ${\vec r}$ is integrated in the amplitude, the phase $\phi({\vec
r}\,)$ is combined with those of the pion and muon fields and disappears
at the end. We will see that 
in a process of  detecting the neutrino with a detector at a  long
distance, an
interference phenomenon of waves due to this slow phase occurs. This
interference  is almost equivalent to 
diffraction  of light through a hole of a finite size. A
light observed at a screen perpendicular to the wave vector shows a
diffraction pattern, while a diffraction pattern is formed  in a parallel
direction to the momentum of the neutrino produced in
the decay of pions. Since angular velocities of  these phases  are extremely
slow  and inversely proportional to the neutrino energy, they
make a diffraction  pattern macroscopic.
From  a weak energy dependence at a high energy,
the phase at the energy $E_{\nu}$ and $E_{\nu}+\Delta E_{\nu}$ is almost 
identical and is stable under a change of the initial state. So the
neutrino diffraction has the same pattern in a broad energy spectrum 
and has the stability.

The slow phase $\bar{\phi_c}$  shows  a characteristic 
feature of
the neutrino wave packet, and  is
   an  intrinsic property of the neutrino wave function at the light
   cone. This phenomenon is independent from the detail of wave packet
   and it is shown that wave packets  in general cases, including those
   of spreading wave packets, have the phase.
 Since the phase in the
   longitudinal direction is the origin of the diffraction,        
   spreading of the wave packet   does not change the behavior in the 
longitudinal direction and does not modify the diffraction pattern. So the
spreading effect   has been  ignored for 
simplicity in this section  and will be studied in the latter section and
   Appendix. It will be shown there that the spreading  in the transverse 
direction modifies the ${\vec k}_{\nu}$ integration but the final result 
turns actually into the same.

It is worthwhile  to compare  a  neutrino velocity  with the
light velocity for a later convenience. A
 neutrino of energy  $1~[\text{GeV}]$ and  a mass 
$1~[\text{eV}/c^2]$    has a velocity
\begin{eqnarray}
v/c=1-2\epsilon, \ \epsilon=\left({m_{\nu}c^2 \over E_{\nu}}\right)^2=5\times 10^{-19},
\end{eqnarray} 
hence the neutrino propagates a  distance $l$, where  
\begin{eqnarray}
l=l_0(1-\epsilon)=l_0-\delta l, \ \delta l= \epsilon l_0,
\end{eqnarray}
while  the light propagates  a distance $l_0$. This difference of distances,
$\delta l$, becomes
\begin{eqnarray}
& &\delta l=5\times 10^{-17}~[\text{m}]; ~l_0=100~[\text{m}], \\ 
\label{neutrino-ovelapp}
& &\delta l=5\times 10^{-16}~[\text{m}]; ~l_0=1000~[\text{m}] ,
\end{eqnarray}
which are much smaller than the sizes of the neutrino  wave packets.
Since the  difference of velocities is small,  the neutrino amplitude 
at the nuclear  or atom
targets  show interference. The geometry of the neutrino interference 
is shown in
Fig.~\ref{fig:geo}. The neutrino wave produced at a time $t_1$ arrives
at one nucleus or atom in the detector and is superposed  to the wave
produced  at $t_2$ and arrives to the same    nucleus or atom same
time. A constructive interference of waves is shown in the text.

\section{Position-dependent  probability and interference }
  Because the wave function at a finite time is a
superposition of waves of broad kinetic energies, that retains the wave
natures and reveals a  diffraction phenomenon. The  probability to
detect the neutrino reflects this  property and receives a 
large finite-size correction  peculiar to tiny masses  in a wide area. 
The finite-size 
correction  is computed rigorously  with a
use of the light-cone singularity of a two point correlation  function 
of a decay process. The correction has a universal
property of the relativistically invariant system and is determined by 
the absolute neutrino mass.

\subsection{Transition probability }
We investigate  the  case of large $\sigma_{\pi}$ later and  study
the amplitude and probability of  $\sigma_{\pi}=\infty$  qualitatively first.
Features of the amplitude to detect the neutrino at a
finite distance are elucidated.  
Integrating  over the  momentum ${\vec k}$ in Eq.\,$(\ref{amplitude})$, we   
obtain the ${\vec x}$-dependent amplitude of the form 
Eq.\,$(\ref{neutrino-position-amplitude})$.  The integrand is a Gaussian
function around the center ${\vec x}_0(t)={\vec
v}_{\nu}(t-\text{T}_{\nu})+{\vec x}(0)$ and is invariant under 
\begin{eqnarray}
& &{\vec x} \rightarrow {\vec x}+{\vec v}_{\nu} \delta t, \\
& &t \rightarrow t+\delta t \nonumber.
\end{eqnarray}
Thus  a shifted energy
\begin{eqnarray}
H_0-{\vec v}_{\nu}\cdot{\vec P},
\end{eqnarray} satisfies
\begin{eqnarray}
[S,H_0-{\vec v}_{\nu}\cdot{\vec P}]=0
\end{eqnarray}
and is conserved.

 Integrating over ${\vec x}$
further, we have the amplitude,
\begin{align}
&T=Ce^{i\phi_0}\bar u(p_{\mu})(1-\gamma_5)u(p_{\nu})  e^{-\frac{\sigma_{\nu}
 }{2}{\delta {\vec p}}^{\,2}} e^{-i\omega \text T/2}2 \frac{\sin (\omega
 \text T/2)}{\omega} 
\label{integrated-amplitude-honbun}
,\\
&\omega=\delta E-\vec{v}_\nu\cdot\delta {\vec p},\,
\delta{\vec p}={\vec p}_{\pi}-{\vec p}_{\mu}-{\vec p}_{\nu},\, \delta E=E({\vec p}_{\pi})-E({\vec p}_{\mu})-E({\vec p}_{\nu}), \nonumber
\end{align}
where $\phi_0$ is a constant. Because the center ${\vec x}_0(t)$ moves with the velocity ${\vec
v}_{\nu}$, the angular velocity in
Eq.\,$(\ref{integrated-amplitude-honbun})$
is different from the energy difference $\delta E$ of the rest system, 
but $\delta E-{\vec v}_{\nu} \cdot \delta{\vec p}$ of the moving system. This is a feature of 
the present amplitude.

In Eq.\,$(\ref{integrated-amplitude-honbun})$, the momentum is
 approximately conserved due to the Gaussian factor  and  $|\delta {\vec p}\,|$ has a finite
 uncertainty. Hence the angular
velocity $\omega $
 behaves differently from $\delta E$. At $\delta {\vec
p}=0$,  $\omega=0$ is the same as the usual case $\delta E=0$, 
 whereas at $\delta {\vec
p} \neq 0$, $\omega=0$  gives the relation  
$\delta E={\vec
 v}_{\nu} \cdot\delta {\vec p}\neq0$. Kinetic energy takes broad range 
 because $\omega$ is different from $\delta E$ and the amplitude is
 broad in $\omega$ at a finite T. Thus, kinetic energy is not conserved
 $\delta E \neq 0$. As is shown in  Appendix B, a shape of the
  configuration of the momentum satisfying $\omega=0$ is a large  ellipse  of the muon momentum where 
 the normal solution of $\delta {\vec p}=0$ and the
 solution of large $|\delta {\vec p}|$ are on the curve. $\omega$ varies
 fast  around the former momentum and  $2{\sin {(\omega \text{T}/2)} \over
 \omega}=2\pi\delta(\omega)$ \cite{Schiff and Landau} can be applied. This
 gives the normal term which satisfies the
 energy-momentum conservation well. On the other hand, $\omega$ varies
 extremely slowly  around the latter
 momentum, and the states of $\omega \approx 0$ lead the slow convergence
 at large T,  and give the finite-size correction.  Since $|\delta {\vec
 p}\,|$ and $\delta E$ are not small, the
 spectrum at the ultraviolet region, which exists  in the wave function 
at a finite time,
 gives a contribution to  the 
finite-size correction.  We 
will study this point further in Appendix B.
   
Although the reason for the large finite-size correction became  clear, 
 it is  not straightforward to compute it  using  
Eq.\,$(\ref{integrated-amplitude-honbun})$. Instead, an expression 
of $|T|^2$ with a correlation function of  coordinates  is convenient 
 for a  rigorous computation, and the finite-size
correction is found with it.   
 
  A transition probability   
of a pion of  a momentum ${\vec
p}_{\pi}$ located at a space-time position $(\text{T}_{\pi},{\vec
X}_{\pi})$, decaying to the  neutrino of the  momentum ${\vec p}_{\nu}$ at  
a space-time position $(\text{T}_{\nu},{\vec X}_{\nu})$ and a muon of momentum
${\vec p}_{\mu}$,
is expressed in the form 
\begin{eqnarray}
|T|^2 &=& g^2 m_{\mu}^2 
|\tilde N|^2 \int d^4x_1 d^4x_2
S_{5}(s_1,s_2)
\nonumber\\
 &\times& e^{i( \phi(x_1) -\phi(x_2))}  
e^{-{1 \over 2\sigma_{\nu} }\sum_i \left({\vec x}_i-{\vec X}_{\nu} -{\vec
			     v}_{\nu}(t_i - \text{T}_{\nu})\right)^2}
\nonumber \\
&\times& e^{-i\left(E({\vec p}_{\pi})(t_1 - t_2)-{\vec p}_{\pi}\cdot({\vec x}_1-{\vec
x}_2)\right)}
\times e^{i\left(E({\vec p}_{\mu})(t_1-t_2)-{\vec p}_{\mu}\cdot({\vec
	     x}_1-{\vec x}_2\right))}
\nonumber \\
&\times&e^{-{1 \over 2 \sigma_{\pi} }\sum_j \left({\vec x}_j-{\vec X_{\pi}}-{\vec v}_{\pi}(t_j-\text{T}_{\pi})\right)^2}
\label{probability}
\end{eqnarray}
where $S_{5}(s_1,s_2)$ stands for  products of Dirac
spinors and their  complex conjugates,   
\begin{eqnarray}
S_{5}(s_1,s_2)=\left(\bar u({\vec p}_{\mu})
 (1 - \gamma_5) \nu ({{\vec p}_{\nu}})\right)\left(\bar u({\vec p}_{\mu})
 (1 -  \gamma_5) \nu ({{\vec p}_{\nu}})\right)^{*},
\label{spinor-1}
\end{eqnarray}
and its spin summation is  given by
\begin{eqnarray}
S^{5}&=&\sum_{s_1,s_2}S^{5}(s_1,s_2)
=\frac{2}{m_{\nu}m_{\mu}}(p_{\mu}\!\cdot\! p_{\nu}).\label{spinor-2}
\end{eqnarray}

The
probability is finite and an order of integrations are interchangeable. 
Integrating  momenta of the final state 
and taking average over the initial momentum, we have the total probability 
 in the form 
 \begin{align}
&\int d{\vec p}_{\pi}\rho_{exp}({\vec p}_{\pi}) d{\vec X}_{\nu}d{\vec
 p}_{\mu}d{\vec p}_{\nu}  \sum_{s_1,s_2}|T|^2 
 \label{probability-correlation1}\nonumber \\
&= g^2 m_{\mu}^2 |N_{\pi\nu}|^2\frac{2}{(2\pi)^3}\int d{\vec X}_{\nu} d{\vec p}_{\nu} \rho_{\nu}^2 d^4x_1 d^4x_2
e^{-{1 \over 2\sigma_{\nu} }\sum_i\left({\vec x}_i-{\vec X}_{\nu} -{\vec
v}_\nu(t_i-\text{T}_{\nu})\right)^2}
\nonumber \\
&  \times 
\Delta_{\pi,\mu}(\delta t,\delta {\vec x})
e^{i \phi(\delta x_{\mu})} 
e^{-{1 \over 2 \sigma_{\pi} }\sum_j \left({\vec x}_j-{\vec X_{\pi}}-\bar{\vec
 v}_{\pi}(t_j- \text{T}_{\pi})\right)^2} 
\nonumber \\
&
N_{\pi\nu} =
\left(\frac{4\pi}{\sigma_{\pi}}\right)^{\frac{3}{4}}\left(\frac{4\pi}{\sigma_{\nu}}\right)^{\frac{3}{4}},~\rho_{\nu}=\left(\frac{1}{2E_{\nu} (2\pi)^3 }\right)^{\frac{1}{2}},~\delta  x= x_1-x_2,
\end{align}
with a  correlation function $\Delta_{\pi,\mu}(\delta t,\delta {\vec x})$. The correlation
function is defined with   a pion's momentum distribution $\rho_{exp}({\vec p}_{\pi})$, by
\begin{align}
\Delta_{\pi,\mu} (\delta t,\delta {\vec x})=
 {\frac{1}{(2\pi)^3}}\int
{d {\vec p}_{\pi} \over E({\vec p}_{\pi})}\rho_{exp}({\vec p}_{\pi})
{d {\vec p}_{\mu} \over E({\vec p}_{\mu})}  (p_{\mu}\!\cdot\! p_{\nu})
 e^{-i\left(\{E({\vec
 p}_{\pi})-E({\vec p}_{\mu})\}\delta t-({\vec p}_{\pi}-{\vec
 p}_{\mu})\cdot \delta {\vec x})\right)}.
\label{pi-mucorrelation1}
\end{align} 

 In the above equation, the final states are integrated over a
 complete set \cite{Ishikawa-Shimomura}.   The muon and neutrino momenta are integrated over  
entire positive energy  regions, and the neutrino position  is 
integrated over the region of the detector. The pion in the initial
 state is assumed to be the statistical ensemble of the 
distribution $\rho_{exp}({\vec p}_{\pi})$.
If the momentum distribution is narrow around the central value, the velocity ${\vec v}_{\pi} $ in the pion Gaussian factor was
replaced with its average $\bar {\vec v}_{\pi}$. This is verified from
the large spatial size of the pion wave packet discussed in the 
previous section. 
For    the probability of  a fixed pion momentum, the correlation function 
 \begin{align}
\tilde \Delta_{\pi,\mu} (\delta t,\delta {\vec x})=
 {\frac{1}{(2\pi)^3}}
{1 \over E({\vec p}_{\pi})}
\int {d {\vec p}_{\mu} \over E({\vec p}_{\mu})}  (p_{\mu}\!\cdot\! p_{\nu})
 e^{-i\left(\{E({\vec
 p}_{\pi})-E({\vec p}_{\mu})\}\delta t-({\vec p}_{\pi}-{\vec
 p}_{\mu})\cdot \delta {\vec x}\right)},
\label{pi-mucorrelation2}
\end{align} 
is used instead of Eq.\,$( \ref{pi-mucorrelation1})$.    
\subsection{Light-cone singularity   }
The expression  Eq.\,$(\ref{probability-correlation1})$ shows that  the
probability gets a finite $\text T (\text{T}=\text{T}_{\nu}-\text{T}_{\pi})$, 
correction  from the
integration over $t_1$ and $t_2$ at  $|t_1-t_2| \rightarrow
\text{T}  $, if the $ \Delta_{\pi,\mu}(\delta t,\delta {\vec
x})$ decreases slowly in this region. The correlation function $\tilde \Delta_{\pi,\mu}(\delta t,\delta {\vec
x})$ is a standard form of Green's function and has the light-cone 
singularity that is real and decreases very slowly
along the light cone. The singularity  is generated by the states
at the ultraviolet energy region. So   the singularity near 
the light-cone region
\begin{align}
\lambda=\delta t^2-{\left|\delta\vec x\right|}^2 = 0,
\end{align}
which is extended in a large  $|\delta {\vec x}|$ and is independent
of ${\vec p}_{\pi}$,   plays an important role 
for  the probability Eq.\,$(\ref{probability-correlation1})$ at a    
finite  T. 
We find, in fact, that the
light-cone singularity of    $\tilde \Delta_{\pi,\mu}
(\delta t,\delta {\vec x})$ \cite{Wilson-OPE}  gives a large finite-size
correction in the following.

\subsubsection{ Separation of singularity  }

If the particles are expressed by plane waves,  the integration is made
over infinite-time interval and the  
energy is strictly conserved  and 4-dimensional  momenta satisfy  
\begin{eqnarray}
p_{\pi}=p_{\mu}+p_{\nu},\ 
(p_{\pi}-p_{\mu})^2=m_{\nu}^2 \approx 0.
\label{lightlike}
\end{eqnarray}
An integral over the momentum in the region where   the 
momentum difference $p_{\pi}-p_{\mu}$ is almost
light-like leads   $\tilde{\Delta}_{\pi,\mu} (\delta
t,\delta {\vec x})$ to have  a singularity around  the light cone,
$\lambda=0$. 
 In order to extract the singular term from  $\tilde \Delta_{\pi,\mu} (\delta
t,\delta {\vec x})$, we write  the integral in a four-dimensional form   
\begin{align}
&\tilde \Delta_{\pi,\mu} (\delta t,\delta {\vec x})=
 {\frac{1}{(2\pi)^3}} {1 \over E({\vec
 p}_{\pi})}I(p_{\pi},\delta x),\nonumber\\
&I(p_{\pi},\delta x)={2 \over \pi} \int d^{4}p_{\mu} \, \theta(p_{\mu}^0)
(p_{\mu}\!\cdot\! p_{\nu}) \text {Im}\left[1 \over p_{\mu}^2-m_{\mu}^2-i\epsilon\right]
 e^{-i\left(\{E({\vec
 p}_{\pi})-E({\vec p}_{\mu})\}\delta t-({\vec p}_{\pi}-{\vec
 p}_{\mu})\cdot \delta {\vec x}\right)}, 
\end{align}
first, and change the integration variable   from $p_{\mu}$ to 
$q=p_{\mu}-p_{\pi}$ that is conjugate to $\delta x$. Next,  we separate the
integration region into two parts,  $0 \leq q^0$ and $-p_{\pi}^0 \leq
q^0 \leq 0$, 
 and have the expressions,     
\begin{align}
&I(p_{\pi},\delta x)=I_1(p_{\pi},\delta x)+I_2(p_{\pi},\delta x), 
\label{seperation-region}\\
&I_1(p_{\pi},\delta x)=\left\{p_{\pi}\! \cdot\! p_{\nu}+p_{\nu}\!\cdot\! \left(-i{\partial \over
 \partial \delta x}\right)\right\} \tilde I_1,\nonumber\\
&\tilde I_1={2 \over \pi}\int d^4 q \,  \theta(q^0)\text {Im}\left[1 \over
 (q+p_{\pi})^2-m_{\mu}^2-i\epsilon\right] e^{iq \cdot \delta x }, \nonumber  \\
&I_2(p_{\pi},\delta x)= {2 \over \pi} \int_{-p_{\pi}^0}^{0}d^4 q\, p_{\nu}\!\cdot\! (p_{\pi}+q)\text {Im} \left[1 \over
 (q+p_{\pi})^2-m_{\mu}^2-i\epsilon\right] e^{iq \cdot \delta x }.
\end{align}
$I_1(p_{\pi},\delta x)$ is the integral over the infinite region 
 and has the light-cone singularity and
$I_2(p_{\pi},\delta x)$ is the integral over  the finite region  and 
is regular. 

  $I_1(p_{\pi},\delta x)$ comes from the states of  non-conserving  kinetic
  energy and does not contribute to the total probability at an 
infinite-time interval.  $I_2(p_{\pi},\delta x)$, on the other hand, 
contributes to that at the infinite-time and finite-time intervals.   So
  the physical 
quantity at the finite distance is computed using the 
most singular term of $I_1$.   

Next we compute  $\tilde I_1$.  Expanding the integrand with 
$p_{\pi}\!\cdot\! q$, we have  $\tilde I_1$  in the form 
\begin{align}
&\tilde I_1(p_{\pi},\delta x)={2 \over \pi}\int d^4 q \, \theta(q^0)~ \text {Im}\left[{1 \over
q^2+m_{\pi}^2-m_{\mu}^2+2q\!\cdot\! p_{\pi}-i\epsilon}\right] e^{iq \cdot \delta x } \nonumber\\
&={2 \over \pi}\int d^4 q  \,\theta(q^0)\left\{1+2p_{\pi}\!\cdot\! \left(i{\partial \over \partial \delta
 x}\right) {\partial  \over \partial {\tilde m}^2}+\cdots \right\}\,\text {Im}\left[ {1 \over
 q^2+{\tilde m}^2-i{\epsilon}} \right]e^{iq \cdot\delta x } \nonumber\\
&=2  \left\{1 +2p_{\pi} \!\cdot\!\left(i{\partial \over \partial \delta
			  x}\right) {\partial  \over \partial {\tilde m}^2}+\cdots \right\}
\int d^4 q \, \theta(q^0)\delta (q^2+{\tilde m}^2) e^{iq \cdot\delta x },\label{singular-function}
\end{align}
where 
\begin{eqnarray}
{\tilde m}^2=m_{\pi}^2-m_{\mu}^2.
\end{eqnarray}
The expansion in $2p_\pi\!\cdot\! q$ of 
Eq.\,$(\ref{singular-function})$  converges  in the region
\begin{eqnarray}
{2p_{\pi}\!\cdot\! q \over q^2+{\tilde m}^2} < 1.
\end{eqnarray}
Here $q$ is the integration variable and varies. So we  evaluate  
the series after the integral and find a condition for its  convergence. 
 We  find later that the series after the momentum integration 
converges in the region
${2p_{\pi}\cdot p_{\nu} \over {\tilde m}^2} \leq 1$.

     $\tilde I_1(p_{\pi},\delta
x)$  is written in the form 

\begin{align}
\tilde I_1(p_{\pi},\delta x)
=  2(2\pi)^3i\left\{1 +2p_{\pi} \!\cdot\!\left(i{\partial \over \partial
 \delta x}\right) {\partial  \over
 \partial {\tilde m}^2}+\cdots \right\}\left( {1 \over 4\pi}\delta(\lambda)\epsilon(\delta
 t)+f_{short}\right),
\end{align}
where $f_{short}$ is written by Bessel functions and a formula for a
relativistic field     
\begin{align}
\int d^4 &q  \, \theta(q^0) \delta(q^2+{\tilde m}^2)e^{iq \cdot\delta x }
= (2\pi)^3i\left[{1
 \over 4\pi}\delta(\lambda)\epsilon(\delta t) +f_{short}\right],\nonumber \\
f_{short}&=-{i {\tilde m} \over
 8\pi \sqrt{-\lambda}} \theta(-{\lambda})\left\{N_1\left(\tilde m \sqrt{
 -\lambda}\right)-i\epsilon(\delta t) J_1\left(\tilde m \sqrt{ -\lambda}\right)\right\} \nonumber \\
&-\theta(\lambda){i
 \tilde m \over
 4\pi^2\sqrt{\lambda}}K_1\left(\tilde
 m\sqrt{\lambda}\right),~\lambda={\delta x}^2,\delta t=\delta x^0\label{singular-function-f},
\end{align}
where $N_1$, $J_1$, and $K_1$ are Bessel functions, is used.
More details are presented in I.

Next  $I_2$ is evaluated. For $I_2$, we use a momentum $\tilde
q=q+p_{\pi}$ and write in the form
\begin{align}
& I_2(p_{\pi},\delta x)=\frac{2}{\pi} \int_{0< \tilde
 q^0<p_{\pi}^0} d^4 \tilde q \, (p_{\nu} \!\cdot\!\tilde q)  \text {Im}\left[\frac{1}
 {\tilde q^2-m_{\mu}^2-i\epsilon}\right] e^{i(\tilde q-p_{\pi}) \cdot\delta x } 
\nonumber\\
&=e^{i(-p_{\pi}) \cdot\delta x }\left\{p_{\nu} \!\cdot\!\left(-i{\partial \over \partial
 \delta x}\right)\right\}  {2 \over \pi}\int_{0< \tilde q^0
 <p_{\pi}^0 } d^4 \tilde{q}\,  \pi \delta( q^2-m_{\mu}^2)  e^{i\tilde q \cdot\delta x } 
\nonumber \\
&= e^{-i p_{\pi} \cdot\delta x}\left\{p_{\nu} \!\cdot\!\left(-i{\partial \over \partial
 \delta x}\right)\right\} \int\frac{ d\vec{q}}{
  \sqrt{\vec{q}^{\,2}+m_{\mu}^2}} \theta\left(p_{\pi}^0-\sqrt{\vec{q}^{\,2}+m_{\mu}^2} \right)
 e^{iq \cdot\delta x }.\label{normal-term}
\end{align}
The regular part $I_2$ has no singularity because the integration domain is
finite and becomes short-range.  

Thus the first term in $\tilde I_1$ gives the most singular 
term  and the rests, the second
term in $I_1$ and $I_2$, give regular terms.  
The correlation function, $\tilde \Delta_{\pi,\mu}(\delta t,\delta {\vec x})$ 
 is written in the form 
\begin{align}
&\tilde \Delta_{\pi,\mu}(\delta t,\delta {\vec x})={1 \over (2\pi)^3} {1 \over E(p_{\pi})}\Biggl[\left\{p_{\pi}\! \cdot\! p_{\nu}-p_{\nu}\!\cdot\!\left(i\frac{\partial}{
 \partial \delta x}\right)\right\}2(2\pi)^3i\nonumber\\ 
& \left\{1 +2p_{\pi} \!\cdot\! \left(i{\partial \over \partial
 \delta x}\right) {\partial  \over
 \partial {\tilde m}^2}+\cdots \right\} \left( \frac{1}{4\pi}\delta(\lambda)  
\epsilon(\delta t)+f_{short}\right) + I_2 \Biggr],\label{muon-correlation-total}
\end{align}
where the dots stand for the higher order terms.

\subsection{Integration over spatial coordinates   }
Next, we integrate over the coordinates ${\vec x}_1$ and ${\vec x}_2$  in
\begin{align}
&\int d{\vec x}_1 d{\vec x}_2e^{i\phi(\delta
 x)}e^{-\frac{1}{2\sigma_{\nu} } \sum_i
\left({\vec x}_i-{\vec X}_{\nu} -{\vec
v}_\nu(t_i-\text{T}_{\nu})\right)^2}
\tilde \Delta_{\pi,\mu}(\delta t,\delta{\vec x}).\label{lightcone-integration1}
\end{align}

\subsubsection{Singular terms: long-range correlation}
The most singular term of $\tilde \Delta_{\pi,\mu}(\delta t,\delta{\vec x})$ 
is substituted, then   Eq.~$(\ref{lightcone-integration1})$ becomes    
\begin{eqnarray}
\label{singular-correlation}
J_{\delta(\lambda)}&=&\int d{\vec x}_1 d{\vec x}_2e^{i\phi(\delta
x)}e^{-{1 \over 2\sigma_{\nu} }\sum_i \left({\vec x}_i-{\vec X}_{\nu} -{\vec
v}_\nu(t_i-\text{T}_{\nu})\right)^2} {1 \over
4 \pi}\delta(\lambda)\epsilon(\delta t)  ,
\end{eqnarray}
and is computed easily  using a center coordinate $R^\mu=\frac{
x_1^\mu+x_2^\mu}{2}$ and a relative coordinate
$\vec{r}=\vec{x}_1-\vec{x}_2$.
After the center coordinate ${\vec R}$ is integrated,
$J_{\delta(\lambda)}$ 
becomes the integral of the   transverse and longitudinal component $({\vec
r}_T,r_l)$ of the relative coordinates,   
\begin{align}
\epsilon(\delta t) (\sigma_{\nu}\pi)^{\frac{3}{2}} \int d{\vec r}_Td r_l \, e^{i\phi(\delta t,{\vec
 r})-\frac{1}{4\sigma_{\nu} }({{\vec r}_T}^{\,2}  +(r_l-{
 v}_{\nu}\delta t)^2)}\frac{1}{4\pi}\delta (\delta t^2-{{\vec r}_T}^{\,2} -{{r}_l}^{2}).
\label{lightcone-integration-s}
\end{align}
The transverse coordinate ${\vec r}_T$ is integrated using the Dirac
delta function and $r_l$ is integrated next.
Finally we have  
\begin{eqnarray}
J_{\delta(\lambda)}&=&{(\sigma_{\nu}\pi)}^{\frac{3}{2}}
 \frac{\sigma_{\nu}}{2}{1 \over |\delta t|
 }\epsilon(\delta t)e^{i\bar \phi_c(\delta t)-\frac{m_{\nu}^4}{
 16\sigma_{\nu} E_{\nu}^4} {\delta t}^2}\nonumber\\
 &\approx& {(\sigma_{\nu}\pi)}^{\frac{3}{2}} \frac{\sigma_{\nu}}{2}
  \frac{1}{|\delta t|
  }\epsilon(\delta t)e^{i\bar \phi_c(\delta t)}\label{lightcone-integration2-2}. 
\end{eqnarray}

The next term  of $\tilde \Delta_{\pi,\mu}(\delta t,\delta{\vec x})$, of
the form ${1 \over \lambda}$, in Eq.\,$(\ref{lightcone-integration1})$
leads       
\begin{eqnarray}
J_{1/\lambda}&=&\int d{\vec x}_1 d{\vec x}_2e^{i\phi(\delta x)}e^{-{1
 \over 2\sigma_{\nu} }\sum_i \left({\vec x}_1-{\vec X}_{\nu} -{\vec
v}_\nu(t_1-\text{T}_{\nu})\right)^2}
{i  \over 4\pi^2 \lambda},
\label{lightcone-integration4}
\end{eqnarray}
which becomes 
\begin{eqnarray} 
J_{1/\lambda}&\approx& {(\sigma_{\nu}\pi)}^{\frac{3}{2}} \frac{\sigma_{\nu}}{2} \left(\frac{1}{
 \pi \sigma_{\nu} |\vec{p}_{\nu}|^2}\right)^{\frac{1}{2}} e^{-\sigma_{\nu}|\vec{p}_{\nu}|^2}
\frac{ 1}{|\delta t| }e^{i\bar
\phi_c(\delta t)}.
\label{lightcone-integration4-2} 
\end{eqnarray} 
This term also has the universal $|\delta t|$ dependence but its magnitude is much
smaller than that of $J_{\delta(\lambda)}$ and is negligible in the present decay mode.  

From Eqs.\,$(\ref{lightcone-integration2-2})$
and $(\ref{lightcone-integration4-2})$, the singular terms
$J_{\delta(\lambda)}$ and $J_{1/\lambda}$ have the slow  phase
$\bar \phi_c(\delta t)$ and the magnitudes that are inversely proportional
to $\delta t$. Thus these terms are long-range with the small
angular velocity and are
insensitive to the ${\tilde m}^2$.  These properties of the
time-dependent correlation functions $J_{\delta(\lambda)}$ 
 hold
for  the general wave packets,
and the following theorem is proved.

{\bf Theorem}

The singular part $J_{\delta(\lambda)}$ of the correlation function has
the slow  phase that is determined with  the absolute value of the neutrino
mass
and the 
magnitude inversely proportional
to $\delta t$,  of the form 
Eq.\,$(\ref{lightcone-integration2-2})$, at the large distance. The phase is given in the form
of a sum of $\bar \phi_c(\delta t)$ and small corrections, which are 
inversely proportional to the neutrino energy in general systems 
and become $1/E^2$ if the neutrino wave
 packet is invariant under the time inversion.  

{\bf (Proof: General cases including spreading of wave packet
)}

We prove the theorem for general wave packets. 
$J_{\delta(\lambda)}$  is written in the form,
\begin{eqnarray}
\label{singular-correlation-centerG}
J_{\delta(\lambda)}=\int  d{\vec r}\, e^{i\phi(\delta x)}
\tilde w \left({\vec r} -{\vec
v}_\nu\delta t\right) \times \frac{1}
{4 \pi}\delta(\lambda)\epsilon(\delta t),  
\end{eqnarray}
where $\tilde w({\vec x}-{\vec v}t)$ is expressed with  a  wave packet in the
coordinate representation $w({\vec x}-{\vec v}t)$ and its complex
conjugate as,
\begin{align}
\tilde w(r_l-v_{\nu}\delta t,{\vec r}_T)&=\int d{\vec R} w\left({\vec R}+\frac{\vec r}{2}\right)w^{*}\left({\vec R}-\frac{\vec r}{2}\right) \nonumber\\
&=\int dk_l d{\vec k}_T e^{ik_l(r_l-v_{\nu}\delta t)+i{\vec k}_T\cdot{\vec
 r}_T+ic_0({\vec k}_T^2)\delta t} |w(k_l,{\vec k}_T)|^2.
\end{align}
The wave function $w({\vec x}-{\vec v}t)$ that includes 
the spreading effect is expressed in the following form 
\begin{eqnarray}
& &w({\vec x}-{\vec v}t)=\int dk_l d{\vec k}_T \,e^{ik_l(x_l-v_{\nu}t)+i{\vec k}_T\cdot{\vec
 x}_T+iC_{ij}k_T^ik_T^jt} w(k_l,{\vec k}_T),\\
& &C_{ij}=C_0 \delta_{ij},~C_0={1 \over 2E},
\end{eqnarray}
instead of the Gaussian function of Eq.\,$(\ref{singular-correlation})$.
A  quadratic form in ${\vec
k}$ in an expansion of $E\left({\vec p}+{\vec k}\right)$ is included
and this makes the wave packet spread with time. The
coefficient $C_{ij}$ in the longitudinal direction is negligible for 
the neutrino and is neglected. Expanding the delta function in the form,
\begin{eqnarray}
\delta({\delta t}^2-r_l^2-{\vec r}_T^{\,2})=\sum_l{1 \over l!}(-{\vec r}_T^{\,2})^l\left({\partial \over
 \partial {\delta t}^2}\right)^l \delta (t^2-r_l^2),
\end{eqnarray}
we have  the correlation function  
 \begin{align}
\label{singular-correlation-centerG4}
J_{\delta(\lambda)}&=\int dr_l d{\vec r}_T e^{i\phi(\delta t,r_l)}
\tilde w( r_l -v_\nu\delta t,{\vec r}_T){1 \over
4 \pi} \left\{1+\sum_{n=1} {1 \over n!}(-{{\vec r}_T}^{\,2})^n \left(\frac{\partial}
{\partial (\delta t )^2}\right)^n\right\}\nonumber\\
&\  \times \delta(\delta t^2-r_l^2) 
\epsilon(\delta t) \nonumber\\
&=  \int dr_l  d{\vec r}_T dk_l d{\vec k}_T e^{i\phi(\delta
  t,r_l)+ik_l(r_l-v_{\nu} \delta t)+i{\vec k}_T\cdot{\vec r}_T+iC_0{\vec k}_T^2 \delta
  t}|w(k_l,{\vec k}_T)|^2 \nonumber \\
&\ \times {1 \over
4 \pi} \left\{1+\sum_{n=1} {1 \over n!}(-{{\vec r}_T}^{\,2})^n \left(\frac{\partial}
{\partial (\delta t )^2}\right)^n\right\}  \delta(\delta t^2-r_l^2) 
\epsilon(\delta t)
\nonumber \\
&=  \int dr_l dk_l  e^{i\phi(\delta
  t,r_l)+ik_l(r_l-v_{\nu} \delta t)} d{\vec r}_T  d{\vec k}_T
e^{+i{C_0{\vec k}_T^2 \delta
  t}}|w(k_l,{\vec k}_T)|^2 \nonumber \\
&\ \times{1 \over 4 \pi} \left\{1+\sum_{n=1} {1 \over n!}\left({\partial^2 \over
		    (\partial {\vec k}_T)^{\,2}}\right)^n \left(\frac{\partial}
{\partial (\delta t )^2}\right)^n\right\} e^{i{\vec k}_T\cdot{\vec r}_T}  \delta(\delta t^2-r_l^2) 
\epsilon(\delta t).
\end{align}
The variable ${\vec r}_T$ is  integrated first and ${\vec k}_T$ is integrated
next. Then we have the expression 
\begin{align}
J_{\delta(\lambda)}&=  \int dr_l dk_l  e^{i{\phi}(\delta
  t,r_l)+ik_l(r_l-v_{\nu} \delta t)}  
|w(k_l,0)|^2  \nonumber \\
& \times {1 \over 4 \pi} \left\{1+\sum_{n=1} {1 \over n!}(-2iC_0 \delta t)^n \left(\frac{\partial}
 {2 \delta t \partial \delta t }\right)^n\right\} (2\pi)^2 \delta(\delta t^2-r_l^2) 
\epsilon(\delta t).
\end{align}
Using the following identity 
\begin{eqnarray}
(2\delta t)^n\left({\partial \over 2\delta t \partial \delta t}\right)^n=\left({\partial
 \over \partial \delta t}\right)^n+O\left({1 \over \delta t}\right) \left(\frac{\partial}{\partial \delta t}\right)^{n-1},
\end{eqnarray}
and taking a leading term in ${1/\delta t}$, we have the final
expression of the correlation function at the long-distance region
\begin{eqnarray}
& &
\label{singular-correlation-centerF}
J_{\delta(\lambda)}= \pi e^{-C_0 p } \epsilon(\delta t){e^{i\bar{\phi}_c(\delta
  t)} \over 2\delta t} \int  dk_l  e^{k_l(i(1-v_{\nu}) \delta t +C_0)}  
|w(k_l,0)|^2.   
\end{eqnarray}

Hence $J_{\delta (\lambda)}$ in Eq.\,$(\ref{singular-correlation-centerF}  )$
becomes  almost the same form as 
Eq.\,$(\ref{lightcone-integration2-2})$ and the slow phase
$\bar \phi_c(\delta t)$ is modified slightly and the magnitude that is 
inversely proportional
to the time difference. $J_{\delta(\lambda)}$ has the universal form for
the general wave packets. By expanding the exponential factor and taking
the quadratic term of the exponent, the above   integral is written in 
the form 
\begin{align}
\label{singular-correlation-centerFc}
& \int  dk_l  (1 +{k_l(i(1-v_{\nu}) \delta
  t +C_0)}+{1 \over 2!}{(k_l(i(1-v_{\nu}) \delta
  t +C_0) )^2})  |w(k_l,0)|^2  \nonumber \\
&~=w_0\left(1+C_0d_1+{d_2 \over
 2!}C_0^2-(1-v_{\nu})^2{\delta t}^2\right)+i(d_1(1-v_{\nu})\delta t
 +d_2C_0(1-v_{\nu}) \delta t ),
\end{align}
where 
\begin{eqnarray}
& &\delta={d_1 \over E}+\frac{d_2}{2}{1 \over E^2},~
\gamma={d_1 \over 2E}+{d_2\over
 2!}\left({1 \over 2E}\right)^2-(1-v_{\nu})^2{\delta t}^2,  \nonumber\\
& &d_1= \frac{1}{w_0}\int d k_l k_l |w(k_l,0)|^2,~d_2= \frac{1}{w_0}\int d k_l
 k_l^2|w(k_l,0)|^2.
\end{eqnarray}
We substitute this expression into the correlation function and   have    
\begin{align}
J_{\delta(\lambda)}=\pi e^{-C_0 |\vec{p}| }\omega_0(1+\gamma) \epsilon(\delta t){e^{i\bar{\phi}_c(\delta
  t)(1+\delta)} \over 2\delta t}
, \ w_0= \int d k_l |w(k_l,0)|^2,
\end{align}

In  wave packets of time reversal invariance, $|w(k_l,0)|^2$ is the
even function of $k_l$. Hence $d_1$ vanishes 
\begin{eqnarray}
d_1=0,
\end{eqnarray}
 and the
correction are 
\begin{eqnarray}
\delta=\frac{d_2}{2}{1 \over E^2},\ 
\gamma={d_2 \over 2!}\left({1 \over 2E}\right)^2-(1-v_{\nu})^2{\delta t}^2.
\end{eqnarray}
{\bf Q.E.D.}

The light-cone region ${\delta t}^2-|{\delta {\vec x}}|^2=0$ is so close 
to  neutrino orbits  that it gives a
finite   contribution to the integral
Eq.\,$(\ref{lightcone-integration1})$.  Since the light-cone singularity is
real, the integral is sensitive  only to the slow neutrino phase and shows  
interference of the neutrino. This theorem is applied to quite general
systems, where the neutrino interact with a nucleus in a target.  
 
\subsubsection{Regular terms: short-range correlation}
Next, we study  regular terms of $\tilde \Delta_{\pi,\mu}(\delta
t,\delta{\vec x})$  in Eq.\,$(\ref{lightcone-integration1})$.  Regular
terms 
are  short-range and the
spreading effect is ignored and the Gaussian wave packet is
studied. First term is  $f_{short }$ in
$I_1$ and is composed of  Bessel functions. We  have   
\begin{eqnarray}
L_1=\int d{\vec x}_1 d{\vec x}_2\, e^{i\phi(\delta
 x)}e^{-\frac{1}{2\sigma_{\nu} } \sum_i \left({\vec x}_i-{\vec X}_{\nu} -{\vec
v}_\nu(t_i-\text{T}_{\nu})\right)^2}
f_{short}. 
\label{lightcone-integration2-1}
\end{eqnarray}
$L_1$ is evaluated at a large $|\delta t|$ in the form 
\begin{align}
L_1 = ({\pi \sigma_{\nu}})^{\frac{3}{2}}e^{iE_{\nu}\delta t}
\int d{\vec r} \, e^{-i\vec p_{\nu} \cdot \vec r-{1 \over 4\sigma_{\nu} }({\vec r} 
- {\vec v}_\nu\delta t)^2}  f_{short},~{\vec r}={\vec x}_1-{\vec x}_2.
\end{align}
Here the integration is made in the space-like region $\lambda <0$. We
write
\begin{eqnarray}
r_l=v_{\nu}\delta t +\tilde r_l,
\end{eqnarray}
and   rewrite $\lambda$ in the form 
\begin{align}
\lambda =\delta t^2-{r}_l^{\,2}-{\vec r_T^{\,2}} =\delta
 t^2-(v_{\nu}\delta t + \tilde r_l)^{2}-{\vec r_T^{\,2}}
\approx -2 v_{\nu}\tilde r_l \delta t -\tilde r_l^2-{\vec
 r_T^{\,2}}.
\end{align}
The $L_1$ for the large $|\delta t|$ is written
with these variables. Using  the asymptotic expression of the Bessel
functions, we  have 
\begin{eqnarray}
L_1&=&({\pi
 \sigma_{\nu}})^{\frac{3}{2}}e^{i(E_{\nu}-|\vec{p}_{\nu}|v_{\nu})\delta t}
\int d{\vec r}_T d \tilde r_l \, e^{-i(  |\vec{p}_{\nu}| \tilde r_l)-\frac{1}{4\sigma_{\nu} }(
\tilde r_l^2+ {\vec r}_T^{\,2})} {i\tilde m \over 4\pi^2}\left({\pi \over
						       2\tilde m}\right)^{\frac{1}{2}}
\nonumber \\  
& & \times\left({ 1  \over {2 v_{\nu}\tilde r_l |\delta t| +\tilde r_l^2+
{\vec r_T^{\,2}} }}\right)^{\frac{3}{4}}
e^{i\tilde m  \sqrt{2 v_{\nu}\tilde r_l |\delta t| +\tilde r_l^2+{\vec r_T^{\,2}}
}} . 
\end{eqnarray}
The Gaussian integration around ${\vec r}_T={\vec
0}$, $\tilde r_l=-2i\sigma_{\nu}|\vec{p}_{\nu}|$ give  the asymptotic expression of
$L_1$ at a large $|\delta t|$
\begin{align}
L_1&=({\pi \sigma_{\nu}})^{\frac{3}{2}}\tilde L_1,\nonumber\\
\tilde L_1&=
e^{i(  E_{\nu}-|\vec{p}_{\nu}|
 v_{\nu})\delta t}  e^{-  \sigma_\nu |\vec{p}_{\nu}|^2}\frac{i \tilde m}{4\pi^2}\left(\frac{\pi}
 {2 \tilde m}\right)^{\frac{1}{2}}
 \left(\frac{ 1 }{{4 v_{\nu}\sigma_{\nu} |\vec{p}_{\nu}|
|\delta t| }}\right)^{{\frac{3}{4}}}
e^{i\tilde m \sqrt{2 v_{\nu}\sigma_{\nu} |\vec{p}_{\nu}||\delta t|}}.\label{asymptotic-expansionL_1}
\end{align}
Obviously $L_1$ oscillates  fast as $e^{i\tilde mc_1|\delta t|^{\frac{1}{2}}}$ where $c_1$
 is determined by $|\vec{p}_{\nu}|$ and $\sigma_{\nu}$ and is short-range.
 The integration carried out with a different  stationary value of $r_l$ 
which takes into account the last term in the right-hand side gives  almost
 equivalent result. The integration  in the time-like region, 
$\lambda >0$,
 is carried in a similar manner and $L_1$ decreases  with time as
 $e^{-\tilde mc_1|\delta t|^{\frac{1}{2}}}$ and 
final  result is almost the same as that of 
the space-like region.
 It is noted that the long-range term which appeared 
from the isolated ${1/\lambda}$ singularity in 
Eq.\,$(\ref{lightcone-integration4-2}) $ does  not exist  in $L_1$ in
 fact. The reason for its absence is that the Bessel function decreases 
much faster in the  space-like region than ${1/\lambda}$ and
 oscillates much faster 
 than ${1/\lambda}$ in the time-like region. Hence the long-range
 correlation is not generated from the $L_1$  and 
 the light-cone singularity $\delta(\lambda) \epsilon(\delta t)$ and 
$1/{\lambda}$ are the
 only source of the long-range  correlation. 

Second term  of Eq.\,$(\ref{lightcone-integration1})$ is from $I_2$, Eq.\,$(\ref{normal-term})$. We have this
 term, $L_2$, 
\begin{align}
L_2&=2 p_{\nu}\!\cdot\!(p_{\pi}-p_{\nu})(\pi\sigma_{\nu})^{\frac{3}
{2}}(4\pi\sigma_{\nu})^{\frac{3}{2}}\frac{1}{
  \left(2\pi\right)^3}\tilde L_2,\nonumber\\
\tilde L_2&= \int {d\vec{q} \over 2\sqrt{\vec{q}^{\,2}+m_{\mu}^2}} 
 e^{-i\left(E_\pi-E_{\nu}-\sqrt{\vec{q}^{\,2}+m_{\mu}^2}-{\vec v}_{\nu}\cdot({\vec
 p}_{\pi}-{\vec q}-{\vec p}_{\nu})\right)\delta t}  \nonumber \\
&\times e^{ -{ \sigma_{\nu} ({\vec p}_{\pi}-{\vec q}-{\vec p}_{\nu})^2}} 
\theta \left(E_\pi-\sqrt{\vec{q}^{\,2}+m_{\mu}^2} \right).\label{lightcone-integration5}
\end{align}

 The angular velocity of the integrand  in $L_2$
 varies with ${\vec q}$ and the integral $L_2$  has  a short-range
 correlation of the length, $2 \sqrt{
 \sigma_{\nu}}$, in the time direction.
So the $L_2$'s contribution to the total probability comes from the small
 $|\delta t|$ region and   corresponds to the short-range component.

Thus the integral over the coordinates  is 
written in the form
\begin{align}
&\int d{\vec x}_1 d{\vec x}_2\,e^{i\phi(\delta x)}e^{-{1 \over
 2\sigma_{\nu} } \sum_i
\left({\vec x}_i-{\vec X}_{\nu} -{\vec
v}_\nu(t_i-\text{T}_{\nu})\right)^2}
\tilde \Delta_{\pi,\mu}(\delta
t,\delta{\vec x})\nonumber \\
& =2i  \frac{p_{\pi}\! \cdot\! p_{\nu}}{E_\pi} \left[ \left(1 +2p_{\pi}
 \!\cdot\! p_{\nu} \frac{\partial}{\partial {\tilde m}^2}+\cdots
 \right)e^{i\bar{\phi}(\delta t)}(J_{\delta(\lambda)}+L_1)+L_2\right]
\nonumber\\
&\approx 2i (\pi\sigma_{\nu})^{\frac{3}{2}} \frac{ p_{\pi} \!\cdot\! p_{\nu}}{E_\pi} \left[\left(1
 +2p_{\pi} \!\cdot\! p_{\nu} 
{\partial  \over \partial {\tilde m}^2}+\cdots \right)\right. \nonumber \\ 
&\left.~~~~\times\left(\frac{\sigma_{\nu}}{2} e^{i\bar \phi_c(\delta t)} 
{\epsilon(\delta t) \over |\delta t|}
 +\tilde L_1\right) -i\left( \frac{\sigma_\nu}{\pi}\right)^{\frac{3}{2}}\tilde L_2\right].\label{lightcone-integration1-1}
\end{align}

In the above equation, $p_{\nu}^2=m_{\nu}^2$ is negligibly small
  compared to $\tilde{m}^2$, $p_{\pi}\!\cdot\! p_{\nu}$ and
${\sigma_{\nu}}^{-1}$, and  is  neglected   in most  places 
except the slow phase $\bar \phi(\delta t)$. The 
first term in the right-hand side of
Eq.\,$(\ref{lightcone-integration1-1})$ is  long-range  and 
the second term is short-range. The long-range term is  separated
from others in a clear manner.

\subsubsection{Convergence condition}

At the end of this section, we find  a condition for our method to be 
valid. In Eq.\,$(\ref{seperation-region})$, the integration region was
split into the one of finite
region $-p_{\pi}^0 \leq q^0 \leq 0$ and the region $0 \leq q^0
$. Accordingly, 
the correlation
function is written into a sum of the singular term and the regular 
term. The singular term is written with  the light-cone
singularity and the power series in 
Eq.\,$(\ref{singular-function})$. Hence this series must converge
 for the present  method of extracting the light-cone singularity 
to be applicable.

We study  the power series 
\begin{eqnarray} 
\label{power-series}
\sum_n (-2p_{\pi} \!\cdot\! p_{\nu} )^n {1 \over n!} 
\left({\partial  \over \partial {\tilde m}^2}\right)^n \tilde L_1,
\end{eqnarray}
 using the asymptotic expression of $\tilde
L_1$, Eq.\,$(\ref{asymptotic-expansionL_1})$, first.  The most
weakly converging  term
in $\tilde L_1$, is from ${\tilde m}^{\frac{1}{2}}$ and other terms converge
when this converges.  The series  
\begin{eqnarray}
S_1=\sum_n (-2p_{\pi} \!\cdot\! p_{\nu} )^n {1 \over n!} 
\left({\partial  \over \partial {\tilde m}^2}\right)^n ({\tilde m}^2)^{\frac{1}{4}}.
\end{eqnarray}
 becomes  the form,
\begin{align}
S_1&=\sum_n \left({-2p_{\pi} \!\cdot\! p_{\nu} \over {\tilde m}^2 }\right)^n {1 \over
 n!} \left(n-\frac{1}{4}\right)!(-1)^n ({ \tilde m})^{\frac{1}{2}}  \nonumber\\
&\approx \sum_n \left(-{2p_{\pi} \!\cdot\! p_{\nu} \over {\tilde m}^2} \right)^n
 (-1)^n n^{-\frac{5}{4}}({ \tilde m})^{\frac{1}{2}} =\sum_n \left({2p_{\pi} \!\cdot\! p_{\nu}\over {\tilde m}^2} \right)^n n^{-\frac{5}{4}} 
({ \tilde
 m})^{\frac{1}{2}}.
\end{align} 
Hence the series converges if the geometric ration is less than 1. At $2p_{\pi} \cdot p_{\nu}={\tilde
m}^2 $ $S_1$ becomes finite, and the value is expressed by the zeta function,
\begin{eqnarray}
S_1=\sum_n n^{-\frac{5}{4}} ({ \tilde m})^{\frac{1}{2}}=\zeta\left(\frac{5}{4}\right)  ({
 \tilde m})^{\frac{1}{2}}.
\end{eqnarray}
Hence in the region,
\begin{eqnarray}
{2p_{\pi}\!\cdot\! p_{\nu} \over {\tilde m}^2} \leq 1,
\label{convergence-condition-ratio}
\end{eqnarray}
 the series converges and 
correlation function has the singular terms.
Outside  this region, the power series diverges and our method
does not work. $I$ is evaluated directly and agree with 
the $I_2$.  

The power series Eq.\,$(\ref{power-series})$ oscillates with
time $\sqrt{|\delta t|}$ rapidly. The series
\begin{eqnarray}
S_2=\sum_n (-2p_{\pi} \!\cdot\! p_{\nu} )^n {1 \over n!} 
\left({\partial  \over \partial {\tilde m}^2}\right)^ne^{i\tilde m \sqrt{2
 v_{\nu}\sigma_{\nu} |\vec{p}_{\nu}||\delta t| }},
\end{eqnarray}
is computed in the form,
 \begin{eqnarray}
S_2=
e^{i\tilde m\left|\sqrt{2
 v_{\nu}\sigma_{\nu} |\vec{p}_{\nu}||\delta t| }\right|\left(1-\frac{p_{\pi} \cdot p_{\nu}}
{{\tilde m}^2}\right)},
\end{eqnarray}
and  oscillates with $\sqrt{|\delta t|}$. So the present method of
separating the light-cone singularity  from the correlation function and
of evaluating  the finite-size correction of the probability is valid in
the kinematical region Eq.\,$(\ref{convergence-condition-ratio})$. 
\subsection{Time-dependent probability}

Substituting Eq.\,(\ref{lightcone-integration1-1}) into Eq.\,(\ref{probability-correlation1}), we have 
the probability for detecting  the neutrino at
a space-time position $(\text{T}_{\nu},{\vec X}_{\nu})$, when the pion momentum distribution $\rho_{exp}({\vec
p}_{\pi})$ is known, in the following form  
\begin{align}
&\int d{\vec p}_{\pi} \rho_{exp}({\vec
p}_{\pi})  d{\vec p}_{\mu} d{\vec X}_{\nu} d{\vec p}_\nu  \sum_{s_1,s_2}|T|^2 
=g^2 m_{\mu}^2
 |N_{\pi\nu}|^2{(\sigma_{\nu}\pi)}^{\frac{3}{2}}\frac{\sigma_{\nu}}{(2\pi)^6} \int {d
 \vec{p}_{\pi} \over E_\pi}\rho_{exp} \label{total-probability2}({\vec p}_{\pi})\nonumber \\
&\times\int d{\vec X}_{\nu}{d{\vec
 p}_\nu  \over E_{\nu}}   p_{\pi} \!\cdot\! p_{\nu} \int dt_1 dt_2
\left[    e^{i {m_{\nu}^2  \over 2E_{\nu}}\delta t} 
\frac{\epsilon(\delta t)}{|\delta t|}
 +{2 \tilde L_1 \over \sigma_{\nu}}-i{2 \over \pi}\left( {\sigma_\nu \over \pi}\right)^{\frac{1}{2}}\tilde L_2\right]\nonumber\\
&\times  e^{-{1 \over 2 \sigma_{\pi} } \left({\vec X}_{\nu}
-{\vec X}_{\pi}+({\vec v}_{\nu}-\bar {\vec v}_{\pi})(t_1-\text{T}_{\nu}) + 
\bar{\vec{v}}_{\pi}(\text{T}_{\pi}- \text{T}_{\nu})\right)^2-{1 \over 2 \sigma_{\pi} }\left({\vec X}_{\nu}-{\vec X}_{\pi}+({\vec
v}_{\nu}-\bar{{\vec v}}_{\pi})(t_2-\text{T}_{\nu}) + \bar {\vec{v}}_{\pi}(\text{T}_{\pi}
- \text{T}_{\nu})\right)^2}.
\end{align}
From a pion mean free path obtained in the Appendix of I, the
coherence condition, Eq.\,$(\ref{coherence-condition})$, is satisfied
and  the pion Gaussian parts are regarded as constant in $t_1$ and $t_2$,
 \begin{align}
& e^{-\frac{1}{2\sigma_{\pi}}\left(\vec{X}_{\nu} - \vec{X}_{\pi} +
			     (\vec{v}_{\nu} - \bar {\vec{v}}_{\pi})(t_1 -
			     \text{T}_{\nu}) + \bar {\vec{v}}_{\pi}(\text{T}_{\pi} -
			     \text{T}_{\nu})\right)^2}\approx \text{constant in
}t_1,\nonumber   \\
& e^{-\frac{1}{2\sigma_{\pi}}\left(\vec{X}_{\nu} - \vec{X}_{\pi} +
			     (\vec{v}_{\nu} - \bar {\vec{v}}_{\pi})(t_2 -
			     \text{T}_{\nu}) + \bar {\vec{v}}_{\pi}(\text{T}_{\pi} -
			     \text{T}_{\nu})\right)^2}\approx \text{constant in
}t_2,\label{coherence-conditions}
 \end{align}
when an integration over  $t_1$ and $t_2$ are made in a distance of 
our interest which is of the order of a few $100$ [m].  
The integration over  $t_1$ and $t_2$ will be made in the next section.

 When the above  conditions Eq.\,$(\ref{coherence-conditions})$ are
 fulfilled, an area where the neutrino is produced is inside of a
 same pion and  neutrino waves are treated coherently and are capable
 of showing interference. In a much larger distance where this 
condition is not satisfied,  two positions can not be in the same pion
 and the interference disappears.

\subsubsection{Integrations over times}
Integrations over  the times $t_1$ and $t_2$ are carried
and a probability at a finite  T is obtained here. An  integral  of 
the slowly decreasing   term over $t_1$ and $t_2$ is  
\begin{eqnarray}
& &i  \int_0^{\text{T}} dt_1 dt_2  {e^{i {\omega_{\nu}}\delta t }
 \over |\delta t|}\epsilon(\delta t)   
= \text{T} \left\{\tilde g(\text{T},\omega_{\nu})+g(\infty,\omega_{\nu})\right\},\ 
\omega_{\nu}={m_{\nu}^2 \over 2E_{\nu}},\label{probability1} 
\end{eqnarray}
where $\tilde g(\text T,\omega_{\nu})$ vanishes at $\text T \rightarrow \infty$.
We understand that the short-range part $L_1$ cancels with
$g(\infty,\omega_\nu) $ and write the total probability with 
$\tilde g(\text{T},\omega_\nu) $ and the short-range term from $ I_2$.

An integral over   times  of the short-range term, $ \tilde L_2$, is  
\begin{align}
&\frac{2}{\pi}\sqrt{\sigma_\nu \over \pi}\int dt_1 dt_2 \tilde
 L_2(\delta t) \nonumber\\
&= \frac{2}{\pi}\sqrt{\frac{\sigma_\nu}{\pi}} \int_0^{\text{T}} dt_1 dt_2 
 \int \frac{d\vec{q}}{2\sqrt{\vec{q}^{\,2}+m_{\mu}^2}} 
 e^{-i\left(E_{\pi}-E_{\nu}-\sqrt{\vec{q}^{\,2}+m_{\mu}^2}-{\vec v}_{\nu}\cdot({\vec
 p}_{\pi}-{\vec q}-{\vec p}_{\nu})\right)\delta t}  \nonumber \\
&\ \ \ \times e^{ -{ \sigma_{\nu} ({\vec p}_{\pi}-{\vec q}-{\vec p}_{\nu})^2}}
 \theta\left(E_{\pi}-\sqrt{\vec{q}^{\,2}+m_{\mu}^2}\right)
    \nonumber \\
&=\text{T} G_0,
\end{align}
where the constant $G_0$ is given in the integral   
\begin{eqnarray}
& &G_0=2 \sqrt{\sigma_\nu \over \pi}  \int \frac{d\vec{q}}{\sqrt{\vec{q}^{\,2}+m_{\mu}^2}} \delta \left(E_{\pi}-E_{\nu}-\sqrt{\vec{q}^{\,2}+m_{\mu}^2}-{\vec v}_{\nu}\cdot({\vec
 p}_{\pi}-{\vec q}-{\vec p}_{\nu})\right)  \nonumber
 \\
& &\times e^{ -{ \sigma_{\nu} ({\vec p}_{\pi}-{\vec q}-{\vec p}_{\nu})^2}}
 \theta\left(E_{\pi}-\sqrt{\vec{q}^{\,2}+m_{\mu}^2}\right), 
\end{eqnarray}
and is estimated numerically. Due to the rapid oscillation in $\delta t$,
$ \tilde L_2$'s contribution  to the probability comes from the 
small $|\delta t|$ region and the integrations over the time becomes 
constant in T.  Hence this has no finite-size  correction. The regular 
term $\tilde L_1$ is also the same.  
\begin{figure}[t]%
\begin{center}
\includegraphics[angle=-90,scale=.5]{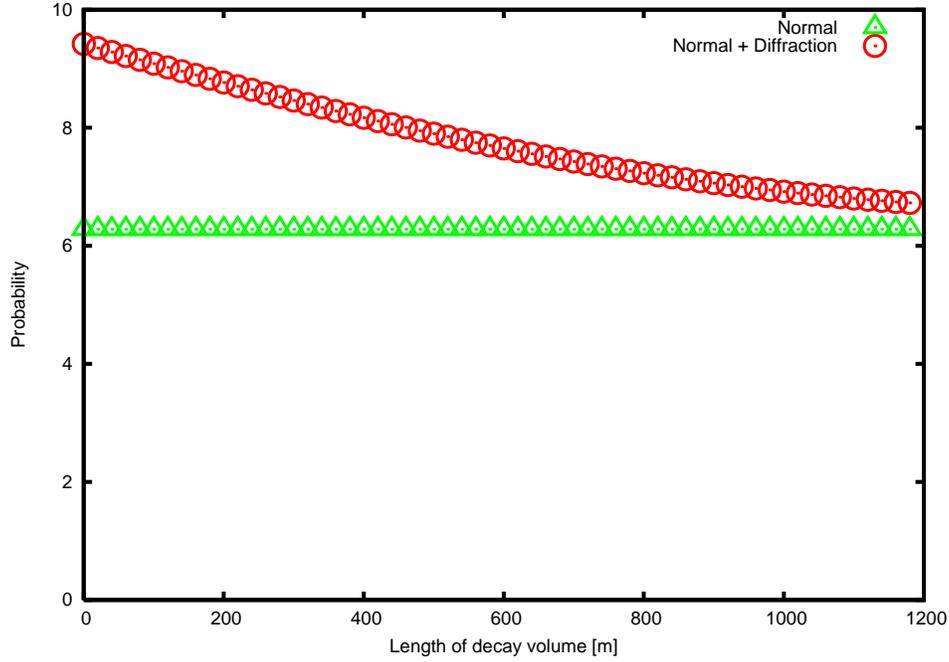}
   \end{center}
\caption{The  probability to detect the neutrino per time in the 
forward direction
 at a distance L is 
given. The constant shows the short-range normal term and   the
 long-range diffraction term is written on top of the normal term.  The horizontal axis 
shows the distance in~[m] and the probability  of the normal term is
 normalized to 2$\pi$.  Clear excess of more than $2/5$ of the normal
 term
 is seen in the 
 distance below 1200~[m]. The neutrino mass, pion energy, neutrino energy are
  1~[eV/$c^2$], 4~[GeV], and 800~[MeV]. Target is ${}^{16}$O.}
\label{fig:virtual-pi-singular}
\end{figure}%

\subsubsection{Total transition probability    }

Adding   the slowly decreasing part  and the short-range part,  we have 
the final expression of the total 
probability. The neutrino coordinate ${\vec X}_{\nu}$  is integrated in Eq.\,($\ref{total-probability2})$ and a factor  $({\sigma_{\pi}\pi})^{\frac{3}{2}}$ emerges. 
This factor is cancelled  with   $({4\pi/\sigma_{\pi}})^{\frac{3}{2}}$
of   the
normalization in Eq.\,$(\ref{probability})$ and a final result is 
independent of
$\sigma_{\pi}$.  The total transition 
probability is expressed in the form,  
\begin{eqnarray}
& &P=\text{T}g^2 m_{\mu}^2
 D_0  \sigma_{\nu}
\int {d\vec{p}_{\pi} \over E_{\pi}}\rho_{exp}({\vec
p}_{\pi})\int {d\vec{p}_{\nu} \over E_{\nu}}  (p_{\pi}\! \cdot\! p_{\nu}) 
 [\tilde g(\text{T},\omega_{\nu}) 
 +G_0 ], 
\label{probability-total}
\nonumber\\
& &
D_0=|N_{\pi\nu}|^2{(\sigma_{\nu}\pi)}^{\frac{3}{2}}
{(\sigma_{\pi}\pi)}^{\frac{3}{2}}{1 \over (2\pi)^6}={1 \over (2\pi)^3},
\label{probability-31}
\end{eqnarray}
where $\text{L} = c\text{T}$ is the length of decay region. 
The first term in the right-hand side of Eq.\,$(\ref{probability-total})$
depends on the time interval T, and the neutrino wave packet 
size $\sigma_{\nu}$, but the second
term does not.

At  a finite $\text{T}$, the first term, which we call a diffraction
 term, does not vanish and
the probability Eq.\,$(\ref{probability-total})$ has the finite-size correction. Its  relative ratio over  the normal 
term $G_0$ is  independent   of detection process. So we compute  
$\tilde g(\text{T},\omega_{\nu})$ and $G_0$ of 
Eq.~($\ref{probability-31}$) at the forward direction $\theta=0$ and the 
energy dependent total probability that is integrated over the neutrino
angle in the following.  

The probabilities per unit  time  in the forward direction 
are plotted in 
Fig.\,\ref{fig:virtual-pi-singular} for the mass of neutrino,
 $m_{\nu}=1\,[\text{eV}/c^2]$,  the pion of the sharp energy $E_\pi = 4$\,[GeV], and the neutrino
 energy $E_\nu = 800$\,[MeV]. For the wave packet size of the neutrino, the size
 of the nucleus of the mass number $A$, $\sigma_{\nu}=
 A^{\frac{2}{3}}/m_{\pi}^2$ 
is used. The value becomes  $\sigma_{\nu}= 6.4/m_{\pi}^2$ for the ${}^{16}$O nucleus and this is used for the following
 evaluations.   From 
this figure it is seen   that there
 is an excess of the flux at short distance region $\text{L}<600$ [m] and the
 maximal excess is about $0.4$ at $\text{L}=0$. The slope at  $\text{L}=0$ is
 determined by $\omega_{\nu}$.
The slowly decreasing   term has  the  finite magnitude and 
 the finite-size correction is large. 

\section{Neutrino spectrum}

\subsection{Integration over neutrino angle}
In Eq.\,($\ref{probability-31} $), the diffraction term $\tilde g(\text
T,\omega_{\nu})$  has a different dependence on the 
angle  from that of  the normal term $G_0$.
In the normal term $G_0$,  the cosine of
neutrino angle $\theta$ is
determined approximately from a mass-shell condition,
\begin{eqnarray}
(p_{\pi}-p_{\nu})^2=p_{\mu}^2=m_{\mu}^2,
\end{eqnarray}
because the energy and momentum conservation is approximately well satisfied 
in the normal term.  
Hence the product of the momenta is expressed with the masses
\begin{eqnarray}
p_{\pi}\!\cdot\! p_{\nu}={m_{\pi}^2-m_{\mu}^2 \over 2}, 
\label{on-shell-angle}
\end{eqnarray}
and  the cosine of the angle satisfies 
\begin{eqnarray}
1-\cos \theta= {m_{\pi}^2-m_{\mu}^2 \over 2|\vec{p}_{\pi}||\vec{p}_{\nu}|}-{m_{\pi}^2
 \over 2|\vec{p}_{\pi}|^2}.
\label{angle-energy-relation}
\end{eqnarray}
The $\cos \theta $ is very close to 1 in a high energy region.
On the other hand,  the diffraction  component, $\tilde
g(\text{T},\omega_{\nu})$ of Eq.\,($\ref{probability-31} $), 
is present in the  domain of the momenta
Eq.\,$(\ref{convergence-condition-ratio})$ i.e., in the kinematical region,  
\begin{eqnarray}
\label{long-kinematical}
|\vec{p}_{\nu}|(E_{\pi}-|\vec{p}_{\pi}|)\leq p_{\pi}\!\cdot\! p_{\nu} \leq {m_{\pi}^2-m_{\mu}^2 \over 2}. 
\end{eqnarray}
Since  the angular region of Eq.\,($\ref{long-kinematical}$) is 
slightly different from Eq.\,$(\ref{on-shell-angle})$ and it is impossible
to distinguish the latter  from the former region
experimentally, the neutrino angle is integrated.   
We integrate over the neutrino angle of both terms separately.
We have  the normal term, $G_0$, in the form
\begin{align}
& \int \frac{d\vec{p}_\nu}{E_\nu} 
(p_\pi\cdot p_\nu) G_0  \nonumber\\
 &\simeq \int \frac{d\vec{p}_\nu}{E_\nu}
 (p_\pi\!\cdot\! p_\nu
 )2\sqrt{\frac{\sigma_\nu}{\pi}}\left(\frac{\pi}{\sigma_\nu}\right)^{\frac{3}{2}}\int \frac{d\vec{q}}{\sqrt{\vec{q}^{\,2} + m_\mu}} \nonumber \\
&\times \delta\left(E_\pi - E_\nu - \sqrt{\vec{q}^{\,2} + m_\mu^2}\right)\delta^{(3)}\left(\vec{p}_\pi - \vec{p}_\nu -
 \vec{q}\right)\theta\left(E_\pi - \sqrt{\vec{q}^{\,2} + m_\mu^2}\right) \nonumber\\
 &=\frac{(2\pi)^2}{\sigma_\nu}\left({m_{\pi}^2-m_{\mu}^2 \over
 2}\right){1 \over |\vec{p}_{\pi}|}\int_{E_{\nu,min}}^{E_{\nu},max} dE_{\nu},
\end{align}
where 
\begin{eqnarray}
E_{\nu,min}={m_{\pi}^2-m_{\mu}^2 \over
 2(E_{\pi}+|\vec{p}_{\pi}|)},\ E_{\nu,max}={m_{\pi}^2-m_{\mu}^2 \over
 2(E_{\pi}-|\vec{p}_{\pi}|)},
\end{eqnarray}
and  the Gaussian function is approximated by the delta function for
the computational convenience. The angle is determined uniquely. 

 We compute the diffraction term next. There are two cases depending on
 the  minimum angle of satisfying the convergence  condition Eq.\,$(\ref{convergence-condition-ratio})$.
 In the
 first energy region, 
\begin{align}
-1 \leq {\frac{E_\pi E_\nu - \frac{1}{2}(m_\pi^2 - m_\mu^2)}{|\vec{p}_\pi||\vec{
 p}_\nu|}} ,
\end{align}
the convergence  condition is satisfied  in a limited region of the angle. We
integrate over this region of the angle and  have  
 the diffraction term
in the form
\begin{align}
 &\int \frac{d\vec{p}_\nu}{E_\nu} (p_\pi\!\cdot\! p_\nu)
 \tilde{g}(\text{T},\omega_\nu) \nonumber\\
&= 2\pi\int \frac{|\vec{p}_\nu|^2d|\vec{p}_\nu|}{E_\nu} \int_{\frac{E_\pi E_\nu - \frac{1}{2}(m_\pi^2 - m_\mu^2)}{|\vec{p}_\pi||\vec{
 p}_\nu|}}^{1}d\cos\theta(E_\pi E_\nu
 - |\vec{p}_\pi||\vec{ p}_\nu|\cos\theta)\tilde{g}(\text{T},\omega_\nu) \nonumber\\
 & = 2\pi\int_{E_{\nu,min}}^{E_{\nu},max}  \frac{dE_\nu}{2|\vec{p}_\pi|}\left\{\frac{1}{4}\left(m_\pi^2 -
 m_\mu^2\right)^2 - (E_\pi E_\nu -
 |\vec{p}_\pi||\vec{p}_\nu|)^2\right\}\tilde{g}(\text{T},\omega_\nu).
\end{align}
Here  the angle is very close to the former value but is not uniquely
determined.
In the second  
 region, the convergence condition is satisfied in arbitrary angle, 
\begin{align}
& {\frac{E_\pi E_\nu - \frac{1}{2}(m_\pi^2 - m_\mu^2)}{|\vec{p}_\pi||\vec{
 p}_\nu|}} \leq -1 ,
\end{align} 
and we have  the diffraction term in the form
\begin{align}
 &\int \frac{d\vec{p}_\nu}{E_\nu} (p_\pi\!\cdot\! p_\nu)
 \tilde{g}(\text{T},\omega_\nu) \nonumber\\
&= 2\pi\int \frac{|\vec{p}_\nu|^2d|\vec{p}_\nu|}{E_\nu} 
\int_{-1}^{1}
d\cos\theta(E_\pi E_\nu
 - |\vec{p}_\pi||\vec{ p}_\nu|\cos\theta)\tilde{g}(\text{T},\omega_\nu) \nonumber\\
 & = 4\pi\int_{0}^{E_{\nu},min} dE_\nu  E_\pi E_{\nu}^2
\tilde{g}(\text{T},\omega_\nu),
\end{align}
\begin{figure}[t]%
\centering{\includegraphics[scale=.6,angle=-90]{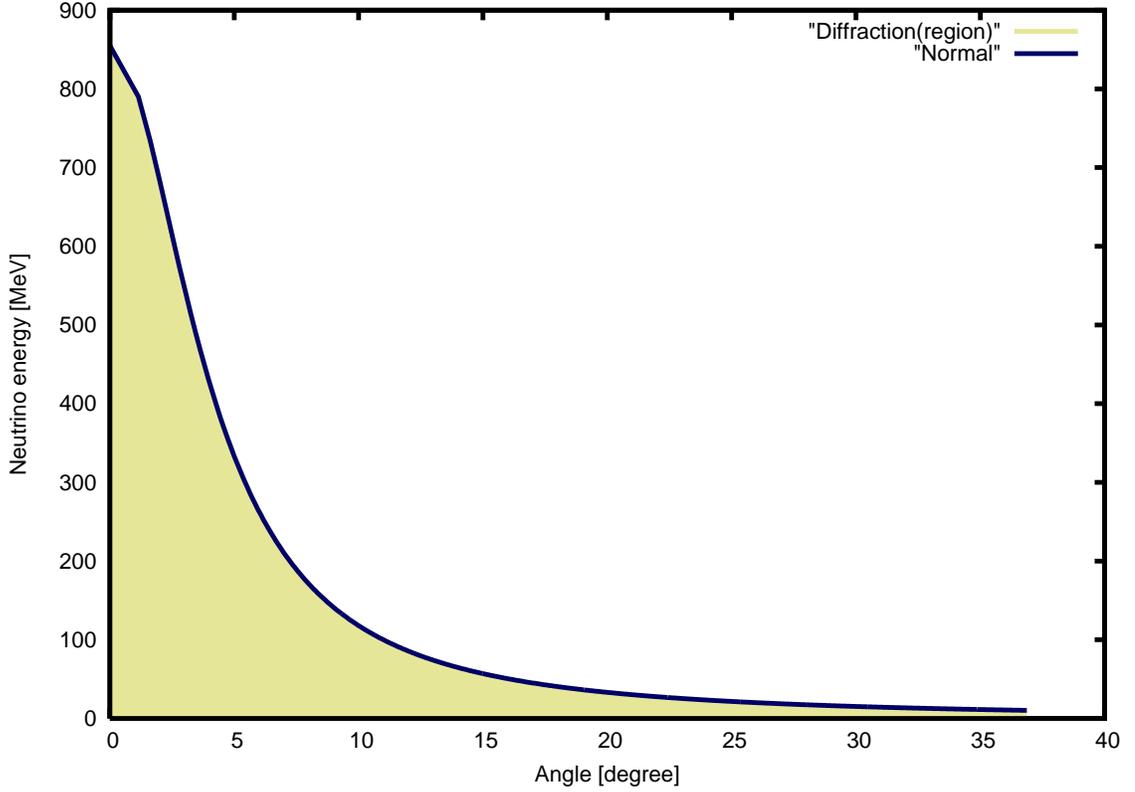}
\caption{The relation between the neutrino angle and energy is
 shown. The energy is determined uniquely with a value on the line in 
the normal component and takes a finite range under the line  in the 
diffraction
  term.   The
 energy of the pion is $E_\pi=2$ [GeV].}}
\label{figure:angle-energy}
\end{figure}%

In the diffraction term, the  angle of the neutrino is different from that
 of the normal term.  The angle  
 dependences of the energy  of normal and diffraction terms are given in
  Fig.\,3. The angle is fixed to one value in the normal term and is
 in a continuous range in the diffraction term.

  Finally we have the energy dependent probability
\begin{align}
\frac{dP}{dE_\nu}&=\text{T}g^2 m_{\mu}^2
 D_0  \int {d
 \vec{p}_{\pi} \over E_{\pi}}\rho_{exp}({\vec
p}_{\pi}) 
 \frac{2\pi}{|\vec{p}_\pi|}\times\biggl[ {\pi}( m_\pi^2
 - m_\mu^2)  \nonumber\\
&+\frac{ \sigma_{\nu}}{2}\biggl(\theta(E_{\nu}-E_{\nu,min})\left\{ \frac{1}{4}\left(m_\pi^2 -
 m_\mu^2\right)^2 - (E_\pi E_\nu - |\vec{p}_\pi|
 |\vec{p}_\nu|)^2 \right\} \nonumber\\
&+\theta(E_{\nu,min}-E_{\nu})
 2E_{\pi}E_{\nu}^2\biggr)\tilde{g}(\text{T},\omega_\nu)\biggr]
\label{total-probability-energy}.
\end{align}
\subsection{Neutrino spectrum }
\subsubsection{Sharp pion momentum}
When the initial pion has a discrete momentum ${\vec P}_{\pi}$, the
$\rho_{exp}({\vec p}_{\pi})$ is given as  
\begin{eqnarray}
\rho_{exp}({\vec p}_{\pi})=\delta({\vec p}_{\pi}-{\vec P}_{\pi}),
\end{eqnarray}
and the probability is expressed in the form,
\begin{eqnarray}
& &\frac{dP}{dE_\nu}=\text{T}g^2 m_{\mu}^2
 D_0   {1 \over E_{\pi}} 
 \frac{2\pi}{|\vec{P}_\pi|}\biggl[ {\pi}( m_\pi^2
 - m_\mu^2) \label{total-probability-energy2}  \nonumber\\
& &+{ \sigma_{\nu} \over 2}\biggl(\theta(E_{\nu}-E_{\nu,min}) \left\{\frac{1}{4}(m_\mu^2 -
 m_\pi^2)^2 - (E_\pi E_\nu - |\vec{P}_\pi|
 |\vec{p}_\nu|)^2\right\}\nonumber\\
& &+\theta(E_{\nu,min}-E_{\nu})
 2E_{\pi}E_{\nu}^2\biggr)\tilde{g}(\text{T},\omega_\nu)\biggr].
\end{eqnarray}

Eq.\,$(\ref{total-probability-energy2})$ is independent of the position
${\vec X}_{\pi}$ and an average over  ${\vec X}_{\pi}$ is easily made.
The result is  obviously the same as Eq.\,$(\ref{total-probability-energy2}  )$.

The probability depends upon the momenta and
the time interval $\text{T}=\text{T}_{\nu}-\text{T}_{\pi}$.
At $\text{T} \rightarrow \infty$, $\tilde{g}(\text{T},\omega_\nu)$ vanishes
and the probability per unit time, a decay rate,  of the energy $E_{\pi}$ is 
given  from the
first term of  Eq.\,$(\ref{total-probability-energy2})$ as
 \begin{eqnarray}
P/\text{T}&=&g^2 m_{\mu}^2
 D_0   {1 \over E_{\pi}}
 \frac{2\pi}{|\vec{P}_\pi|} {\pi}( m_\pi^2
 - m_\mu^2) \int_{E_{\nu,min}}^{E_{\nu,max}} dE_{\nu}\nonumber\\
&=&g^2 m_{\mu}^2 {1 \over 4\pi}\frac{m_{\pi}^2}{E_{\pi}}\left(1-{m_{\mu}^2
 \over m_{\pi}^2}\right)^2.
\end{eqnarray} 
This  value is independent of  the wave packet size and is
consistent with \cite{Stodolsky}. This,  furthermore, agrees with  the standard 
value obtained using the plane waves. Hence the large T limit of our 
result is equivalent to the known result obtained with the standard S-matrix.


 \begin{figure}[t]%
\begin{center}
  \includegraphics[angle=-90,scale=.5]{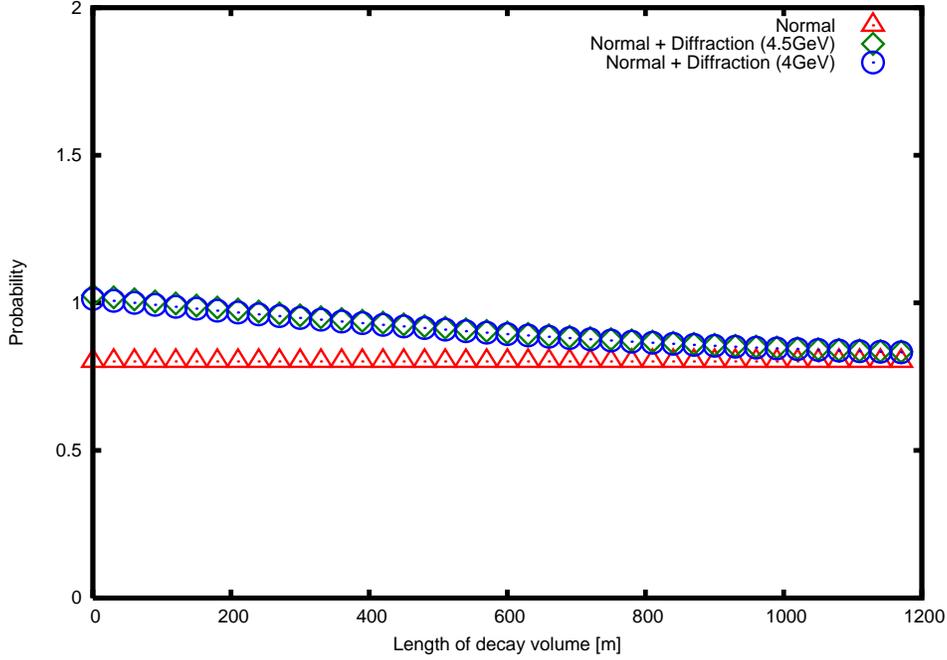}
   \end{center}
  \caption{The  total probability per time  integrated over  the neutrino
  angle  at a finite distance L is 
given. The constant shows the  normal term and   the diffraction term is 
written on top of the normal term.  The horizontal axis 
shows the distance in~[m] and the total probability  is normalized to a 
unity at $\text{L}=0$. The excess becomes less clear than the forward
  direction, but is seen in the distance below 1200\,[m]. The neutrino mass, pion energy, neutrino energy are
  1.0~[eV/$c^2$], 4~[GeV] and 4.5~[GeV] , and 800~[MeV]. Target is ${}^{16}$O.}
 \label{fig:total-int-1}
\end{figure}%
 \begin{figure}[t]%
\begin{center}
  \includegraphics[angle=-90,scale=.5]{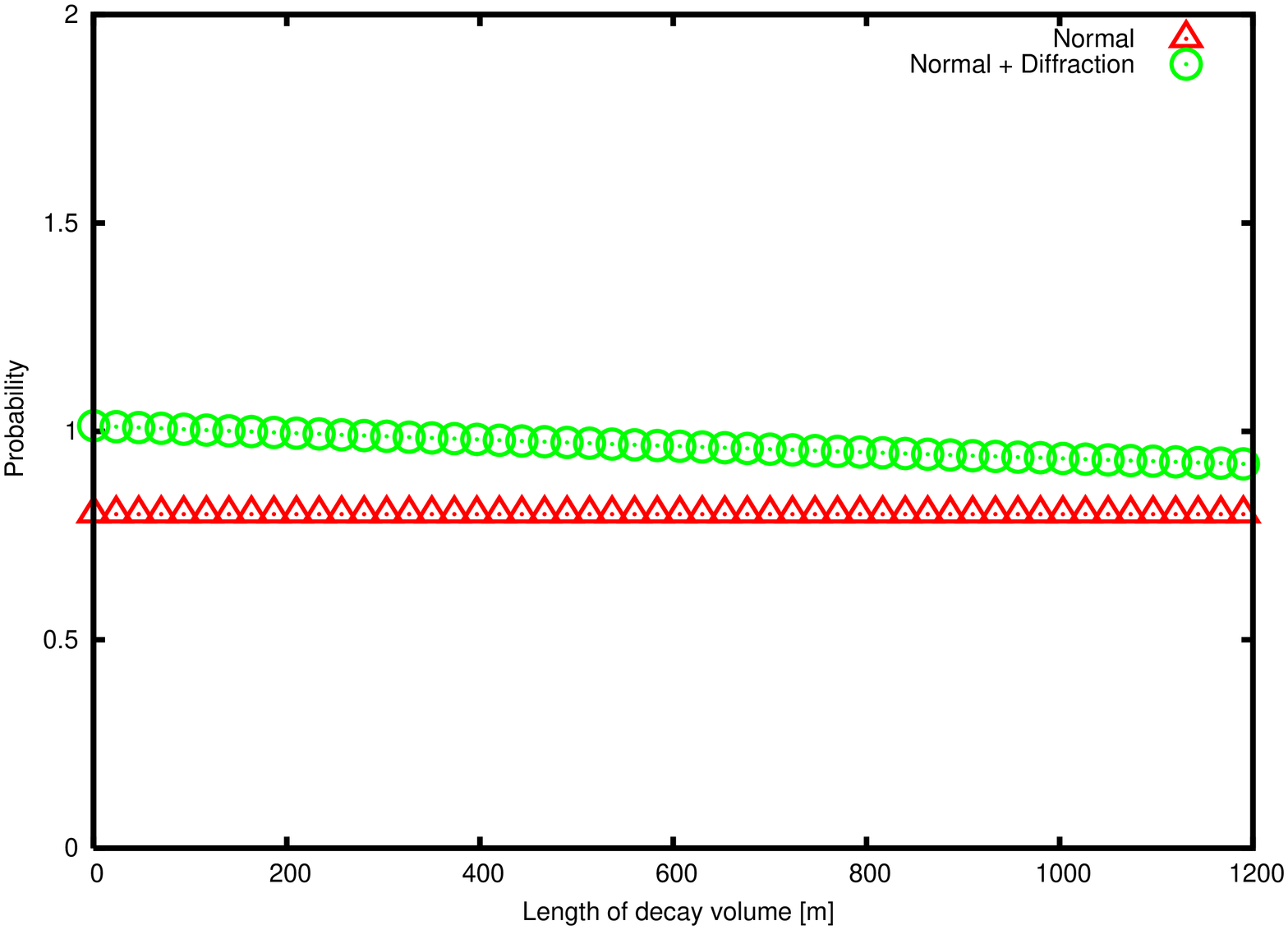}
   \end{center}
  \caption{The  total   probability integrated over  the  neutrino angle per time at a finite distance L is 
given. The constant shows the  normal term and   the diffraction  term is written on top of the normal term.  The horizontal axis 
shows the distance in~[m] and the probability of the normal term  is
  normalized to 0.8. Clear uniform excess is seen in the
 distance below 1200\,[m]. The neutrino mass, pion energy, neutrino energy are
  0.6~[eV/$c^2$], 4~[GeV], and 800~[MeV]. Target is ${}^{16}$O.}
 \label{fig:total-int-.6}
\end{figure}%
\subsubsection{Position dependence}

\begin{figure}[t]
   \begin{center}
   \includegraphics[angle=-90,scale=.5]{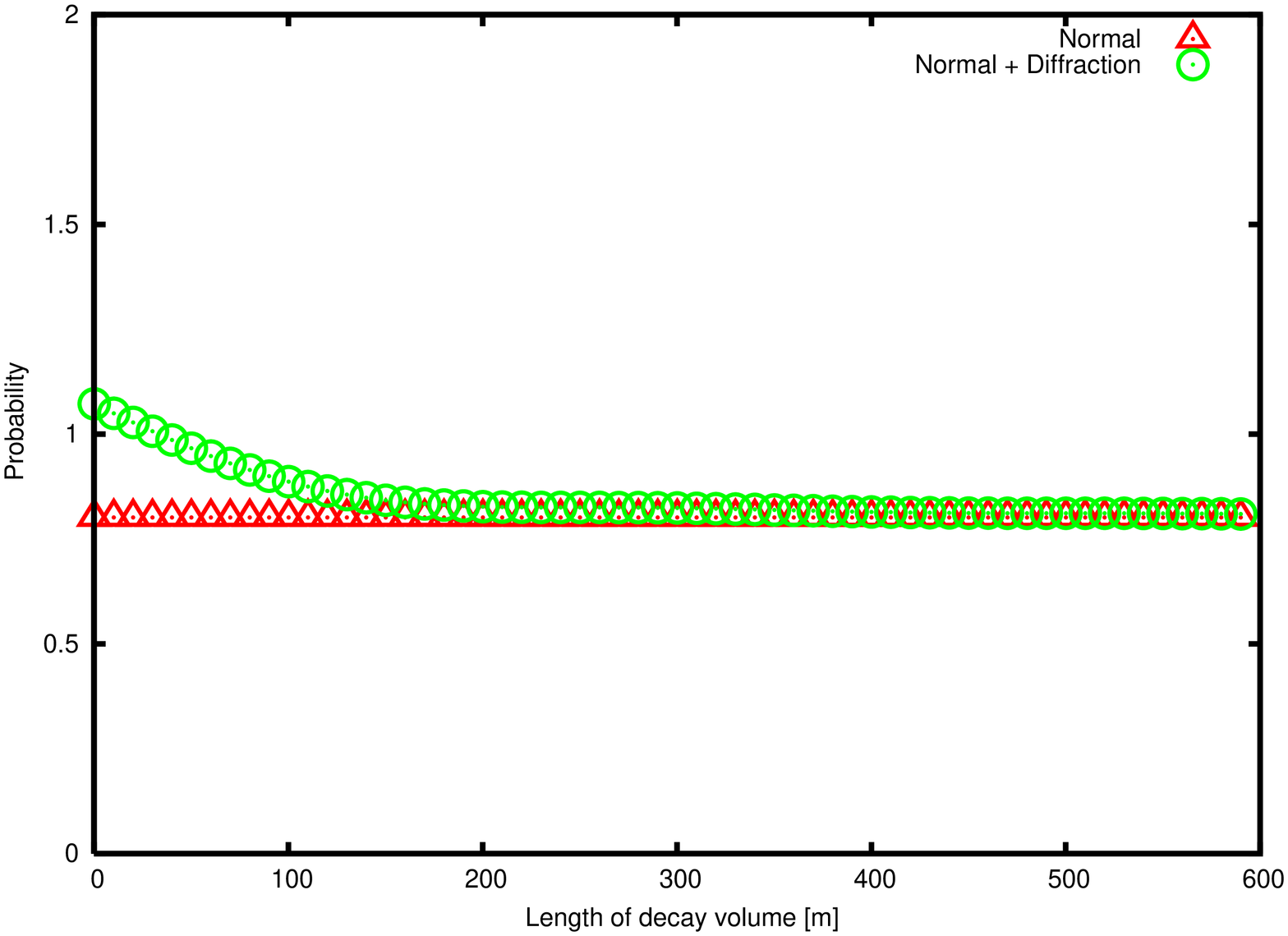}
   \end{center}
\caption{The  total  probability integrated over  the angle per time at a 
finite distance L is 
given. The constant shows the  normal term and   the diffraction 
 term is written on top of the normal term.  The horizontal axis 
shows the distance in~[m] and the probability of the normal term  is
 normalized to 0.8.  Clear excess and decreasing behavior are  seen in the
 distance below 600~[m]. The neutrino mass, pion energy, neutrino energy are
  1~[eV/$c^2$], 4~[GeV], and 100~[MeV]. Target is ${}^{16}$O.}
 \label{fig:total-int-1-100}
\end{figure}

The position dependence of the total probability per time $P/\text{T}$,
 Eq.\,$(\ref{total-probability-energy})$ is
 presented next. $\tilde g(\text T,\omega_{\nu})$ varies with the
 distance L defined by $\text{L}=c\text T$, whereas  $G_0$ is
 constant. The probability depends upon the magnitude $\omega_{\nu}\text{T}$ 
and for the  neutrino mass
 $m_{\nu}=1.0\,[\text{eV}/c^2]$ and the pion energy  $4$\,[GeV] and
 $4.5$\,[GeV] are given in  Fig.\,$\ref{fig:total-int-1}$, and for the 
smaller neutrino mass 
$m_{\nu}=0.6\,[\text{eV}/c^2]$ is given in
 Fig.\,$\ref{fig:total-int-.6}$. 
 $\tilde
 g(\text{T},\omega_{\nu})$ of a lighter mass decreases  more slowly 
with the  distance than that  of
 $m_{\nu}=1\,[\text{eV}/c^2]$.  A longer distance is necessary
to see a signal  if the neutrino mass  is
 even smaller. For the detection of the muon neutrino, the neutrino energy
 should be larger than the muon mass, hence the experiment of the energy
 lower than  $100$\,[MeV] is impossible. For this energy, the electron 
neutrino is used then.  We present the total probability for the lower
 energies next. The probability for $m_{\nu}=1.0\,[\text{eV}/c^2]$ with the energy  $100$\,[MeV]
is given in  Fig.\,$\ref{fig:total-int-1-100}$.
The slowly decreasing component of the probability becomes more
 prominent with lower values.  Hence to observe this component, the 
experiment of the lower neutrino energy is more convenient.   
 \begin{figure}[t]
  \begin{center}
   \includegraphics[angle=-90,scale=.5]{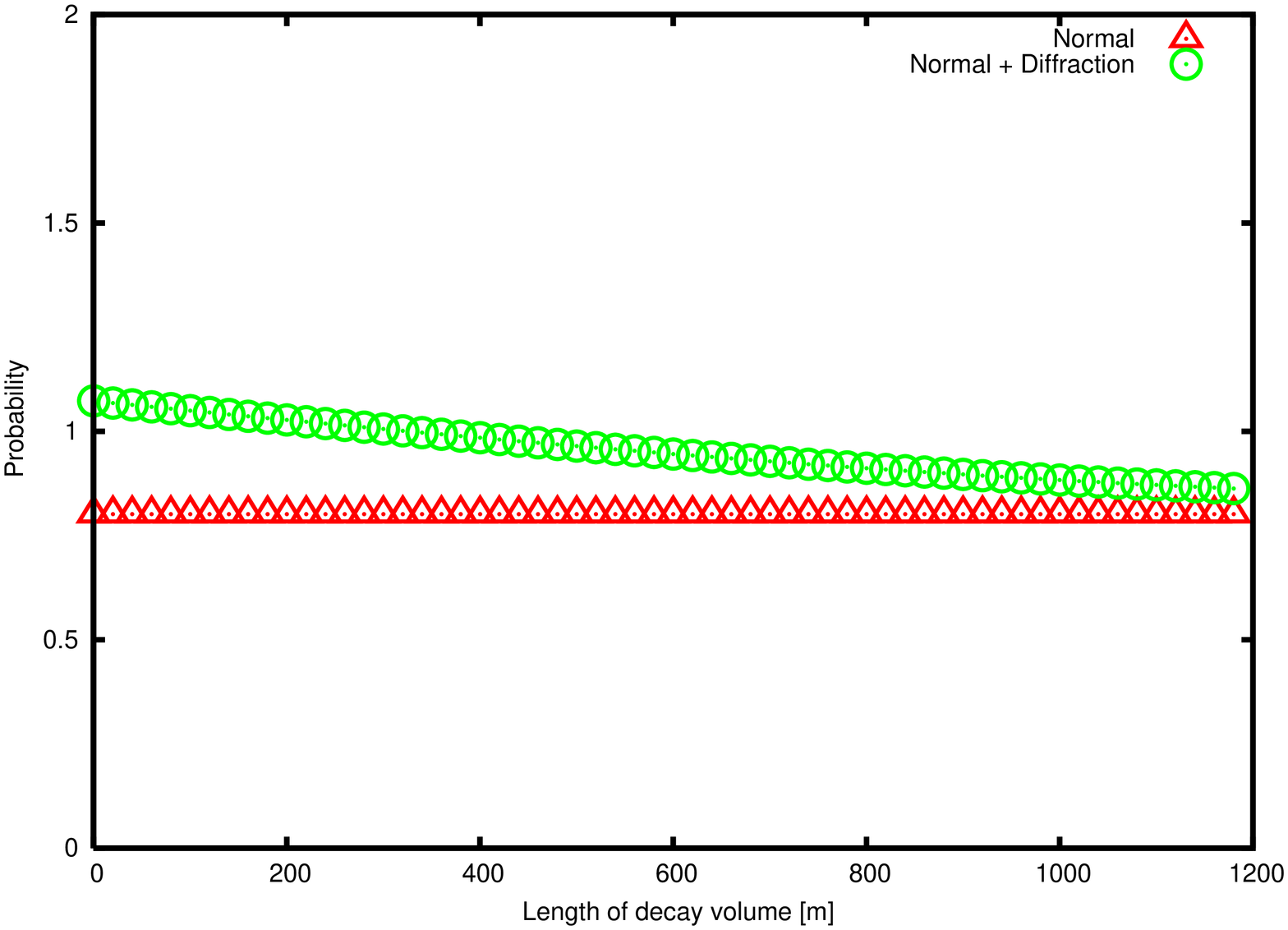}
  \end{center}
  \caption{The  probability integrated over the neutrino angle 
per time at a finite distance L is 
given. The constant shows the normal term and   the diffraction 
 term is written on top of the normal term.  The horizontal axis 
shows the distance in~[m] and the probability  is normalized to
  0.8. Clear 
excess is seen in the
 distance below 1200~[m]. The neutrino mass, pion energy, neutrino energy are
  0.1~[eV/$c^2$], 4~[GeV], and 10~[MeV]. Target is ${}^{16}$O.}
 \label{fig:total-int-.1-10}
\end{figure}%

From Eq.\,$(\ref{probability-31})$, and $\tilde g(\text T,\omega_{\nu})=\frac{c\omega_{\nu}}{
2\text{T}}$ at a large T, the typical length $l_0$ of the diffraction  term is  
\begin{eqnarray}
l_0~[\text{m}] ={2E_{\nu} \hbar c \over m_{\nu}^2 }= 400{E_{\nu}[\text{GeV}] \over
 m_{\nu}^2[\text{eV}^2/c^4]}.
\end{eqnarray}
The observation of this component together with the neutrino's energy
would make a  
determination of the neutrino absolute mass   possible. The neutrino's 
energy is measured with uncertainty $\Delta E_{\nu}$, which is of the 
order of $0.1 \times E_{\nu}$. This uncertainty is $100$\,[MeV] for the energy
$1$\,[GeV] and is accidentally same order as that of the minimum uncertainty 
$\hbar/\delta x$ derived from  the nuclear size $\delta x$. The total probability for a larger
value of energy uncertainty is easily computed using
Eq.\,(\ref{total-probability-energy}).   Figs.\,4-7 show the
distance dependence of the probability. If the mass is around $1\,[\text{eV}/c^2]$ the
excess of the neutrino flux of
about $20$ percent at the distance less than a few hundred meters is
found. 
We use mainly $m_{\nu}=1\,[\text{eV}/c^2]$ throughout this section.
Because the probability has a constant  term and the T-dependent 
term, the T-dependent   term is extracted  easily 
by subtracting the constant term from the total probability. The slowly 
decreasing  component decreases with the scale determined by the
neutrino's mass and the energy.

We plot the figure for $m_{\nu}=0.1
\,[\text{eV}/c^2]$, $E_{\nu}=10\,[\text{MeV}]$ in
Fig.\,$\ref{fig:total-int-.1-10}$.
A decreasing behavior  is clearly seen. So
in order to  observe    the slowly decreasing behavior for the small
neutrino  mass less than or about the same as $0.1\,[\text{eV}/c^2]$, the electron
neutrino should be used. The decay of the muon and others will be
studied in a forthcoming paper.

 \begin{figure}[t]
  \begin{center}
   \includegraphics[angle=-90,scale=.40]{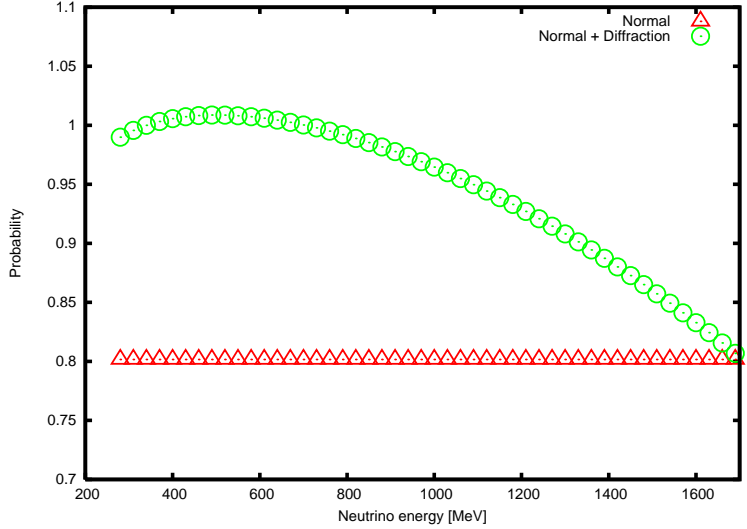}
  \end{center}
  \caption{The energy dependence of the probability integrated 
  over the angle at distance $\text{L} = 100$ [m] is 
given. The lower curve  shows the  normal term and   the diffraction
  term is added  on top of the normal term.  The horizontal axis 
shows the neutrino energy in~[MeV] and the probability of the normal
  term  is
 normalized to 0.8.  The neutrino mass and pion energy are
  1.0~[eV/$c^2$] and 4~[GeV]. Target is ${}^{16}$O.}
 \label{fig:total-ene-1}
\end{figure}%

\subsubsection{Energy  dependence}

The energy spectrum of the neutrino from the high-energy pion is
  studied next. Since the diffraction term has the origin in the final
  states that do not conserve  the kinetic energy, that should
  show unusual behavior. 
   In    Fig.\,$\ref{fig:total-ene-1}$, the spectrum for the neutrino mass 
and pion energy, 1.0~[eV/$c^2$] and 4~[GeV], are given. The  spectrum
  of the normal term is flat because the energy in the rest system is 
fixed to one value from the energy-momentum conservation, whereas  that 
of the  diffraction is not fixed to one value at the rest system and 
is not flat   but has a maximum at the energy 
$E_{\nu}\approx { E_{\nu,max}/3}$. The  diffraction term becomes
  much larger in much  higher energy.

A unique property of the neutrino diffraction is identified by
 its  energy spectrum. 
\begin{figure}[t]
\begin{center}
\includegraphics[scale=.40,angle=-90]{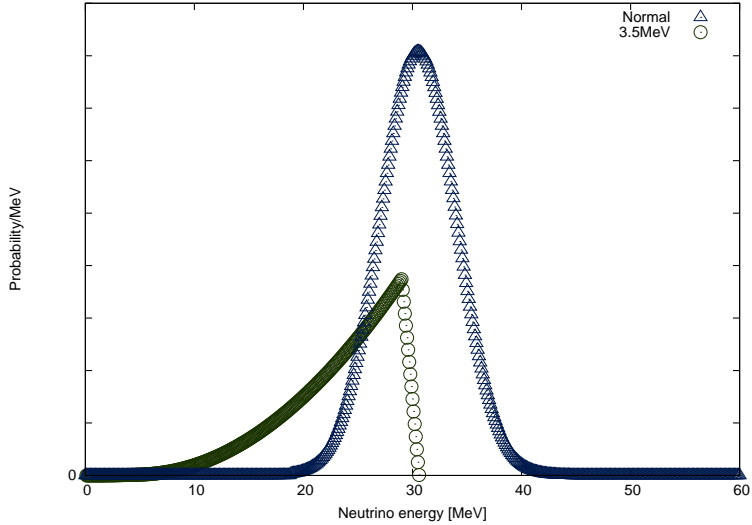}
\end{center}
\caption{The neutrino energy spectrum of the normal and 
 diffraction terms in the rest
 system of the pion are  given for the wave packet size of $5$ [MeV]. The 
 former spectrum becomes  wide due to the
 wave packet effect and the latter spectrum becomes  wider  than the
 normal component and is in the lower
 energy region.  The magnitude of the diffraction 
term is arbitrary.  The neutrino mass is 1.0~[eV/$c^2$]. The length is
 $\text{L}=10$ [m].
  }
\label{figure:rest-pion1}
\end{figure}
The energy spectrum of the normal term from a pion at rest  for the wave
packet size of the momentum width $5$ [MeV] is
given in  Fig.\,\ref{figure:rest-pion1}. The spectrum has 
a  peak at the value  derived from the  energy-momentum conservation,
\begin{eqnarray}
m_{\pi}=E_{\nu}+E_{\mu},\ {\vec p}_{\nu}+{\vec p}_{\mu}=0,
\end{eqnarray} 
of  the two body decay.
The neutrino energy is uniquely determined to the
value 
\begin{eqnarray}
E_{\nu}={m_{\pi}^2-m_{\mu}^2 \over 2m_{\pi}}.
\label{energy-rest}
\end{eqnarray}
The spectrum becomes broad due to the finite wave packet effect.  

The same figure shows another  component in the low energy region, which
 corresponds to the
  diffraction term. In the low energy region of the neutrino, the 
light-cone singularity is  not dominant and next to leading terms 
contribute. This figure does not include these non-leading terms, hence  
the magnitude of the diffraction 
term is arbitrary.  The length is $\text{L}=10$ [m].

\begin{figure}[t]
\begin{center}
\includegraphics[scale=.40,angle=-90]{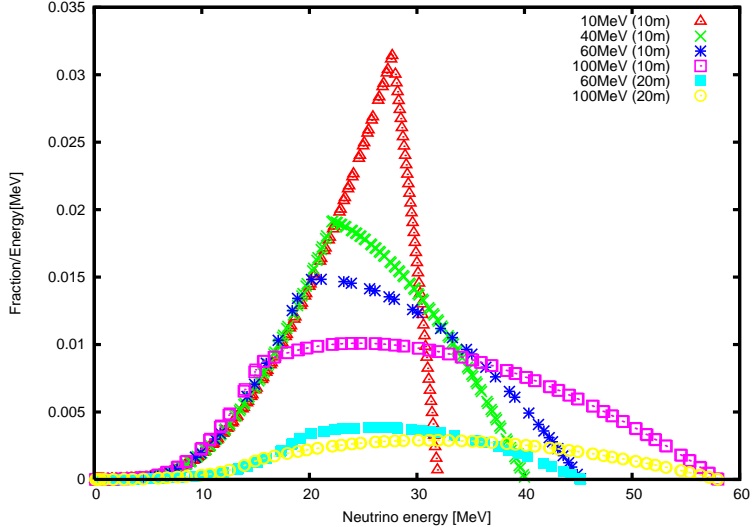}
\end{center}
\caption{The neutrino energy spectrum of the fraction of the 
 diffraction term are given for the energy of the pion 10, 40, 100 [MeV/c]
 and the length $\text{L}=10$ [m], and for the energy of the pion 60,
 100 [MeV/c]
 and the length $\text{L}=20$ [m]. Target is ${}^{56}$Fe. The fraction is small in lower energy and  
is larger in higher  energy.  The diffraction term may be observable in
 these energy regions too. The neutrino mass is 1.0~[eV/$c^2$].  }
\label{figure-intermediate-pion}
\end{figure}

 Fig.\,$\ref{figure-intermediate-pion}$  shows the energy spectrum of 
the fraction of the diffraction term over the normal term,
which are
  computed with the $(V-A)\times(V-A)$ current interaction and is 
represented in a  latter section(Sec.7.3) for ${}^{56}$Fe. Its energy   is different from
  Eq.\,$(\ref{energy-rest})$ and spreads over  in a wide   energy
  regions. Only the leading 
  term is taken into account in this figure.
Because the energy is away  from the normal term, this  component  may 
look like  a background noise which is
  uncorrelated with the system.  The magnitude becomes  small 
in a lower energy region, because  the neutrino spreads  uniformly 
in all angle there. 
It is noted that
  the finite-size correction is not invariant under the Lorentz
  transformation and the magnitude of the diffraction term becomes 
  larger in  higher energy. 
Thus the fraction of the electron mode varies with the pion's energy. 
{\bf This unusual
behavior is a characteristic feature of the diffraction component.} Our 
result, in fact, shows that this  background becomes  larger as  the 
pion's energy becomes larger but  has
   the universal property.

\subsubsection{Wide distribution of pion momentum  }
When a momentum distribution $\rho_{exp}({\vec p}_{\pi})$ of 
initial pions is known,  an energy-dependent probability 
is computed  using  the expression   Eq.\,$(\ref{total-probability-energy})$.
 Eq.\,$(\ref{total-probability-energy})$ is also independent of the position
${\vec X}_{\pi}$ and depends upon  a pion momentum and 
a neutrino momentum  and the time interval  $\text{T}=\text{T}_{\nu}-\text{T}_{\pi}$. In
experiments, a position of a pion is not measured and an average over 
a position is made. 
An  average probability  agrees with  
Eq.\,$(\ref{total-probability-energy})$. This  probability varies slowly 
with  the pion's momentum and is regarded constant in the energy range 
of the order of $100$\,[MeV]. So the experimental observation of the diffraction  
term is quite easy.
   
\subsection{On the universality of the diffraction term }
 The finite-size correction of the 
probability to detect the neutrino has various unique properties. 
This component is decreasing with the time interval T, hence the total 
probability is not
proportional to T in this region. In classical particle's decay, the 
decay process occurs randomly and follow Markov process. Hence 
an average number of   decay products is necessary proportional
to T. Now due to the finite-size correction, this property does not
hold. This is not surprising in the interference region $\text{L} \leq l_0$,
because the quantum mechanical interference effect modifies the
probability.

The finite-size correction  is expressed with the universal function $\tilde
g(\omega_{\nu},\text {T})$, where $\omega_{\nu}={m_{\nu}^2 \over 2E_{\nu}}$. This  
is  determined  only with  the mass and energy of 
the neutrino and is  independent of details of other parameters of 
the system such as the size, shape, and position  of the wave packets
and others. Hence the diffraction component has    
the genuine property of the wave function 
$|\text {muon,~neutrino}(t)\rangle$, of Eq.\,$(\ref{state-vector} )$, and is capable of  
experimental measurements.

\subsubsection{Violation of conservation laws}
The probability is computed with $S[\text {T}]$ and reflects the wave
function at a finite time $t$, hence  the states of 
non-conserving  kinetic energy contribute to the 
finite-size correction.  
  So conservation  laws that are connected with the space-time symmetry 
get  modified and various probabilities become different from those of
$\text {T} \rightarrow \infty$.
The leading finite-size corrections have, nevertheless, universal forms
that are proportional to $\tilde g(\omega,\text{T})$. 
\subsubsection{Comparisons on the neutrino diffraction with
   diffraction of 
classical waves through a hole}

{\bf 1  Inelastic channel}. 

The neutrino diffraction is the quantum mechanical phenomenon. The neutrino 
 produced in the weak decay of the pion is expressed with the many-body 
wave function composed of the pion, muon, and neutrino. Hence the
 probability to detect the neutrino  is computed with
this many-body wave function. Since this  neutrino  inside  the
 coherence length is very different from  
the free isolated neutrino, its  probability receives  the large finite-size
correction of the universal behavior. Its  magnitude, however, depends
on the wave packet size which is determined with the nucleus that 
the neutrino interacts with. So the finite-size correction is determined
by the many-body wave function and the out-going wave.
We should
note that quantum mechanical probability is determined with  the overlap 
of the in-coming  state with the out-going  state and depends on the
 both  states.

In a classical wave phenomenon, on the other hand, an intensity is 
determined with only the in-coming wave.   A magnitude of the 
in-coming wave is directly observed. 
Hence the finite-size correction and  the interference
pattern  are determined only by  the  in-coming
wave. Thus interference of the quantum mechanical wave is different
from that of the classical wave. 
 \\
{\bf 2 Pattern in longitudinal  direction}

The neutrino diffraction is a part of the finite-size correction that
results from the wave natures of the wave function at a finite time $t$.
They are generated by the states that are orthogonal to the states at $t
\rightarrow \infty$ hence  its
magnitude is positive semi-definite  and 
depends on the time interval
T. Hence the neutrino flux has the excess that decreases with the
distance in the  direction to the neutrino momentum and vanishes at the
infinite distance. 

The diffraction pattern of  light through a hole or the interference 
pattern of light  in a double slit experiment are different. The
intensity 
have modulations in the perpendicular direction to the wave vector. The 
interference term is a product of two waves of different phases and 
so oscillates.   
 Integrating the  intensity
 over the whole screen, the oscillating interference terms cancel and
 the total intensity  is constant. 

Thus  the diffraction pattern of the
 neutrino is very different from that of light.
\\
{\bf 3   $\omega_{\nu}$ is  Einstein minus  de Broglie  
frequencies  }

The pattern of the neutrino diffraction is determined with the angular
velocity $\omega_{\nu,diff}=\omega_{\nu,E}-\omega_{\nu,dB}$. Since $\omega_{\nu,E}$ and 
$\omega_{\nu,dB}$ are almost the same for the neutrino, they are almost
cancelled and $\omega_{\nu,diff}$ becomes extremely small and stable in $E_{\nu}$.

The interference pattern of the light on the screen of the double slit 
experiment, on the other hand,  
  is determined with the angular
velocity $\omega_{\gamma,dB}$. Since  $\omega_{\gamma,dB}$ is large and proportional
to the energy, the pattern varies rapidly with the position in the
screen and with the energy.   The diffraction
pattern of the light passing through the hole varies   rapidly.
\subsubsection{Muon in the pion decay}
In experiments of observing  the muon in the pion decays,  the neutrino
is not observed.
In this situation, the muon's diffraction  term has a magnitude that 
 is determined by  the ratio of the  mass and energy,
${m_{\mu}^2/(2E_{\mu})}$.
 \begin{figure}[t]
  \begin{center}
   \includegraphics[angle=-90,scale=.50]{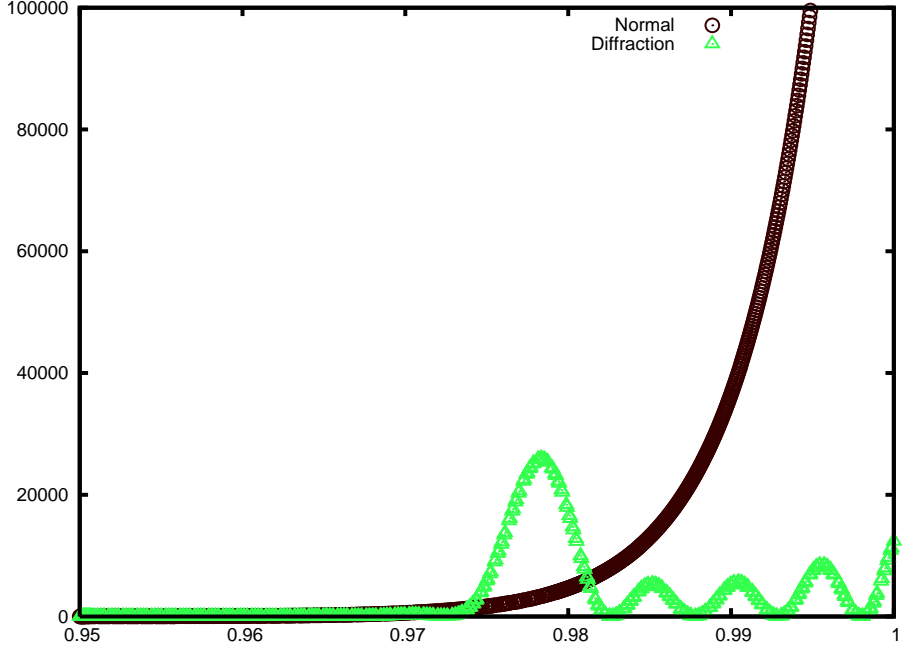}
  \end{center}
  \caption{The angle  dependence of the diffraction  
and  normal components of probability  is given. The large peak shows  the
  normal component and the small peaks at the tail of the previous peak 
 show   the diffraction component.  The horizontal axis 
shows the cosine of the angle between ${\vec p}_{\nu}$ and ${\vec
  p}_{\pi}-{\vec p}_{\mu}$ and the vertical axis shows the
  probability.   The  pion energy,  muon energy, and time interval 
  are  250~$m_{\nu}$,  210~$m_{\nu}$, and  30 $m_{\nu}^{-1}$.}
 \label{fig:omega}
\end{figure}%
Since  the muon mass is  larger than the neutrino mass
by $10^8$,  the value $m_{\mu}^2/(2E_{\mu})$ for the muon is much 
larger  than that of 
the neutrino by $10^{16}$. For
the muon of energy 1\,$[\text{GeV}]$, the length is of the order of $l_0=10^{-14}$\,[m].
This value is a microscopic size and $\tilde g(\text{T},\omega_{\mu})$ vanishes
at a macroscopic length. 
Hence the probability of detecting the muon at the
macroscopic distance 
becomes constant.  The
muon from the pion decay shows  neither the finite-size correction nor
the diffraction effect.  This  probability agrees
with the production
probability.  The muon and neutrino behave  differently at the finite distance.

If  the muon is observed under  a condition that the neutrino
is  detected at the finite $\text {T}$, $S[{\text{T}}]$ is applied and
the probability to detect the muon has the contribution from the
neutrino
diffraction. The diffraction component gives a wide energy spectrum for
the muon since that comes from the tail of the
distribution function.  Fig.\,${\ref{fig:omega}}$ shows   a 
probability integrated over  the neutrino energy in this condition 
that both the muon and neutrino are detected,  
which is obtained from Eq.\,$(\ref{integrated-amplitude-honbun})$. In this
figure, we use units $c=1,\hbar=1$  and express the energy and time with
the neutrino mass $m_{\nu}$. 
Energy of the pion is $250$ $m_{\nu}$ and the muon has the energy $210$
$m_{\nu}$ 
and has an angle with the pion of $\cos \theta =0.95-1$. The  cosine of 
the angle between ${\vec p}_{\pi}-{\vec p}_{\mu}$ and ${\vec p}_{\nu}$ 
is in the  horizontal axis. T is $30$  $m_{\nu}^{-1}$. The neutrino 
mass of an unphysical magnitude  of the order of MeV
 and the value of T are chosen in such manner  that the numerical 
computation of diffraction component is easily made. Qualitative
features  of  Fig.\,$\ref{fig:omega}$ are that there exist  a large peak at 
$\cos \theta \approx 1$ and small peaks at the tail region.
 The former is the peak from the root of $\omega=0$ at $\delta {\vec p} \approx 0$ and
$\delta E =0$ and the latter are the peaks  from  the roots of
$\omega=0$ of $\delta {\vec p} \neq 0$ and $\delta E \neq 0$. The latter
 peaks, which  do not exist in the
probability of detecting only the muon,  show the feature of the 
diffraction component of the probability when  the neutrino is
detected. Thus the  diffraction component is observed in the muon also 
when the neutrino is detected simultaneously.   Experimental 
verification of the diffraction term of this situation using the muon   
may be made in future.

As was shown in Section 3, the production rate is common to the muon and
neutrino, since they  are produced in the same decay process.  However the
rates measured by apparatus depend on the condition of the measurement.
If one particle is measured and other is un-measured, its rate for the
neutrino  receives the large  finite-size
correction, but the rate for the muon receives no correction. Hence they 
become different each other.  If both particles are measured
simultaneously, the rates for both are the same. 

Thus the neutrino
is in the non-asymptotic region of the finite  correlation in wide area,
and  the transition amplitude for  a neutrino that is observed at a
finite distance by a nucleus becomes different from that of the infinite
distance. The neutrino wave is 
 a superposition of those waves that are  produced at 
different positions and the probability  
gets  an additional  contribution and the  
probability  is modified by the diffraction term. The
overlap between the 
neutrino wave that is detected with a nucleus in a detector and those
that are produced  from  a pion decay shows the neutrino diffraction
of  unique properties.
So
the neutrino flux measured with  its collisions with a nucleus in targets are  
different from that defined from the norm of wave function. 
\section{Implications}
In this section, various physical quantities of neutrino processes which 
are modified by the  neutrino diffraction are studied. Particularly 
 neutrino nucleon total cross sections,  quasi-elastic cross sections,
electron-neutrino production anomaly, a proton target enhancement, and
an anomaly in atmospheric neutrino  are such processes that 
have significant contributions from the neutrino diffraction.

\subsection{Total cross sections of $\nu_{\mu}$-N scattering}

 Neutrino collisions with hadrons in high energy regions  are understood 
well with the  quark-parton model. A total cross section of a  high
energy neutrino is proportional to
the  energy and  is written in the form 
\begin{eqnarray}
\label{total-crossection}
& &\sigma^{\nu}={M_N E_{\nu}G_F^2 \over \pi }(Q+1/3 \bar Q),
\end{eqnarray}
using integrals of quark-parton distribution functions $q(x)$ and $\bar
q(x)$ and  $Q=\int_0^1 dx xq(x),\bar Q=\int_0^1 dx x\bar q(x)$. The cross
section is
proportional to the neutrino energy and a current value is
$\sigma_{\nu}/E= 0.67 \times 10^{-38}[\text{cm}^2/\text{GeV}]$. 

Now the rate of process of the neutrino  produced by a decay of a pion and 
interacting    with a nucleus has   a finite-size correction. It
modifies  the probability of the neutrino collision 
and  the  cross section. We estimate its effect
hereafter. Including 
the diffraction term, the effective neutrino flux becomes     
the sum of the normal and
diffraction terms  
\begin{eqnarray}
f=f_{normal}(1+r_{diff}),
\end{eqnarray}
where $r_{diff}$ is the rate of the diffraction component over the
normal component, and is a function of the combination  
$({m_{\nu}^2 \over 2cE_{\nu}}\text{L})$,
\begin{eqnarray}
r_{diff}=d_{0} \tilde g(\frac{m_{\nu}^2}{2cE_{\nu}}\text{L}),
\label{E-depedent-correction}
\end{eqnarray}
where  L is a length of the decay volume and the coefficient $d_0$
is determined from  geometries of experiments. 

When a detector is
located at the end of the decay volume, the correction
factor Eq.\,$(\ref{E-depedent-correction})$ is used. In actual case, the
detector is located in a distant region from the decay volume. There are 
material or soil between them and pions are stopped in beam dump. The neutrino
is produced in the decay region  and  propagates freely afterward. Since 
the wave packets of one $\sigma_{\nu}$ form the complete
set \cite{Ishikawa-Shimomura}, the wave packet of the size at the 
decay volume  is the $\sigma_{\nu}$
determined with  the detector.
The 
neutrino flux at the end of the decay volume is computed with the
diffraction term of the decay volume 's length L and the wave packet
size of the detector. 
Wave packets of this $\sigma_{\nu}$  propagate
 freely from the end of decay volume to the detector. The final value of 
neutrino flux  at the detector is found combining both effects.
When neutrino changes flavor  in
this period, the final probability for each flavor is written with  a
usual formula of flavor oscillation. 
 
The true  neutrino events   in experiment is converted to the cross
 section, $\sigma^{exp}(E)$, that includes 
 the diffraction component and is connected
with the cross section computed  with only the normal component $\sigma^{the}(E)$ by  
the rate   
\begin{eqnarray}
\sigma^{the}(E)
=\sigma^{exp}(E) { 1 \over1+ r_{diff} }.\nonumber 
\end{eqnarray}
Conversely the experimental  cross section is written as 
\begin{eqnarray}
\sigma^{exp}(E)/E= ( 1+ r_{diff} )(\sigma^{the}(E)/E).
\label{energy-dependece}
\end{eqnarray}
$\sigma^{the}(E)/E$ is  constant from Eq.\,($\ref{total-crossection}$) so  the E-dependence of 
$\sigma(E)^{exp}/E$ is due to E-dependence of $ r_{diff}$,
Eq.\,$(\ref{E-depedent-correction})$.  

The correction $ r_{diff}$ depends  on the geometry of 
experiments and the material of the detector.  
We compute $ r_{diff}$ using the experimental conditions 
of  MINOS \cite{excess-total-detectorMino}  and NOMAD \cite{excess-total-detectorNOMAD} and the total cross sections. 
\begin{figure}[t]
\includegraphics[scale=.3,angle=-90]{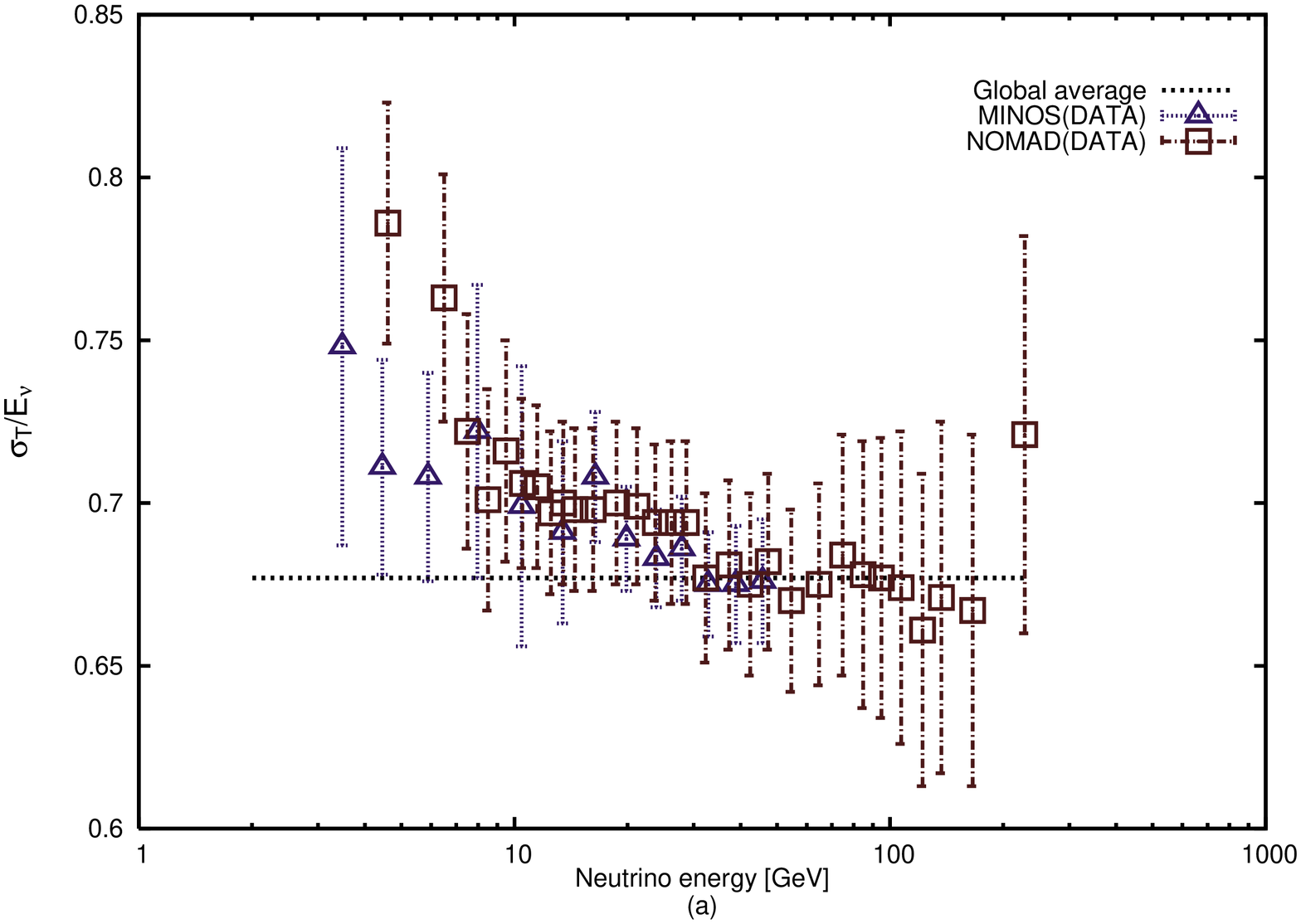}
\includegraphics[scale=.3,angle=-90]{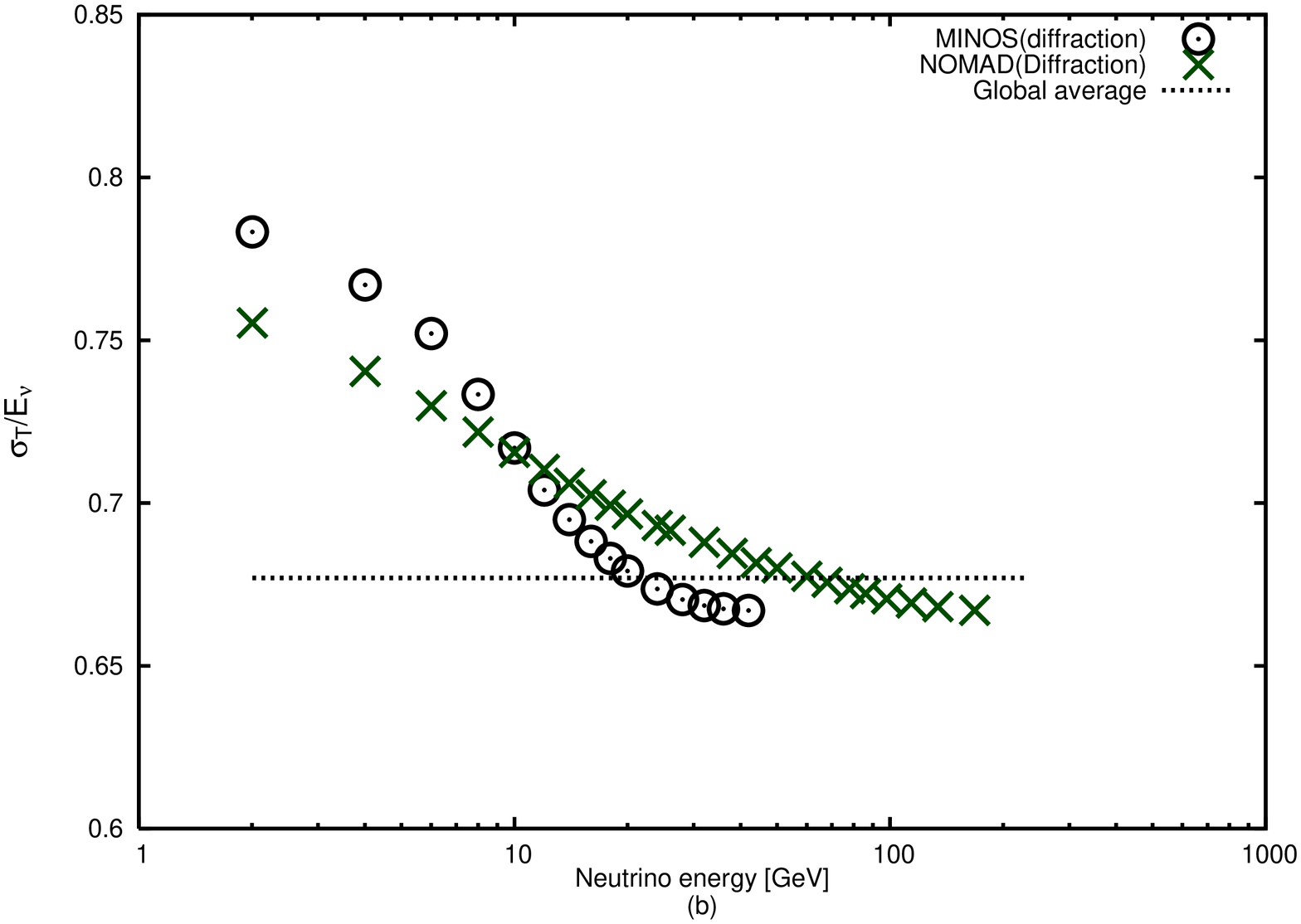}
\caption{Neutrino-Nucleon total cross section of MINOS and NOMAD  (a) and total cross sections of the sums of normal
 and diffraction terms in geometries  of MINOS and NOMAD (b) 
are given. The horizontal axis shows the neutrino energies  in [GeV] and the vertical
 axis shows the ratio of the cross section over the energy.}
\label{NOMAD and NOMAD:fig}
\end{figure}
The geometry of MINOS and  NOMAD are the following. The lengths between the
pion source and the neutrino detector, $\text{L}_{det-so}$, and those  
of the decay region, $\text{L}_{decay-reg}$, 
 are:
\begin{eqnarray}
NOMAD~&:&  \text{L}_{det-so}=835\,[\text{m}],\ \text{L}_{decay-reg}=290\,[\text{m}], \\
MINOS~&:&\text{L}_{det-so}=1040\,[\text{m}],\ \text{L}_{decay-reg}=675\,[\text{m}].
\end{eqnarray} 
Also  pion beam spreading was included from  
 angle of initial pion;   $0$ to $10\,[\text{mrad}]$ for NOMAD and
 $0$ to $15\,[\text{mrad}]$ for MINOS.

The wave packet size is estimated  with  the size of target nucleus. 
From the size of the nucleus of the mass number $A$, we have  
$\sigma_{\nu}= A^{\frac{2}{3}}/m_{\pi}^2$. 
For various material   the value are
\begin{eqnarray}
& &  \sigma_{\nu}= 5.2/m_{\pi}^2; {}^{12}C~ nucleus.\nonumber
\label{wave-packet-size}\\
& & \sigma_{\nu}= 14.6/m_{\pi}^2; {}^{56}Fe ~nucleus. 
\end{eqnarray}

Including the geometries, beam spreadings, and wave packet sizes, we
   computed the total cross sections and compared with the experiments
   in Fig.\,$\ref{NOMAD and NOMAD:fig}$. 
   These cross sections  computed theoretically  slowly 
decrease  with the energy in the geometry dependent manner 
   and agree with the experiments. Since the experimental parameters
   such as the neutrino energy and others are different in two
   experiments, the agreements of the theory with the experiments are
   highly non-trivial.
So the large cross sections at low energy regions may be attributed to 
   the diffraction component. 

We have compared only NOMAD and MINOS here. Many experiments are listed 
in particle data \cite{particle-data} and most of them have similar
energy dependences and agree qualitatively with the presence of the 
diffraction components. It is important to notice 
that the magnitude of diffraction component is sensitive to geometry.
Furthermore  if a kinematical constraint
Eq.\,($\ref{angle-energy-relation}$) on 
the angle between ${\vec p}_{\pi}$ and  ${\vec p}_{\nu}$ was required,
only the events of the normal term was selected. Then  the cross section 
should agree with that of the normal term.

\subsection{Quasi-elastic  cross sections}

Quasi-elastic or one pion production
processes are understood relatively well theoretically. The diffraction modifies 
the total events of these processes also.

The cross sections for  
\begin{eqnarray}
& &\nu+n \rightarrow \mu^{-}+p(+\pi^0),\\
& &\nu+p \rightarrow \mu^{-}+p+\pi^{+},\\
& & \bar \nu+p \rightarrow \mu^{+}+n(+\pi^0) ,
\end{eqnarray}
and the neutral current
 process  
\begin{eqnarray}
\nu+N \rightarrow \nu+N(+\pi^0),
\end{eqnarray}
are known well  using CVC, PCAC, and 
vector dominance and are studied recently by MiniBooNE \cite{excess-qenear-detectorMini}. The parameter is the axial vector meson $M_A$ and
higher mass contributions. So these cross section are used to study the 
diffraction terms.

\subsection{Electron neutrino anomaly}

In pion decays, a branching ratio  of an electron mode is smaller
than that of a muon mode  by 
factor $10^{-4}$ due to the helicity suppression of the decay of a
pseudo-scalar particle caused by the charged current
interaction.  
This behavior of the total rates  has been  confirmed by the 
observations of charged leptons.

Now the probability  to detect a neutrino  inside the  coherence length,
where the neutrino retains the wave natures,  is affected by the 
finite-size correction. Because this correction  comes from the states
that have different kinetic-energy from the initial value,  the
neutrino in this region  
does not follow the conservation law satisfied in the asymptotic 
region $t \rightarrow \infty$. The rate that electron neutrino  is
detected  is not suppressed. The  ratio of the probability to detect 
the electron neutrino  over that of the muon neutrino becomes  
 substantially larger in near-detector regions.

To compute  the transition probability and the spectra of electron 
and muon neutrinos, we  start from 
the $(V-A)\times(V-A)$ interaction Lagrangian
(Hamiltonian). The result of the probability is almost the same as 
that of Eq.\,$(\ref{weak-hamiltonian})$ in the muon mode but is different 
in the electron mode since the diffraction component does not satisfy 
the rigorous conservation of the kinetic-energy and momentum.
   In I, it was found that the  initial pion is described by  a wave
   packet of a large size. Hence the initial pion of the plane wave is
   studied here.  The 
amplitude $T$   is
written with the hadronic $V-A$ current and  Dirac spinors  in the form
\begin{align}
T = \int d^4xd{\vec k}_{\nu}
\,N\langle 0 |J_{V-A}^{\mu}(x)|\pi \rangle 
\bar{u}({\vec p}_l)\gamma_{\mu} (1 - \gamma_5)\nu({\vec k}_{\nu})\nonumber\\
\times e^{ip_l\cdot x + 
ik_\nu\cdot(x - \text{X}_\nu)
 -\frac{\sigma_{\nu}}{2}({\vec k}_{\nu}-{\vec p}_{\nu})^2},  
\end{align}
where 
$N=ig \left({\sigma_\nu/\pi}\right)^{\frac{4}{3}}\left({m_l m_{\nu}}/{
 E_l E_{\nu}}\right)^{\frac{1}{2}}$, and  the time $t$ is
 integrated in the region $\text{T}_{\pi} \leq t \leq \text {T}_{\nu}$. 
  The transition
probability 
to this final state is
written, after the spin summations are made, with the correlation
function and the neutrino wave function in the form 
 \begin{align}
&\int  \frac{d{\vec
 p}_l}{(2\pi)^3} \sum_{s_1,s_2}|T|^2 =   \frac{C}{E_\nu}\int d^4x_1 d^4x_2 
e^{-\frac{1}{2\sigma_\nu}\sum_i ({\vec x}_i-\vec{x}_i^{\,0})^2} \Delta_{\pi,l}(\delta x)
e^{i \phi(\delta x)},
\label{probability-correlation} 
\end{align}
where $C=g^2
\left({4\pi}/{\sigma_{\nu}}\right)^{\frac{3}{2}}V^{-1}$, $V$ is
a normalization volume for the initial pion, $\vec{x}_i^{\,0} = \vec{\text{X}}_{\nu} + {\vec
v}_\nu(t_i-\text{T}_{\nu})$, $\delta x
=x_1-x_2$, $\phi(\delta x)=p_{\nu}\!\cdot\!\delta x $
  and 
\begin{align}
\Delta_{\pi,l} (\delta x)=
 {\frac{1}{(2\pi)^3}}\int
 \frac{d {\vec p}_l}{E({\vec p}_l)}\left(2(p_{\pi}\cdot p_{\nu})( p_{\pi}\cdot p_l)-m_{\pi}^2 (p_l\cdot p_{\nu})\right)  
 e^{-i(p_{\pi}-p_l)\cdot\delta x }. 
\label{pi-mucorrelation}
\end{align}

The 
probability of detecting  a neutrino of $p_{\nu}$ at
${\vec X}_{\nu}$ and a lepton $l$ of arbitrary momentum is expressed 
as the sum of the normal term $G_0$ and the
diffraction term $\tilde g(\text{T},\omega_\nu)$, 
\begin{align}
P=N_2\int \frac{d^3 p_{\nu}}{(2\pi)^3}
\frac{p_{\pi}\! \cdot\! p_{\nu}(m_{\pi}^2-2p_{\pi}\! \cdot\! p_{\nu}) }{E_\nu}
 \left[\tilde g(\text{T},\omega_{\nu}) 
 +G_0 \right]
\label{probability-leptons}, 
\end{align}
where $N_2 = 8\text{T}g^2 \sigma_\nu$ and $\text{L} = c\text{T}$ is the
length of decay region. In $G_0$ the energy and momentum are conserved
approximately well and 
\begin{eqnarray}
p_l \approx p_{\pi}-p_{\nu},
\end{eqnarray}
is satisfied. Hence from a  square of the both hand sides, the mass shell
condition  
\begin{eqnarray}
m_l^2 \approx m_{\pi}^2-2  p_{\pi} \cdot p_{\nu},
\end{eqnarray}
is obtained. Thus the normal terms are proportional to the square of
lepton masses and the electron mode is suppressed \cite{Sakai-1949,Jack,Ruderman,Anderson}. 
 In $\tilde{g}(\text{T},\omega_{\nu})$, on the other hand, momenta satisfy
\begin{eqnarray}
p_l \neq p_{\pi}-p_{\nu},
\end{eqnarray}
and  the diffraction  terms are not proportional to the square of
lepton masses and the electron mode is not suppressed.

The total probability   of detecting   a neutrino or a charged lepton  in the 
pion decay at macroscopic distance  is written 
 in  the form, 
\begin{eqnarray}
P= P_{normal}+P_{diff}^{l}.
\label{probability-lepton}
\end{eqnarray}
In Eq.\,$(\ref{probability-lepton})$,  $P_{normal}$ is  the normal
term  that is obtained  from the decay probability $G_0$ in
Eq.\,$(\ref{probability-leptons})$ and $P_{diff}^l$ is  the diffraction  term that
is determined from $\tilde g$ in Eq.\,$(\ref{probability-leptons})$. 
The former probability agrees to that obtained using the plane waves and
the  latter one has not been  included before and its effect is
estimated here.
The diffraction term  at  $\text{T}$ is described with  its mass
 and energy   in the universal form
\begin{eqnarray}
& &P_{diff}^l=C\text {T} \tilde g(\text{T},\omega_l).
\label{diffraction} 
\end{eqnarray}
The frequency  $\omega_l$ are small for neutrinos
and large in charged leptons. 
$\tilde g(\text {T},\omega_l)$ is positive definite and decreases slowly with a 
distance $\text{L}$ and  vanishes at infinite  distance. Hence at $\text{L}=\infty$,
the probability agrees with  the normal component, 
\begin{eqnarray}
P= P_{normal}.
\label{probability-lepton-as}
\end{eqnarray}
The length   scale  
${2cE_l /m_l^2}$  is  macroscopic size 
in  neutrinos  but  is $10^{-10}$\,[\text{m}] or less for the electron and
muon. The magnitude of
 $\tilde g(\text{T},\omega_l)$ at the macroscopic distance is given in 
Fig.\,2 of I. At $\text{L}=100\,[\text{m}]$,
 $E=1\,[\text{GeV}]$ for the mass
 $1\,[\text{eV}/{c^2}]\,(\nu)$, $0.5\,[\text{MeV}/{c^2}]\,(e)$, and $100\,[\text{MeV}/{c^2}]\,(\mu)$, the
 values are, 
\begin{eqnarray}
& &\tilde g(\text{T},\omega_{\nu} ) \approx 3, \nonumber\\
& &\tilde g(\text{T},\omega_e ) \approx 0, \nonumber \\
& &\tilde g(\text{T},\omega_{\mu} ) \approx 0.
\end{eqnarray}
In this region, they satisfy
\begin{eqnarray}
\tilde g(\text{T},\omega_l ) \approx {m_{\nu}^2 \over m_l^2} \tilde g(\text{T},\omega_{\nu} ), 
\end{eqnarray}
hence  the diffraction component at a macroscopic distance is finite in 
the neutrino and vanishes in others. 
It
is striking that  the  probability to detect  the neutrino  has an
additional 
term and  is not equivalent to  that of the charged lepton even though they are
 produced in the same decay process.  

The diffraction term
   is generated by the  tiny neutrino mass  and the light-cone
   singularity. Hence the pattern is determined by $(E({\vec
   p}_{\nu})-c|\vec{p}_{\nu}|) t$ and becomes   
 long-range.  Because  $P_{diff}$   is the finite-size correction caused
by the neutrino  interference, it has  
different properties  from $P_{normal}$ in  flavor 
and  momentum dependences. The neutrino diffraction furthermore is  
sensitive to the absolute neutrino mass.   We study implications of the
diffraction term to  the electron mode in the pion decays here.
\begin{figure}[t]
\begin{center}
\includegraphics[scale=.45,angle=-90]{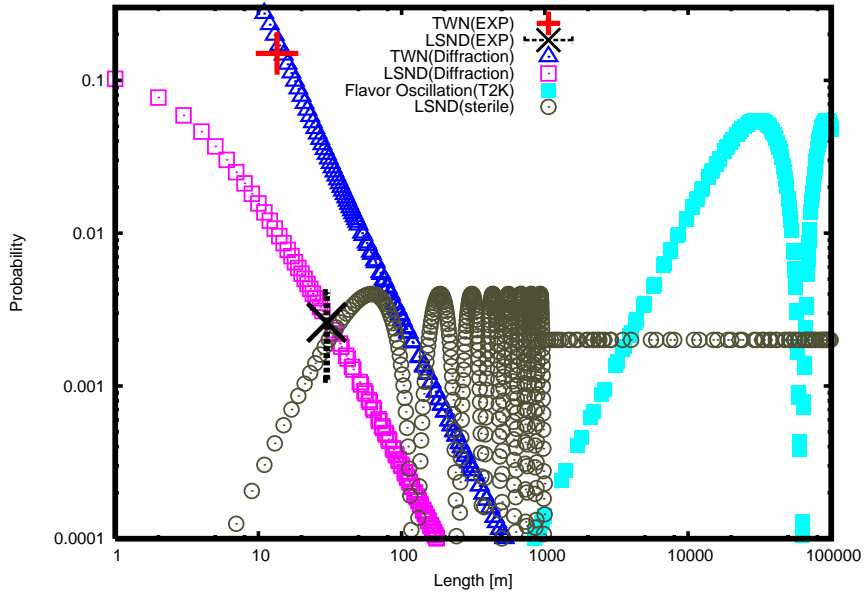}
\end{center}
\caption{Experiments of LSND and TWN are compared with the theoretical
  values of the diffraction terms. TWN(EXP) and  LSND(EXP) show the
  experimental values and  TWN(Diffraction) is computed with the
  parameters $m_\nu=0.2\,[\text{eV}/c^2]$,
	    $E_\nu=250[\text{MeV}]$, $P_\pi=2[\text{GeV}/c]$,
	    LSND(Diffraction) is computed with  $m_\nu=0.2\,[\text{eV}/c^2]$, $E_\nu=60\,[\text{MeV}]$,
	    $P_\pi=300\,[\text{MeV}/c]$. Flavor oscillation
  oscillation(T2K) shows the values for 
	    $\sin^2\theta_{13}=0.11$, $\delta m^2_{23} =
	    2.4\times10^{-3}\,[\text{eV}^2/c^4]$, $E_\nu = 60\,
	    [\text{MeV}]$, and LSND(sterile) shows with
  $\sin^2\theta=0.004$, $\delta
	    m^2 = 1.2\, [\text{eV}^2/c^4]$, $E_\nu = 60\,[\text{MeV}]$.}
\label{LSND :fig}
\end{figure}
In Fig.\,$\ref{LSND :fig}$, experiments of LSND \cite{excess-LSND} and the
two neutrino experiment( TWN) \cite{excess-two-neutrino} are compared with
the diffraction components and the flavor oscillations.  Theoretical
values are obtained including geometries of the experiments. Since those 
of LSND and TWN are different, the theoretical value for the LSND are
smaller than that for TWN. The experimental values plotted with crosses 
 agree with the theoretical values. The  values  from the flavor
 oscillations expected from  the current parameters are also shown. 
The mass-squared differences and mixing angles from the recent ground 
experiments lead negligible values for both experiments. A sterile 
neutrino of the  mass around $1$ $[{\text{eV}}/c^2]$ is necessary to fit 
the data of  LSND with the flavor oscillation. The agreements of 
the values from the neutrino diffraction in  LSND and TWN suggest that 
it is unnecessary to introduce  additional parameters.    

\begin{figure}[t]
\begin{center}
\includegraphics[scale=.45,angle=-90]{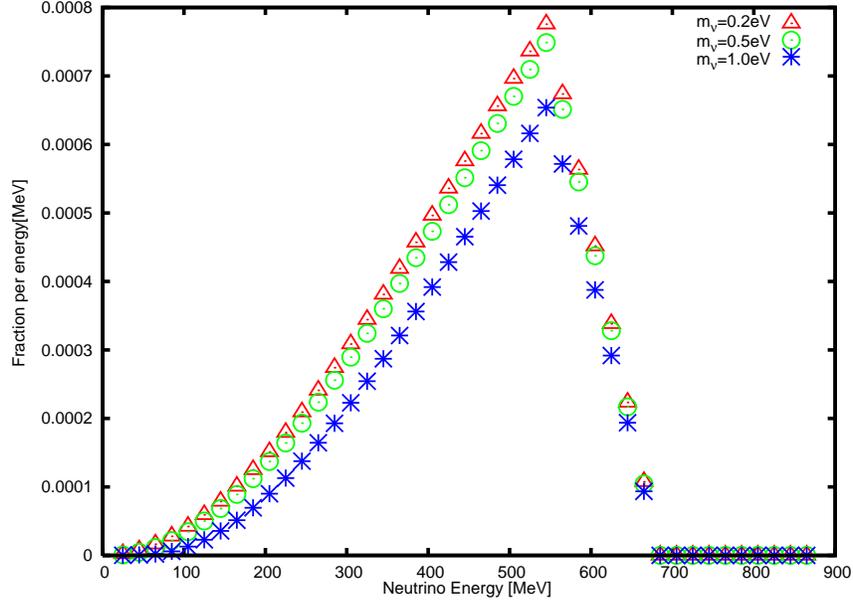}
\end{center}
\caption{Fraction of the electron neutrino of the mass 0.2
  $[\text{eV}/c^2]$, 0.5 $[\text{eV}/c^2]$ and 1$[\text{eV}/c^2
]$
  at L=110\,[m], distance=170\,[m] of T2K geometry and $P_\pi=2$ [GeV/$c$].}
\label{T2K-confuguration}
\end{figure}
In Fig.\,$\ref{T2K-confuguration}$, the maximum possible fraction of 
the electron neutrino in a geometry of T2K experiment is shown. The
 spreading of the pion beam is ignored in this Figure. Since the    
diffraction is sensitive to the pion beam spreading, the real value may
becomes smaller than this figure. 
\subsection{Proton target  anomaly}
Magnitude of  diffraction component depends upon the size of the nucleus
which neutrino interacts with and is expressed by the wave packet size
$\sigma_{\nu}$. It becomes larger with the larger target. It is known
and used in the text that nuclear size is proportional to $A^\frac{2}{3}$ and 
the large $A$ nuclear gives a large diffraction component, generally. Proton has a 
smallest intrinsic size. However    a proton is expressed by a wave function of 
its  position in matter. So  the wave packet size is determined by a size
of this wave function. Since a proton is the lightest nucleus, it has
the largest size. We estimate this size using center of mass gravity effect
between proton and electron.
For the proton's mass $m_{p}$ and the electron mass $m_{e}$, an  
electron's coordinate
${\vec x}_{electron}$ and the proton  coordinate ${\vec x}_{p}$
are expressed as,
\begin{eqnarray}
& & {\vec x}_{electron}={\vec X}+{m_{p} \over
 m_{e}+m_{p}}{\vec r}\approx {\vec X}+(1 +{1 \over 2000}){\vec r},\nonumber \\
& &{\vec x}_{p}={\vec X}-{m_{e} \over
 m_{e}+m_{p}}{\vec r}\approx{\vec X}-{1 \over 2000}{\vec r}.
\end{eqnarray}
If the wave function of the atom is 
\begin{eqnarray}
\Psi({\vec R}) \varphi({\vec r}),
\end{eqnarray}
and the function of the relative coordinate, $\varphi({\vec
r})$, is extended
by an amount $R_{atom}$ which is about $10^{-10} [\text{m}]$ then the proton 
is extended with a radius  
\begin{eqnarray}
R_{p}={1 \over 2000}R_{atom}\approx 5\times 10^{-14} \,[\text{m}].
\label{center-mass-gravity}
\end{eqnarray} 
This  value is much shorter than the atomic scale and is larger than one 
nucleon's size  $1\,[\text{fm}]=10^{-15}\,[\text{m}]$ by factor $50$. 
Thus proton  in  solid  is  extended to  the size ${1 \over
2000} $ of the atomic wave
  function, which is larger than  the nuclear size  of O. 
Hence proton  gives the important role in the neutrino diffraction. 
Its size may be  in the range 
\begin{eqnarray}
l_{proton}(U)=5\times 10^{-14}-10^{-13} \,[\text{m}].  
\label{proton-size}
\end{eqnarray}

An enhancement of  diffraction contribution due to  the proton 
is expected in 
\begin{eqnarray}
\bar \nu+p \rightarrow \mu^{+} +X^{0},\ X^{0}=n,\ p\pi^{-},\ n\pi^{0},\ others.
\end{eqnarray}

\subsection{Atmospheric neutrino}
The neutrino flavor oscillation was  found first with  an atmospheric
neutrino. Neutrinos are produced from decays of  charged pions and muons 
in secondary cosmic rays.  Since the matter density is low in
atmosphere, these 
charged particles travel freely long distance. Thus  neutrinos produced in 
decays of pions or muons   show the diffraction phenomenon and the 
diffraction components  are added to
the neutrino fluxes. These neutrino events  may be  observed  in 
detectors set in the ground, such as Super-KamiokaNDE(SK) if the absolute 
mass is a reasonable value. The minimum mass  allowed 
 from the mass-squared difference is about the value, 
$\sqrt {\delta m^2} \approx 10^{-2} \,[\text{eV}/c^2]$. Then  the length that the
diffraction component is observed becomes  
$\text{L}_0={2E_{\nu} c \over m_{\nu}^2 } \approx 20\,[\text{km}]$ for $E_{\nu}=1 \,[\text{GeV}] $, which is longer than
the height of troposphere. Hence the diffraction 
component  could  be observed  with  the angle-dependent excess of  the 
electron and muon neutrino fluxes. Since the diffraction components from 
pion decays are common to  both  neutrinos,
their ratio is not sensitive to the diffraction.  Instead of this ratio, 
a ratio of the neutrino flux to the flux of charged leptons is good to
see the signal of the neutrino diffraction. 

\section{ Behaviors of the probability suggest  violations  
at first sight but do not so in fact.}

\subsection{ Unitarity}
Probability of detecting the neutrino per time  $P(\text{L})$ decreases 
with the distance L. This behavior of  the probability  appears to  suggest   
that the probability is not preserved and is inconsistent with the
unitarity. However this behavior is derived from the  
 $S[\text T]$ that satisfies $S^{\dagger}[\text{T}]S[\text{T}]=1$ and is 
consistent with the unitarity.  The probability at L is determined with 
S-matrix $S[\text T],\ \text{L}=c\text{T}$ and 
 has two components
$P=P^{(normal)}+P^{diff}(\text{L})$. Both terms are positive semi-definite and 
the latter is decreasing with L because the constraint to the final
state from the energy conservation becomes more stringent with
increasing  L. This decreasing behavior is a natural consequence of
the unusual properties of the finite-size correction and  is consistent 
with the unitarity 
$S^{\dagger}[\text{T}]S[\text{T}]=1$. The unitarity leads that the life time
of the pion becomes larger if the neutrino is detected at a finite T.

\subsection{ Lepton number non-conservation}

The probability of the pion decay process in the situation where 
 the neutrino is detected has the large finite-size correction.  
The charged lepton shows the same behavior in the same situation. This  
will be confirmed if the charged lepton is measured simultaneously with
 the neutrino in
 experiments. This has not been done and is consistent with the lepton number
 conservation.

The probability of the pion decay process in another   situation
where the neutrino is un-detected has no finite-size correction. The
charged lepton in this process does not show the finite-size correction 
and its decay probability is computed with the standard calculation of 
using the plane waves. This situation has been studied well experimentally and
agrees with the theoretical calculations obtained with $S[\infty]$. 

Now the boundary conditions
of the above two cases are different. One  boundary condition  leads 
uniquely one consequence  and the different boundary conditions may lead
the different probabilities. Our results show that the different boundary
condition on the neutrino leads the different result on the decay
probability. Thus  the probability to detect the neutrino  in the first case
is different from that of the charged
lepton in the second
case. It is meaningless to compare the probability for neutrino  
in the first case  with that for the charged lepton in the
second case, because they follow the different boundary conditions.

Since the 
probability of detecting the neutrino at a finite distance deviates from that at
the infinity and the probability of producing the neutrino is defined with the value at
the infinite distance, they are different. 
The fact that the probability of detecting  the neutrino is different 
from that of the charged lepton 
does not mean the violation of the lepton number conservation, but means
that the  probabilities  depend  on 
the boundary condition. The
lepton number is conserved. 
 Thus the
different behavior of the  probability from that of the
production  is similar to that of the retarded electric potential of a moving
charged body.

\subsection{ Dependence on wave packet size}

It is known that the total probability at $\text{T}=\infty$ does not
depend on the
wave packet size \cite{Stodolsky}. The result of the present paper
Eq.\,$(\ref{total-probability-energy})$ in 
fact shows that  the  first  term in the right-hand side   is independent 
of the wave packet size. Now the second term in
Eq.\,$(\ref{total-probability-energy})$, which is the finite-size
correction,   is proportional to $\sigma_{\nu}$.   Since $S[\text T]$ is
determined with the wave packets, its boundary conditions are determined
with $\sigma_{\nu}$ and  the finite-size correction 
  depends on $\sigma_{\nu}$. That  increases with 
$\sigma_{\nu}$   and diverges at $\sigma_{\nu}=\infty$. The 
diverging correction  at $\sigma_{\nu}=\infty$ is consistent, in fact, 
with the fact that the total cross section diverges for
the plane waves \cite{Asahara}.  This occurs because the denominator of 
the neutrino propagator vanishes.
Nevertheless the finite-size correction   has the 
universal properties,   which  were  computed  
with $S[\text T]$ defined by the wave packets.

\section{Unusual features }

\subsection{Energy non-conservation and violation of symmetries} 

The S-matrix   at the finite-time
interval $S[\text T]$ does not commute with the free Hamiltonian $H_0$,
and satisfies 
 Eq.\,$(\ref{commutation-relation-S(T)  })$. In this region, one pion
 and decay products co-exist in a coherent manner that is determined by
 $H_1$. Hence $H_1$ has a finite expectation value and the kinetic 
energy defined by the free part $H_0$ is different from that of $H$. 
$H_0$ is not  conserved in  
$S[\text T]$, despite the fact that the total energy defined by $H$ is
conserved. The contribution to
the transition probability from these  states was computed 
in the text analytically.  They exhibit  the neutrino diffraction 
and give  the finite-size correction to the neutrino flux. 

The finite-size correction is not invariant under Lorenz transformation
either  and the magnitude of the diffraction term in the pion rest system is
  smaller than that in the high energy pion.   
\subsection{
Helicity suppression}
 
Decay rate of the pion to the electron mode is suppressed over that to 
the muon mode 
by the helicity suppression. The helicity suppression hold in a decay of
a pseudo-scalar particle to a neutrino and lepton caused by $V-A$ weak
interaction. Conservations of the kinetic energy  and the angular
momentum enforce vanishing of the amplitude  for the massless 
lepton. Now the kinetic energy is not conserved in the non-asymptotic 
region   and   the helicity suppression is not effective.
The probabilities to detect the neutrino in this region
are not suppressed.  Thus the   normal
terms hold these properties and  the electron mode is 
suppressed, whereas the   finite-size correction does not hold
these properties. 
  The electron
mode is not suppressed in the diffraction component and has the sizable 
magnitude in the spatial region where the normal term is negligible.

Thus when the neutrino is observed in near-detector region, the
electron neutrino is substantially enhanced.

\subsection{Large finite-size correction} 

The finite-size correction of the probability to detect the neutrino  
is finite in the spatial region  $\text{L} \leq l_0$ of the distance L between 
the pion and the neutrino,  where the interaction
energy is finite and the energy defined by the free Hamiltonian $H_0$
deviates from the conserved total-energy. The states of  these 
energies  of $H_0$  that are  different from that of the initial state
interact with $S[\text T]$ in a universal manner that is determined by
$H_1$ and gave the finite-size correction.  Particle spectrum 
at ultra-violet region is universal in  relativistic invariant systems  
and gives the  light-cone singularity to correlation functions. The
light-cone singularity is real and extended to a large area and gives 
the universal finite-size  correction to light particles. It
might be remarkable that the states of the ultraviolet region give the
observable effect to the probability of  the tree diagram.        
 
Since the  probability to detect  the neutrino has the large finite-size
 correction, the neutrino detection affects  strongly the pion 
 decay probability. 
The decay rate of the pion in the situation where 
the neutrino is detected is different from that where the neutrino is 
un-detected, because they are described by the amplitude of the
 different boundary conditions.  

The life time of the pion in 
which the produced neutrino is observed  becomes shorter  than the 
normal value. This phenomenon that the life time is modified by  its 
interaction with matters is known in  the literature as  quantum Zeno
 effect. Neutrinos actually interact extremely weakly 
with matters  and  a majority of  neutrinos are passing
 freely without any interaction  and are not  affected by this effect.  
Consequently  the majority of the pions are  not affected by  the  
finite-size  effect and its life
 time is not modified and has the normal life time. Although the detected
 neutrino receives the large finite-size correction, its effect is
 negligibly small for  observables of the pions.

\subsection{Interference in parallel direction to the momentum}

The diffraction term depends on the time interval $\text T$ and is positive
definite. Its pattern changes with the distance L extremely slowly. 
Thus the interference pattern is along the parallel direction to the
momentum.

\subsection{Dependence on the apparatus}

The neutrino diffraction appears in  the quantum mechanical transition
probability and  depends on both the initial  and final
states. The wave function of the final state is determined by   the 
apparatus and so are the scattering amplitude and probability.   
It is in fact quite reasonable  that the interference pattern depends on the 
apparatus in quantum mechanics. In classical physics, on the other
hand, any physical variables are observable and interference patterns do 
not depend on the apparatus. Hence the fact that interference pattern 
of the probability of the present work depends upon the apparatus and  
is quite different from  that of the classical physics is not surprising.  
\section{Summary and implications}

  We found that the neutrino in the pion decay has a large
 non-asymptotic region of having  wave natures and  the  probability 
to detect  the neutrino  in this region reveals  unusual property.   The 
neutrino   is unique 
and shows the large finite-size correction of the 
diffraction form. The unusual 
diffraction pattern is caused by 
the relativistic invariance and the tiny neutrino mass. 
 Its origin, mechanism, characteristic features, and implications 
are presented. 

The pion, neutrino, and charged lepton are described by a many-body wave
 function that follows the  Schr\"{o}dinger  equation. The wave function 
at a finite $t$ has various unusual
 properties that are different from those at $t \rightarrow
 \infty$. Especially the total kinetic energy is different from that of
 the initial state and takes a wide range of values. Hence the wave functions of 
these particles vary in  space and time. The physical quantities of these
 wave functions are computed by the S-matrix at a finite-time interval
 that satisfies the boundary condition at a finite time.
The  position-dependent probability to  detect neutrino, which  is
 proportional to the flux,  was computed with
 the  $S[\text{T}]$ expressed by wave packets and  
the  large finite-size correction  of  universal property that  is
 insensitive to  the pion's initial conditions was found.   The pion's mean 
 free path  estimated in Appendix of I shows that this is  long  enough for
 the new   term  to be observed in the experiments.
The  diffraction term reflects the finite-size correction,  and
  depends  on the  distance, energy, angle, and the absolute neutrino
 mass in the universal manner.

The wave function at a finite  $t$ has the continuous kinetic energy
that leads the space-time dependent wave nature, which is probed 
with the wave packet localized in space.   
Eq.\,$(\ref{integrated-amplitude-honbun})$  is derived in this manner
and shows a reason why the probability  has  the large finite-size
correction. 
 The
angular velocity $\omega$ is given by  the energy of the moving
frame $\omega=\delta E-{\vec
v}_{\nu}\cdot \delta{ \vec p}$,   consequently 
$\omega=0$ has the root  of $\delta E \neq 0$ and $\delta {\vec p} \neq 0$ in 
addition to that of  $\delta E = 0$ and $\delta {\vec p} = 0$.  The
former root exists  only
in  $S[\text T]$ and gives the finite-size correction. The probability 
around the former root receives also  the finite-size correction, as is 
shown also in Appendix B.  Since the kinetic energy and momentum are 
not conserved in
this root,  the finite-size correction has various unusual properties. 
 Thus the states which violate  the energy and momentum conservation  
give the correction. To compute the  dominant contribution in the 
long-distance region rigorously, it is convenient to interchange the
order of the integrations and introduce  the correlation function  
 $\Delta_{\pi,\mu}({\delta x})$. 
 The states of non-conserving the energy and momentum  form the 
light-cone singularity of $\Delta_{\pi,\mu}({\delta x})$ and 
give the most important contribution to the finite-size 
correction of the probability to detect the neutrino.
 The regular terms  of 
$\Delta_{\pi,\mu}({\delta x})$ give   the  
constant value of  the probability.

The light-cone singularity   is formed from the superposition of waves of the 
infinite momentum, which have always the light velocity and vanishing  
complex phase.  So when it is  combined with  the 
slow phase
 $\bar{\phi}_c$ of the neutrino wave function,
Eq.\,$(\ref{light-phase})$, this slow phase remains to physical
observable.

The slow phase of the neutrino wave packet is  the outcome of the
 cancellation between the time and space parts  of extremely light
 neutrino. The phase is 
determined by the difference of space-time coordinates $(\delta t,\delta
 {\vec x})$ and the
 central values of the energy and momentum,  as
${E({\vec p}\,)\delta t-{\vec p}\!\cdot\! \delta {\vec x}}$, where the energy $E({\vec p}\,)$ is
 given by $\sqrt {{\vec p}^{\, 2}+m^2}$. 
When the  difference of positions $\delta {\vec x}$ is 
moving with the light velocity in the parallel direction to the momentum ${\vec
p}$, ${\vec v}_c=c({\vec p}/|\vec{p}|)$, as $\delta {\vec x}={\vec v}_c \delta t$, then     
the total  phase becomes $E_{\nu}\delta t-{\vec p}_\nu\!\cdot\! \delta {\vec
 x}=(m_\nu^2/2E_{\nu}) \delta t$.
 The angular velocity becomes the small value $m_\nu^2/(2E_{\nu})$
 and makes the interference phenomenon long-range. 
The new term in the probability 
decreases   slowly with the distance in the universal manner determined
by the mass and energy of the neutrino as $m_{\nu}^2/(2E_{\nu})$. 
This  form is independent from the details of wave packet shape and 
parameters as far as the reality of the neutrino wave function $\tilde
w({\vec x})$ is satisfied, which is ensured from the invariance under
the time inversion.  The relative magnitude of this component is not
universal and depends upon  the size of wave packet. 
Based on the estimation   of the size, we found that the magnitude of 
the new universal term is sizable for the measurement.
Since the slope of  the diffraction component  is 
determined by 
the mass and energy of the neutrino, the 
absolute value of the neutrino mass would be found from the neutrino 
interference experiments.

The diffraction  components were  compared with several previous neutrino 
experiments in Section 7. 
First, the slight energy dependence of the total neutrino-nucleon cross
sections at high energy regions,
which is hard to understand in the standard theory,  was
shown to agree with the excess of the effective neutrino flux due to the 
diffraction.
 The excesses of neutrino events  will be observed in other reactions as
well at  macroscopic short distance regions. 
Theoretical calculations 
at distances of the order of a few hundred meters were 
computed and shown in
Figures of Section 6.  From these figures, the excesses are  not 
large but are sizable magnitudes. Hence  these excesses  shall 
be observed in these  distances. 
Actually fluxes measured in the near detectors of the long-baseline 
experiments of K2K \cite{excess-near-detectorK2K} and MiniBooNE
\cite{excess-near-detectorMini}  may show  excesses of about
$10-20$ percent of the Monte Carlo estimations. Monte Carlo estimations of
the fluxes are obtained using naive decay probabilities and do not have
the coherence effects we presented in the present work. So the excess of
these experiments may be related with the excesses due to
interferences. The excess is not clear in MINOS 
\cite{excess-near-detectorMino}. With more
statistics, qualitative analysis  might become  possible to test the new 
universal term on the neutrino  flux at the finite distance.   
 If the mass is in the range from $0.1\,[\text{eV}/c^2]$ to $2\,[\text{eV}/c^2]$, the 
near detectors  at T2K, MiniBooNE, MINOS and other experiments  might 
be able to measure these signatures. 

Second, the suppression of the electron  mode  in pion decay is modified
 in the probability to detect neutrinos. Since the energy-momentum 
conservation does not hold
in the $S[\text{T}]$,  the 
 helicity suppression mechanism does not work
in the diffraction component. 
So the electron mode is enhanced drastically.  
The theoretical value of fraction that includes the finite-size
 correction of the electron mode   was
compared with LSND and TWN, and agreements were obtained.  Further
confirmation of the diffraction component by observing the electron 
neutrino in pion decay will be made using modern version of LSND or similar
experiments. T2K near detector is a possible place for that. Third, anomalies  
in proton target, and atmospheric neutrino would supply also specific 
signature of the neutrino diffraction.   
  The neutrino diffraction is sensitive to the absolute neutrino
mass but is not so to other parameters such as pion and neutrino
energies. Hence the observations of the neutrino diffraction is easy.
The absolute neutrino mass  could be found with these experiments.

At the end, we summarize the reasons why the interference term of the 
long-distance behavior emerges in the pion decay and is computed with 
the wave packet representation. The
connection of the long-distance interference phenomenon of the neutrino with 
the Heisenberg's uncertainty relation is also addressed. 

  Relativistic invariance forces the particle's momentum unlimited and 
makes  the correlation function $\Delta_{\pi,\mu}(\delta t, \delta\vec{x})$
  have  a singularity near  $\lambda=0$,
which is extended to  large distance $|\delta {\vec x|} \rightarrow
\infty$. This is one of the  features of 
relativistic quantum fields in Minkowski space-time and is one reason 
why the long-range correlation emerged.  
For a non-relativistic system, on the other hand,
the same calculation for stationary states is
made by,
\begin{eqnarray}
\int d {\vec k} \langle {\vec x}_1| {\vec k} \rangle \langle {\vec k}|{\vec
x}_2 \rangle = \delta({\vec x}_1-{\vec x}_2),
\end{eqnarray}
and the only one point $\delta {\vec x}=0$ satisfies the condition.
 Long-range correlation is not generated. The  rotational invariant  three-dimensional space 
is compact but the Lorentz invariant  four-dimensional space is non-compact.   
So it is quite natural for the non-relativistic system not to have the 
long-range correlation that the relativistic system has. The light-cone 
singularity is the peculiar  property of the relativistic system. 

Heisenberg uncertainty relation is slightly modified in the wave along
the light cone. The 
neutrino wave function behaves at  large distance along the
light-cone region 
in the form
\begin{eqnarray}
& &\psi_{\nu}(t,{\vec x})=f{e^{i(E_{\nu}t-{\vec p}_{\nu}\cdot{\vec x})} \over x}= f{e^{i{m_{\nu}^2 \over
2E_{\nu}}t}\over ct},
\end{eqnarray}
where $f$ has no dependence on the distance $|\vec{x}|$. Consequently the
uncertainty relation between the energy width $\delta E$ and the time
interval $\delta t$ becomes
\begin{eqnarray}
\delta t \delta{m_{\nu}^2 \over 2E_{\nu}} =\delta t \delta E 
\times{1 \over 2} \left({m_{\nu} \over E_{\nu}}\right)^2 \approx \hbar.
\end{eqnarray}
The ratio $({m_{\nu}/E_{\nu}})^2$ is of the order of $10^{-18}$ 
and $\delta t$ becomes macroscopic  even if the energy width
$\delta E$ is  microscopic of the order of 100 [MeV]. For instance 
if the pion Compton wave length, $\lambda_{\pi}$, is used for the 
microscopic length, then $c \delta
t$ becomes 
\begin{eqnarray}
10^{18}\times \lambda_{\pi} \approx 10^3 \,[\text{m}],
\end{eqnarray} 
 which is about the distance between the pion source and the near 
detector in fact. So  interference effect of the present paper
appears in this distance and is observable using the apparatus of much
smaller size.

In the time interval ${\text T} \leq l_0/c$, the finite-size  correction is
not negligible.   The diffraction term has the finite value and  
contributes to the probability. Hence, the  probability to detect the
neutrino
 deviates from the production probability.    
  In another region  $l_0/c \leq {\text T}$, the neutrino behaves like a
  free isolated particle and is in asymptotic region. 
The diffraction term vanishes and the probability agrees with  the normal
term. The probability to detect the neutrino is computable with the
ordinary S-matrix and agrees with the production probability.
Wave packet
formalism is applicable to both of the asymptotic and non-asymptotic 
regions.

The  characteristic small phase of the neutrino wave function of the
angular velocity ${m_{\nu}^2 \over 2E_{\nu}}$ along the light
cone causes the diffraction phenomenon at the anomalously large area.
There would be similar  phenomena in other light particles or others
where  the scattering matrix of the finite-time interval  $S[\text{T}]$  
are important. Unique properties of $S[\text{T}]$ may give new insights to
those phenomena that are hard to understand with  $S[{\infty}]$.

In this paper we studied the amplitude and probability in the lowest
order in $G_F$ and ignored  higher-order effects such as pion life 
time, pion mean free path, and effects of electroweak gauge theory  
in studying the quantum effects. They do not modify the effect of the
tree diagrams studied in the present work. We will
study these problems  and other large scale physical phenomena 
of low energy neutrinos in subsequent papers.

    
\section*{Acknowledgements}
This work was partially supported by a 
Grant-in-Aid for Scientific Research ( Grant No. 24340043).
Authors  thank Dr. Kobayashi, Dr. Nishikawa, Dr. Nakaya,
and Dr. Maruyama for useful discussions on 
the near detector of T2K experiment, Dr. Asai,
Dr. Kobayashi, Dr. Kawamoto, Dr. Komamiya, Dr. Minowa, Dr. Mori, and Dr. Yamada
for useful discussions on interferences. 
\\
{}

\appendix
\def\thesection{Appendix \Alph{section}}
\def\thesubsection{\Alph{section}-\Roman{subsection}}


\section {Long-range correlation for general wave packets  }\label{App:OPE}

Non-Gaussian wave packets were studied in the general manner  and 
the universal behavior of the phase was obtained in the text. In this
appendix, the explicit forms of the wave packets are studied as concrete
examples.  It is re-confirmed
that the long-range component of the probability  at around  $t={2 \pi
E_\nu/m_\nu^2}$ becomes the universal form.

{\bf type 1}

One way to express the non-Gaussian wave packet is to multiply 
 Hermitian polynomials and to write the amplitude in the form 
\begin{eqnarray}
  \frac{N_{\nu}}{(2\pi)^{\frac{3}{2}}}\int d{\vec
 k}_{\nu}e^{-{\sigma_{\nu} \over 2}({\vec k}_{\nu}-{\vec
 p}_{\nu})^2}H_n(\sqrt{\sigma_{\nu}}({\vec k}_{\nu}-{\vec p}_{\nu}))
e^{i\left(E({\vec k}_{\nu})(t-T_{\nu})-{\vec
 k}_{\nu}\cdot({\vec x}-{\vec X}_{\nu})\right)}, 
\end{eqnarray}
where $H_n$ is assumed to be real  in order for the wave packets 
to preserve the time reversal symmetry  and an even function of ${\vec
k_{\nu}}-{\vec p}_{\nu}$ in order for the wave packets 
to preserve parity, as was shown in Appendix of I

For  the simplest case
\begin{eqnarray}
H_n= {\sigma_{\nu}}({\vec k}_{\nu}-{\vec p}_{\nu})^2, 
\label{quadratic-form}
\end{eqnarray}
the wave packet in the coordinate representation is
\begin{align}
\left({2\pi \over \sigma_{\nu}}\right)^{\frac{3}{2}} e^{i(E({\vec p}_\nu)(t-\text{T}_{\nu})-{\vec p}_\nu\cdot({\vec x}-{\vec
 X}_{\nu})) -{1 \over 2\sigma_{\nu}}({\vec x}-{\vec X}_{\nu}-{\vec v}_\nu(t-\text{T}_{\nu}))^2} \left(3-{1 \over \sigma_{\nu}}\{{\vec x}-{\vec X}_{\nu}-{\vec
		v}_\nu(t-\text{T}_{\nu})\}^2 \right),
\end{align}
and is substituted into  the integral
 Eq.\,({\ref{singular-correlation}}).
\begin{align}
\tilde J_{\delta({\lambda})}=N_{\nu}^2(\sigma_{\nu}\pi)^{\frac{5}{2}}{1 \over
 2r^0}e^{i(E-p_{\nu}) \delta t}[-\frac{13}{4}+\frac{9}{4\sigma_{\nu}}
(1-v_{\nu})^2(\delta t)^2+O(1-v_{\nu})^4(\delta t)^4].
\end{align}
Thus the phase factor has the same universal form as the Gaussian wave 
packet and the correction is determined by the negligible small 
parameter $
{1 \over \sigma_\nu}(1-v_\nu)^2(\delta t)^2=\left({1 \over E_{\nu} \sigma_\nu}\right)^2\left({m_{\nu}^2
 \over 2E_{\nu}}\delta t\right)^2$.

We have proved that  the correlation function of the non-Gaussian wave 
packet has the same slow phase and long-range term as the Gaussian 
wave packet and the small correction becomes negligible for the simplest case
Eq.\,$\ref{quadratic-form}$.  Hence for any polynomials $H_n$ that are
invariant under the time and space inversions,  the correlation function
has the same long-range term and small negligible corrections. 

{\bf type 2}

For another  non-Gaussian wave packet 
\begin{eqnarray}
\label{non-gaussian-alpha}
  \frac{N_{\nu}}{(2\pi)^{\frac{3}{2}}}\int d{\vec
 k}_{\nu}e^{-\alpha({\vec k}_{\nu})+i\left(E({\vec k}_{\nu})(t-\text{T}_{\nu})-{\vec
 k}_{\nu}\cdot({\vec x}-{\vec X}_{\nu})\right)},
\end{eqnarray}
we have the same result.

$\bf type~ 3$

In the type 1 and 2 the time reversal and parity symmetries  are assumed
for the wave packet shape. If these symmetries are not required, the
function $H_n$ or $\alpha$ has an imaginary part. In this case, the
correlation function has a correction term in  the order of $(1-v)(t_1-t_2)$
and  this term is expressed 
\begin{eqnarray}
(1-v_\nu)\delta t={1 \over E_\nu}{m_{\nu}^2 \over 2E_{\nu}}\delta t,
\end{eqnarray} 
hence the correction term vanishes at the high energy. With a suitable
parameter,  the universal form of the slowly decreasing component of the
probability of the present work may become observable even in arbitrary
system. The Lorentz invariant form of the energy dependent phase of the 
wave packet and the light-cone singularity of the pion and muon decay
vertex give this universal behavior.


\section{Origin of the large finite-size correction }
We study the reason why the finite-size correction are large for the
light particles from the amplitude.   Plane waves are assumed for the  
initial pion  and un-detected particle and the wave packet of
$\sigma_{\nu}$ is assumed for the neutrino in the final state. 
 Integrating 
the  coordinate ${\vec x}$ in Eq.\,$(\ref{amplitude})$, for
$\sigma_{\pi}=\infty$,  we have 
the amplitude,
\begin{align}
&T=Ce^{i\phi_0}\bar u(p_{\mu})\gamma_{\mu}(1-\gamma_5)u(p_{\nu})
 \langle 0|J_{\text V-A}^{\mu}(0)|\pi \rangle e^{-{\sigma_{\nu}
 \over 2}{\delta {\vec p}}^{\,2}}\int_0^{\text T}dt e^{-i \omega t},
\label{integrated-amplitude}
\nonumber \\
&\omega=\delta E-{\vec v}_{\nu}\cdot \delta {\vec p},
\, \delta E=E({\vec p}_{\pi})-E({\vec p}_{\mu})-E({\vec
 p}_{\nu}),\,\delta {\vec p}={\vec p}_{\pi}-{\vec p}_{\mu}-{\vec p}_{\nu} , 
\end{align}
where $C=(2\pi\sigma_{\nu})^{3/2}
 e^{-iE_{\nu}\text{T}_{\nu}}$.
The integration over $t$ leads  to   the expression  
\begin{align}
\label{direct-time-amplitude}
T=Ce^{i\phi_0}\bar u(p_{\mu})\gamma_{\mu}(1-\gamma_5)u(p_{\nu})  \langle 0|J_{\text V-A}^{\mu}(0)|\pi \rangle e^{-{\sigma_{\nu}
 \over 2}{\delta {\vec p}}^{\,2}} e^{i  {\omega \text{T}}/2} 
\left[2\frac{\sin (\omega \text{T}/2)}{\omega
 }\right] .
\end{align}
 The fraction in the bracket coincides with the delta function $2\pi
 \delta(\omega)$ of 
representing  the energy conservation at the limit $\text{T} \rightarrow \infty
$ and  its deviation  from the delta function at a finite T is
negligible, if its behavior
near the roots of ${\vec p}_{\mu}$ of $\omega=0$ is normal of having 
a finite derivative. In this situation, the finite-size correction
becomes negligible.   The finite-size corrections of the integral or the
average of probability 
\begin{eqnarray}
|T|^2=|C|^2| \bar u(p_{\mu})\gamma_{\mu}(1-\gamma_5)u(p_{\nu})  \langle
 0|J_{\text V-A}^{\mu}(0)|\pi \rangle |^2 e^{-\sigma_{\nu}{\delta {\vec
 p}}^{\,2}} \left[2{\sin (\omega \text{T}/2) \over \omega}\right]^2,
\end{eqnarray}
are  determined by the roots  of   $\omega=0$ and the behaviors of 
the $\omega$ around the roots.   

The normal root  satisfies 
\begin{eqnarray}
\delta {\vec p}=0,\ \delta E=0,
\end{eqnarray}
and  agrees to the solution of the plane waves.  $\omega$ is expressed
by the momentum of the pion and muon as
\begin{eqnarray}
\omega=E_{\pi}-E_{\mu}({\vec p}_{\mu})-E_{\nu}({\vec p}_{\pi}-{\vec p}_{\mu}),
\end{eqnarray}

The roots are on the ellipse of ${\vec q}={\vec p}_{\pi}-{\vec p}_{\mu}$
\begin{eqnarray}
& &{x^2 \over a^2}+{y^2 \over b^2}=1,\ 
x=  q_x-{1 \over 2}\left\{1-\left(\frac{m_{\mu}}{m_{\pi}}\right)^2p_{\pi}\right\} ,\ y=
  q_y \nonumber ,\\
& &a=\left(1-{m_{\mu} \over m_{\pi}}\right)p_{\pi},\  b= {1 \over 2}\left(1-{m_{\mu}
 \over m_{\pi}}\right)^2p_{\pi},
\end{eqnarray} 
where the direction of ${\vec p}_{\pi}$  is chosen to the x-axis. 
The derivative of the angular velocity with respect to the momentum 
is expressed with the velocities 
\begin{eqnarray}
{\partial \over \partial p_{\mu}^i} \omega=-{v}_{\nu}^i+{v}_{\mu}^{\,i},
\end{eqnarray}
which has the magnitude
\begin{eqnarray}
\sum_i \left({\partial \over \partial p_{\mu}^i} \omega\right)^2=(v_{\nu}-v_{\mu}\cos \theta)^2+v_{\mu}^2(1-\cos^2 \theta),
\end{eqnarray}
where $\vec{v}_{\mu}$, and $\vec{v}_{\nu}$ are velocities of the muon and
neutrino and  $\theta$ is the angle between ${\vec p}_{\pi}$ and ${\vec
p}_{\mu}$.
The $\omega $ varies from $\omega=0$ to $\omega({\vec p}+ds {\vec n}_{normal})$ in the
normal direction and the value is computed  as 
\begin{eqnarray}
\omega({\vec q}+ds{\vec n}_{normal})=ds \sqrt{{q_x^2 \over a^2}{1 \over
 a^2}(1 -{1 \over b^2})+{1 \over b^2} } \geq ds {1 \over b},\ 
{\partial \over \partial s}\omega  \geq  {1 \over b} .
\end{eqnarray}
Since the slope is not small,  the integral  of $|T|^2$   
along the normal direction with the parameter $s$ converges fast at 
the large T. Hence the ordinary 
prescription for the square of the fraction 
\begin{eqnarray}
\left(2\frac{\sin (\omega \text{T}/2)}{\omega
 }\right)^2=2\pi \text{T} \delta ({\omega}),
\label{largeT-dirac-formula}
\end{eqnarray}
is valid. The   finite-size correction vanishes.     

Next we study the roots of $\omega=0$ of $\delta {\vec p} \neq 0 $. The
$\omega $ is expressed in the following form,
\begin{eqnarray}
\omega=E_{\pi}-E({\vec p}_{\mu})-{\vec v}_{\nu}\cdot({\vec
 p}_{\mu}-{\vec p}_{\mu}) -{m_{\nu}^2 \over E_{\nu}}.
\end{eqnarray} 
The condition $\omega=0$ 
is fulfilled  by the momentum ${\vec q}={\vec p}_{\pi}-{\vec p}_{\mu}$
on the  ellipse 
\begin{align}
({ q_x}+{v_{\nu}(E_{\pi}-p_{\pi})} )^2(1-v_{\nu}^2)+\left({
 q_y-(p_{\pi})_y}\right)^2 =\frac{\left(v_{\nu}E_{\pi}-(p_{\pi})_x\right)^2}{
 1-v_{\nu}^2}-{(p_{\pi})_{x}}^2-m_{\pi}^2,
\end{align} 
where the x-axis is chosen in the direction of ${\vec p}_{\nu}$. This ellipse  has a large major axis that is inversely proportional to
$({E_{\nu} \over m_{\nu}})^2$, and a minor axis that is proportional to $({E_{\nu} \over m_{\nu}})$. 

On this curve, there exists a special    solution of, $\delta {\vec
p}=0,\delta E=0$, which has the finite derivative of   $\omega$ along the
normal direction  as was computed in the previous part. There exist
other roots  of $\delta {\vec
p}\neq 0,\delta E \neq 0$, away from the above  root, which  have  different
properties.   Because the ellipse is 
extremely long, the derivative  of $\omega$ along the normal direction 
in the second solution  becomes 
\begin{eqnarray}
{\partial \omega \over \partial s } = {m_{\nu}^2 \over E_{\mu}}.
\end{eqnarray}
 This value is   extremely small and   $|T|^2$  varies extremely slowly   
in the normal direction and the integral  over $s$ deviates from that obtained
from the infinite region by an amount that is determined by the  ${
m_{\nu}^2 \over E_{\nu}} \text T $.    
This gives the finite-size correction in the region $\text T \leq
{E_{\mu} \over m_{\nu}^2}$. The finite-size correction has the origin in
the root of $\omega=0$ at $\delta E \neq 0,\delta {\vec p} \neq 0$, hence
  does not hold   properties that are derived from the energy and
  momentum conservation.   Thus the finite-size correction is computed
uniquely with $S[\text T]$. 
Having tiny mass, the finite-size correction for the neutrino 
appears in the macroscopic area.

For the detection  of the muon, on the other hand, the muon  
velocity ${\vec v}_{\mu}$  replaces   the neutrino  velocity
${\vec v}_{\nu}$ in Eq.\,$(\ref{direct-time-amplitude})$
and  the ellipse for the muon obtained from $\omega'=0$ with
\begin{eqnarray}
\omega'=E_{\pi}-E_{\mu}({\vec p}_{\mu})-E_{\nu}({\vec p}_{\nu})-{\vec
 v}_{\mu}(\delta {\vec p}),
\end{eqnarray}
does not have large major axis. Hence  $\omega'$ has a finite slope on the
normal direction to the ellipse and varies steeply at the large T. The
integral of $|T|^2$ over $s$ converges fast and the   formula of
Eq.\,$(\ref{largeT-dirac-formula})$ can be  applied.

Thus the reason why there is the large finite-size correction in the
probability of detecting the neutrino was elucidated. Nevertheless, it is
not   easy to integrate the muon momentum and to obtain the finite-size 
correction  using the expression Eq.\,$(\ref{integrated-amplitude})$. 
Especially Lorentz invariance is not clear in this method.  
 Instead, the method  of 
integrating   the coordinates in  $|T|^2$ at the end 
 is suitable for a   
rigorous computation of the finite-size correction, which was applied in
the text. 


\begin{thebibliography}{}
\bibitem{J-J} The last form of the interaction Hamiltonian is used in
	the majority parts of the text. If the product of the $(V-A)$ currents 
for  the weak Hamiltonian is used, the
	magnitude of the diffraction term is slightly modified for the
	muon mode. For the electron mode, the $(V-A)\times(V-A)$ weak
	Hamiltonian  must be used because this includes the fluctuation 
correctly for the diffraction term. The diffraction  gives the dominant
	contribution to the electron mode. The detailed account  will be 
presented in
	Section 7.3.
\end{thebibliography}
\end{document}